\documentclass[12pt]{article}	
\usepackage{amsmath,amssymb,amsthm}   
\usepackage{amsfonts}
\usepackage{mathtools}
\usepackage{breqn}              
\usepackage[export]{adjustbox}
\usepackage{algorithm}
\usepackage{algpseudocode}
\usepackage{physics}

\usepackage{comment}
\usepackage{cite}
\usepackage{bm}
\usepackage{graphicx}
\usepackage{epstopdf}
\usepackage{subcaption}
\usepackage[a4paper, total={6.5in, 9in}]{geometry}
\usepackage{authblk}
\usepackage{color,soul}
\usepackage[english]{babel}
\usepackage{stackengine}
\usepackage{xcolor}  
\usepackage{tikz}
\usetikzlibrary{shapes,arrows}
\usepackage[makeroom]{cancel} 

\usepackage{array}
\usepackage{booktabs}
\usepackage{multirow}
\usepackage{colortbl} 

\usepackage{bigints}
\usepackage[title,titletoc,toc]{appendix}
\usepackage{etoolbox}
\appto\appendices{\counterwithin{equation}{section}}

\usepackage{hyperref}
\hypersetup{
    colorlinks=true,
    linkcolor=blue,
    filecolor=blue,      
    urlcolor=blue,
    citecolor=blue
}

\newcommand{\eqnref}[1]{\textcolor{blue}{(\ref{#1})}}

\newcommand{\bs}{\boldsymbol}
\newcommand{\tb}{\textbf}

\newcommand\groupequation[2][30pt]{%
  \setbox0=\hbox{$\displaystyle#2$}%
  \stackengine{0pt}{\copy0}{%
    \makebox[\linewidth]{\hfill$\left.\rule{0pt}{\ht0}\right\}$\kern#1}}
    {O}{c}{F}{T}{L}
}

\newcommand{\bu}{\bs{u}}
\newcommand{\bxi}{\bs{\xi}}
\newcommand{\bumeas}{\tilde{\bs{u}}}

\newcommand{\umeas}{\tilde{u}}
\newcommand{\Fmeas}{\tilde{F}}
\newcommand{\bx}{\bs{x}}

\newcommand{\bw}{\bs{w}}
\newcommand{\bp}{\bs{p}}

\newcommand{\udiff}{\left\{u^n_i - \umeas^n_i \right\}}
\newcommand{\mass}{\left[ M \right]}
\newcommand{\bzero}{\bs{0}}

\newcommand{\nsd}{n_{sd}}

\newcommand{\bh}{\bs{h}}
\newcommand{\bn}{\bs{n}}
\newcommand{\bSigma}{\bs{\Sigma}}

\newcommand{\Gammabh}{\Gamma_{\bh}}

\newcommand{\vfvec}{\hat{v}}
\newcommand{\bvfvec}{\hat{\bs{v}}}

\newcommand{\bC}{\bs{C}}
\newcommand{\bzeta}{\bar{\bs{\zeta}}^e}
\newcommand{\nbzeta}{\bar{\zeta}^e}
\newcommand{\Ie}{\bar{I}^e} 
\newcommand{\bebar}{\bar{\bs{b}}^e}
\newcommand{\F}{\bs{F}}
\newcommand{\f}{\bs{f}}

\begin{document}

\title{A comparative study of calibration techniques for finite strain elastoplasticity: Numerically-exact sensitivities for FEMU and VFM}

\author[1]{Sanjeev Kumar  \thanks{sanjeev.kumar@austin.utexas.edu}}
\author[2]{D. Thomas Seidl \thanks{dtseidl@sandia.gov}}
\author[2]{Brian N. Granzow \thanks{bngranz@sandia.gov}}
\author[1]{Jin Yang  \thanks{jin.yang@austin.utexas.edu}}
\author[1]{Jan N. Fuhg   \thanks{jan.fuhg@utexas.edu (corresponding author)}}
\affil[1]{Aerospace Engineering and Engineering Mechanics, The University of Texas at Austin, TX 78712, USA}
\affil[2]{Sandia National Laboratories, Albuquerque, New Mexico, NM 87185, USA}

\date{}
\maketitle
\begin{abstract}
Accurate identification of material parameters is crucial for predictive modeling in computational mechanics. The two primary approaches in the experimental mechanics' community for calibration from full-field digital image correlation data are known as finite element model updating (FEMU) and the virtual fields method (VFM). In VFM, the objective function is a squared mismatch between internal and external virtual work or power. In FEMU, the objective function quantifies the weighted mismatch between model predictions and corresponding experimentally measured quantities of interest. It is minimized by iteratively updating the parameters of an FE model. While FEMU is seen as more flexible, VFM is commonly used instead of FEMU due to its considerably greater computational expense. However, comparisons between the two methods usually involve approximations of gradients or sensitivities with finite difference schemes, thereby making direct assessments difficult. Hence, in this study, we rigorously compare VFM and FEMU in the context of numerically-exact sensitivities obtained through local sensitivity analyses and the application of automatic differentiation software. To this end, both methods are tested on a finite strain elastoplasticity model. We conduct a series of test cases to assess both methods' robustness under practical challenges.
\end{abstract}

\tb{Key Words:}  finite element model updating, virtual fields method, finite strain elastoplasticity, automatic differentiation, forward sensitivities, adjoint, PDE-constrained optimization

\section{Introduction}
\label{sec:Intro}

In recent decades, numerous constitutive models have been developed to effectively capture both elastic and dissipative deformations, enabling predictions of material responses under various loading conditions in many scenarios of interest in science and engineering. The parameters in the constitutive relations differ for each material, which means that a model's effectiveness in accurately replicating a material's realistic behavior essentially hinges on the quality of these parameters. Accurate calibration of these model parameters is, therefore, crucial \cite{rossi2022testing}. The traditional method for identifying model parameters in elastoplastic deformation relies on a simple uniaxial test, which assumes that stress and strain are uniaxial and homogeneous in the gauge section before necking occurs. Although these assumptions simplify the calibration process, they can be impractical, particularly for complex constitutive models, as they tend to increase the number of tests needed to gather sufficient data for an accurate calibration \cite{bruschi2014testing}. Additionally, the uniaxial stress-strain assumption may become invalid once necking begins. Full-field measurements offer a way to move past the limitations of classical calibration tests by leveraging heterogeneous data such as displacement, temperature, and strain \cite{hild2006digital, sutton2009image}. 

Within the experimental mechanics community, two key approaches have emerged for calibrating model parameters using full-field digital image correlation data: Finite Element Model Updating (FEMU) and the Virtual Fields Method (VFM) \cite{pierron2021towards}. Although both methods aim to correlate experimental observations with model predictions, they differ significantly in their formulation, numerical implementation, and computational demands.

 In FEMU, or PDE-constrained optimization, an objective function is minimized by iteratively updating the parameters to reduce the weighted discrepancy between model predictions and experimentally measured quantities of interest \cite{cottin1984parameter, pagnacco2005inverse, seidl2022calibration}. Full finite element (FE) simulations are needed in the forward problem in FEMU. Quantities of interest, such as strain, displacement, forces, and/or moments are computed from the FE solution and compared with corresponding experimental measurements. For more details, see the recent review article \cite{chen2024finite}.

While FEMU can be highly effective and versatile, it is dependent on solutions of costly forward simulations.
The forward boundary value problem (BVP) requires a complete specification of boundary conditions (BCs) to be well-posed. Practically, a mix of traction-free and displacement BCs from the DIC data are imposed. Lastly, the objective function typically contains both deformation (i.e.\ displacement or strain) and load-matching terms, and balancing the two is often not straightforward in real applications.

VFM, proposed in Ref. \cite{grediac1989principe}, has emerged as a noteworthy inverse solution technique \cite{grediac2006virtual} and has attracted the attention of researchers from various disciplines due to its computational efficiency \cite{jiang2020identifying,  chalal2006experimental, grediac2006applying, pierron2000identification, toussaint2006virtual, berry2014identification, rahmani2014situ, rossi2016application, fu2022vfm, martins2019calibration, valeri2017determining, notta2015innovative, zhang2017verification, notta2013identification}. VFM provides a direct approach to solving inverse problems when the equilibrium equation is a linear function of the material parameters. By avoiding the solution of a partial differential equation (PDE), this method greatly reduces the computational time. However, for nonlinear cases, the process requires an iterative solution through an optimization algorithm. In particular, VFM utilizes full-field strain or displacement data to extract model parameters by minimizing the difference between the internal and external virtual work or power of the entire deformation process \cite{pierron2012virtual}. By incorporating the constitutive equations and choosing specific virtual fields, an equation can be derived in which only ``hidden'' internal state variables and constitutive parameters are unknown \cite{promma2009application, jones2018parameter}. However, the choice of virtual fields has been demonstrated to have a sizable impact on the performance of the VFM \cite{grediac2004numerical}, particularly in nonlinear cases and with noisy data \cite{prates2016inverse}.

Unlike FEMU, the VFM does not require the solution of large-scale linear systems, and therefore it has a considerably lower computational cost. This benefit come at the expense of requiring full-field experimental data across the entire domain, a requirement that is not essential for FEMU. Moreover, VFM results appear to be sensitive to noisy data. To minimize the effect of noise and stabilize the VFM inversion procedure, Avril et al.  \cite{avril2004sensitivity} introduced a stiffness-based virtual field for anisotropic elasticity and later expanded it to small-strain elastoplastic deformation for cyclic loading \cite{pierron2010extension}. However, implementing stiffness-based virtual fields can be cumbersome due to manual calculation of the elastoplastic tangent matrix. Marek et al. \cite{marek2017sensitivity} proposed the concept of sensitivity-based virtual fields to reduce the influence of noise on parameter identification. This approach assigns greater weight to locations on the specimen where more information about a parameter is encoded. Two noise filtering approaches, namely the Gaussian filter and the median filter, are employed to smooth the displacement for VFM in \cite{mei2021introducing}.

In this article, we present a comparative study of FEMU and VFM for calibrating finite deformation elastoplastic constitutive model parameters from full-field deformation data.
Similar studies exist, see e.g., 
\cite{dirisamer2017comparison,martins2018identification}, however, these comparisons rely on finite difference (FD) approximations for computing objective function gradients, which apart from being an approximation has the notable drawback of requiring many solutions of the forward problem for a single estimate of the gradient. Since the number of necessary forward calls also scales with the parameter space dimension, the required computational resources to solve these inverse problems will also increase significantly.
To avoid these issues, we rely on the work of Seidl and Granzow \cite{seidl2022calibration}
and employ a deterministic optimization framework, where the gradient of the objective function is \textbf{numerically-exact} (i.e.\ exact as possible subject to machine precision, solver tolerances, and round-off errors) using local sensitivity analyses paired with automatic differentiation, which can dramatically reduce the cost of solving the inverse problem relative to using
FDs. In particular, Seidl and Granzow investigated the use of forward and adjoint sensitivities-based algorithms for computing the gradient of finite-strain plasticity calibration problems from full-field data using FEMU.
To provide a fair comparison between FEMU and VFM, we are extending their work by applying forward and adjoint sensitivities to also compute the gradient in VFM. 
Therefore, we believe that, for the first time, we can provide a fair assessment of (i) the accuracy and robustness, and (ii) the computational cost of both methods. We perform the the following six experiments (\textbf{E1} - \textbf{E6}) on synthetic data:
\begin{itemize}
    \item[] \textbf{E1}:  The accuracy of solving the inverse problem for FEMU and VFM and their respective computation time on \textbf{noiseless} data.
    \item[] \textbf{E2}:  The sensitivity of both methods to the initial guess of the optimization algorithm.
    \item[] \textbf{E3}: The accuracy of solving the inverse problem in the case of increasingly \textbf{noisy} data in both the measured displacement field as well as in the load measurement.
    \item[]  \textbf{E4}: The dependence of both methods on model form error due to misspecification in the form of the hardening law.
    \item[] \textbf{E5}: The dependence of both methods on model form error due to a mismatch in the discretization error in the synthetic data and computational models, which serves as a proxy for the mismatch in resolution between DIC measurements and computational meshes. 
    \item[]  \textbf{E6}: The dependence of both methods on model form error due to the assumption of plane stress in cases where the thickness of the specimen becomes exceedingly large.
\end{itemize}

We believe that these six points cover a wide range of scenarios that have an influence on the calibration accuracy of finite strain elastoplastic constitutive parameters from full-field data.
We have deliberately decided to use synthetic data in this comparison to have access to a ``ground truth'' parameter set.
We further remark that, even though limited to a particular constitutive behavior, we believe that our analysis in general extends to other elastic and inelastic material models. An implementation of both FEMU and VFM including their numerically-exact gradients with adjoint sensitivities and all the following studies are available in a fork of \texttt{CALIBR8} \cite{calibr8} at \url{https://github.com/sanjais37/calibr8}.

The rest of the article is organized as follows. The variational formulation of the equilibrium equation that provides the foundation for both FEMU and VFM is discussed in section \ref{sec:variational_problem}. Section \ref{sec:inverse_formulations} presents the inverse formulations, which includes the definition of objective functions, the derivation of sensitivity matrices for forward sensitivity (FS) VFM and the system of adjoint equations for adjoint-based VFM. Finally, in section \ref{sec:results} FEMU and VFM are compared using the six studies established above, followed by a discussion in section \ref{sec:discussion}, before concluding the article in section \ref{sec:conclusion}.

 \begin{figure}
\includegraphics[width=0.70\linewidth,center]{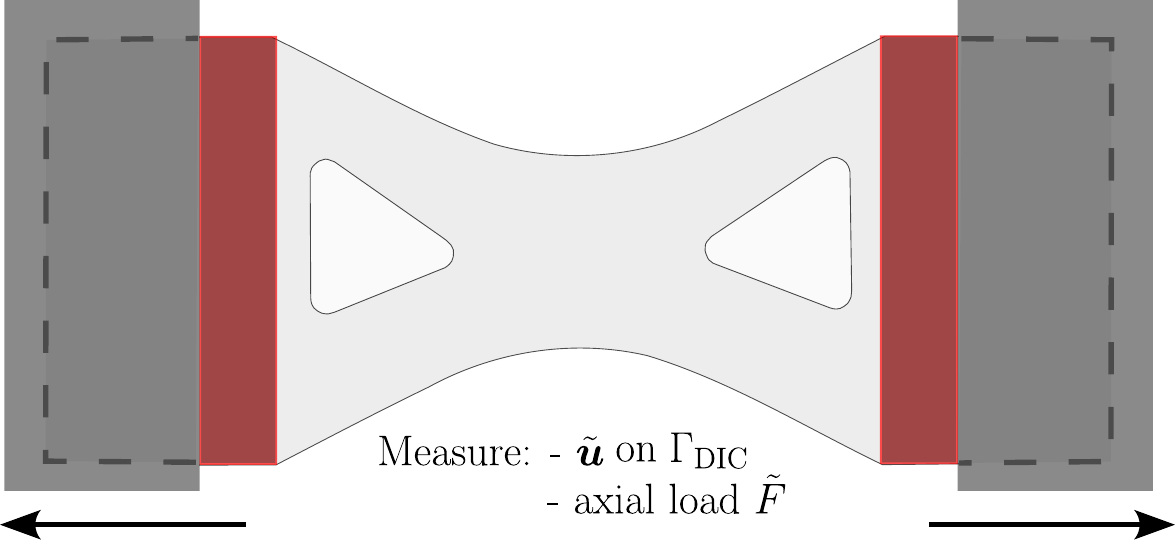}
 \caption{A prototypical characterization experiment. The test sample sits in machine grips (dark gray) and is pulled to failure. Only the light gray part of the visible region is modeled. Digital image correlation provides full-field displacement data over the entirety of this region, and a load cell measures axial load.}
 \label{fig:schematic_full_specimen_with_grips}
 \end{figure}

 \begin{figure}
\includegraphics[width=0.65\linewidth,center]{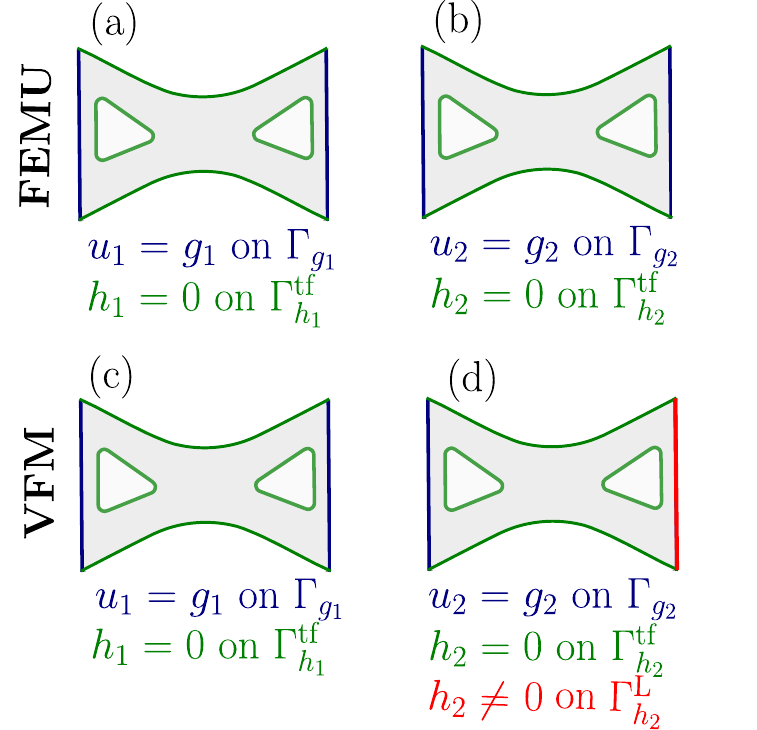}
 \caption{BCs for the modeled geometry in Figure~\ref{fig:schematic_full_specimen_with_grips}. The symbols $\Gamma_{g_i}$, $\Gamma^{\text{tf}}_{h_i}$, and $\Gamma^{\text{L}}_{h_i}$ denote portions of the boundary over which displacement, traction-free, and non-zero traction values are prescribed. Notably, FEMU and VFM necessitate different BCs, as $h_2(\bx)$ is generally unknown, but in VFM its integral is set equal to the measured axial load $\Fmeas$. The values for the displacement BCs come directly from the measured field $\bumeas$.}
 \label{fig:schematic_BC_FEMU_VFM}
 \end{figure}

\section{Variational problem}
\label{sec:variational_problem}

The variational or weak formulation of the equilibrium equation, together with the constitutive relation(s) for the material(s) in the body, forms the basis for both the FEMU and VFM approaches to solving the optimization-based calibration problem. In FEMU, the discretized forms of these equations serve as the constraint (i.e.\ they are enforced exactly). Experimental quantities like displacement and load are predicted and matched in a least-squares objective function.
On the other hand, in VFM the measured displacement field is directly inserted into the variational formulation and a scalarized (through the application of virtual fields) version of the weak form is minimized. For clarity, we will ignore time dependence in this presentation and focus on a single deformed configuration, as the generalization is straightforward.

We write these equations generically using a stress tensor $\bSigma$, which can be either the Cauchy or first Piola-Kirchhoff stress tensor under infinitesimal or finite strain assumptions, respectively, in 3D, or specialized forms of these tensors in plane stress. Generally, stress depends on both a history of the deformation and the constitutive or material model parameters $\bp$. We assume that the displacement \textit{global state variable} $\bu$ and the constitutive model-specific internal or \textit{local state variables} $\bxi$ are sufficient. Details on the specific form of the plane stress specialization of the equilibrium PDE, constitutive model, and evolution equations for $\bxi$ employed in our numerical results are summarized in Appendix~\ref{app:residuals}.

The starting point in our derivation of the weak form
is the equilibrium equation, which may be obtained from the conservation of linear momentum and application of the quasi-static assumption which justifies neglecting the inertial term:

\begin{equation}
    \nabla \cdot \bSigma(\bu, \bxi; \bp) = \bzero \quad \text{in} \ \Omega \subset \mathbb{R}^{\nsd}.
    \label{eq:strong_form}
\end{equation}
\noindent where $\nsd$ denotes the number of spatial dimensions, which will be either two (plane stress) or three in this work. We use the indices $i$ and $j$ for spatial components, which are understood to vary from $1, \ldots, n_{sd}$. In the derivations that follow, we employ the Einstein summation convention unless explicitly stated otherwise.

On its own, the equilibrium equation does not lead to a well-posed problem for the displacement field given $\bxi$ and $\bp$. Displacement and traction BCs 

\begin{align}
u_i &= g_i \ \text{on} \ \Gamma_{g_i}, 
\label{eq:disp_bcs} \\
\Sigma_{ij} n_j &= h_i \ \text{on} \ \Gamma_{h_i},
\label{eq:trac_bcs}
\end{align}
\noindent are required and the conditions $\Gamma = \overline{\Gamma_{g_i} \cup \Gamma_{h_i}}$ and $\Gamma_{g_i} \cap \Gamma_{h_i} = \varnothing$ must be met for a complete specification. It is important to emphasize that there exist multiple, potentially infinite complete combinations of displacement and traction BCs that will yield the same displacement solution, and we will show that FEMU and VFM typically make different choices. The \textit{essential boundary conditions} in Eq. \eqref{eq:disp_bcs} motivate the function spaces

\begin{align}
\mathcal{S} &= \{u_i \in \mathcal{H}^1(\Omega)\ \text{and} \ u_i = g_i \ \text{on} \ \Gamma_{g_i}\}, \\
\mathcal{V} &= \{w_i \in \mathcal{H}^1(\Omega) \ \text{and} \
w_i = 0 \ \text{on} \ \Gamma_{g_i} \}
\label{eq:fspace_v0},
\end{align}

\noindent where $H^1(\Omega)$ is the Sobolev space of square-integrable functions that also possess square-integrable first derivatives.

We obtain the weak form by multiplying Eq. \eqref{eq:strong_form} with a weighting function or variation $\bw \in \mathcal{V}$, and perform integration by parts to obtain

\begin{equation}
\left(\nabla \bw, \bSigma \right) - \left\langle \bw, \bSigma \cdot \bn \right\rangle_{\Gamma} = 0, 
\label{eq:weighted_residual}
\end{equation}

\noindent where $(\cdot, \cdot)$ indicates an inner product over $\Omega$, $\langle \cdot, \cdot \rangle_\Gamma$ is a duality pairing over $\Gamma := \partial \Omega$, and $\bn$ is the outward unit normal on $\Gamma$. The boundary term may be simplified through the application of Eqs. \eqref{eq:trac_bcs} and \eqref{eq:fspace_v0}. We are now ready to state the continuous variational problem: find $\bu \in \mathcal{S}$ such that

\begin{equation}
\int_{\Omega} \partial_j w_i \Sigma_{ij} \ d\Omega 
- \sum_{i=1}^\nsd \int_{\Gamma_{h_i}} w_i h_i \ d\Gamma
= (\nabla \bw, \bSigma) - \langle \bw, \bh \rangle_{\Gamma_{\bh}} = 0 \quad \forall \bw \in \mathcal{V},
\label{eq:continuous_weak_form}
\end{equation}

\noindent which we have written in both index and symbolic notation for clarity. This form of the equilibrium equation is also known as \textit{the principle of virtual work}.

Next, we determine the discrete counterpart of Eq. \eqref{eq:continuous_weak_form}. We use linear shape functions here, but higher-order bases may be utilized \cite{kim_finite_2021}. Let $\mathbb{P}^1(\Omega^e)$ denote the space of piecewise linear polynomials over finite elements $e = 1, \ldots, n_{el}.$
We define the finite-dimensional function spaces

\begin{align}
\mathcal{S}^h &= \{\bu^h \in \mathcal{S} \cap \mathbb{P}^1(\Omega^e)\}, \\
\mathcal{V}^h &= \{\bw^h \in \mathcal{V} \cap \mathbb{P}^1(\Omega^e) \}, \label{eq:Vh_space}
\end{align}
which contain functions of the form

\begin{equation}
\bu^h = \sum_{a=1}^{n_{np}} \bu^a N_a (\bx) = \sum_{a=1}^{n_{np}} u^a_i N_a (\bx) \hat{e}_i,
\end{equation}

\noindent where $n_{np}$ is the number of nodal points, $N_a(\bx)$ are the finite element shape functions, and $\hat{\bs{e}}$ is the standard Euclidean basis. We substitute the finite element expansions $\bu^h$ and $\bw^h$ into Eq. \eqref{eq:continuous_weak_form} to obtain

\begin{equation}
\left(\nabla \bw^h, \bSigma(\bu^h,\bxi;\bp) \right) - \left\langle \bw^h, \bh \right\rangle_{\Gammabh} = 0 \quad \forall \bw^h \in \mathcal{V}^h.
\label{eq:discrete_weak_form}
\end{equation}

\noindent We then express Eq. \eqref{eq:discrete_weak_form} in index notation and substitute in the expansion for the weighting function $\bw^h$ to yield

\begin{equation}
w^a_i \int_{\Omega} \partial_j N_a(\bx)  \bSigma_{ij}(\bu^h,\bxi;\bp) \ d\Omega -  \sum_{i=1}^\nsd w^a_i \int_{\Gamma_{h_i}} h_i N_a(\bx) \ d\Gamma = 0 \quad \forall w^a_i N_a(\bx) \hat{e}_i \in \mathcal{V}^h.
\label{eq:index_discrete_weak_form}
\end{equation}

Crucially, the coefficients in the weighting function appear in both terms and Eq. \eqref{eq:index_discrete_weak_form} must hold for all functions in $\mathcal{V}^h$. Therefore, they are arbitrary and may be removed to obtain the expression for the \textit{discrete global residual}

\begin{equation}
R^a_i := \int_{\Omega} \partial_j N_a(\bx)  \bSigma_{ij}(\bu^h,\bxi;\bp) \ d\Omega - \sum_{i=1}^\nsd \int_{\Gamma_{h_i}} h_i N_a(\bx) \ d\Gamma = 0,
\label{eq:discrete_global_residual}
\end{equation}

\noindent which is a vector with a dimension equal to the number of global degrees of freedom $n_{dof} = n_{sd} n_{np}$.

With the global residual defined, we take a step back and describe the physical experiment and its relation to it. Many characterization tests involve a load frame that holds a test specimen in rigid grips that is pulled to failure. The presence of holes and notches in the specimen will produce non-uniform stress and strain fields.

Figure \ref{fig:schematic_full_specimen_with_grips} shows a prototypical experimental setup where displacement is measured on the visible surface of the sample via DIC and a load cell measures axial (the direction of pulling) force. Typically, DIC data, a dense point cloud, will have a small gap near the edges of the visible region, so the region modeled by finite elements stops some distance away from the grips. This improves the quality of the displacement data that will be used for BCs.

We conclude this section by describing how the global residual fits into the constrained optimization problems that define FEMU and VFM. In what follows, we use a tilde to denote an experimentally measured quantity and assume that the displacement data (often a point cloud from DIC) has been ``remapped" to nodal coefficients in the FE expansion $\bumeas^h$. The \textit{axial load} $\Fmeas$ is the component of the net reaction force in the axial direction.

In FEMU, we introduce a computational model of the experiment (equilibrium PDE + BCs + constitutive model) and adjust its parameters until the model's predictions match the experimental data in a weighted sense. This is accomplished by solving

\begin{equation}
\begin{split}
\underset{\bp}{\text{min}} \quad & J(\bp) \propto \int_{\Gamma_{\text{DIC}}} \| \bu^h - \bumeas^h \|^2 \ d\Gamma + (F - \tilde{F})^2 \\
\text{s.t.} \quad & \bs{R}\left( \bs{u}^h, \bs{\xi}; \bp \right) = \bzero, \\
& \bs{C}(\bu^h, \dot{\bu}^h, \bxi, \dot{\bxi}; \bp) = \bzero,
\end{split}
\label{eq:rough_FEMU_objective}
\end{equation}

\noindent where we have included the \emph{local residual} $\bs{C}$ for the constitutive model evolution equations that close the BVP.
The most logical choice of BCs for FEMU is shown in Figure \ref{fig:schematic_BC_FEMU_VFM} (a) and (b).  They ``drive'' the forward problem, as we have neglected body forces.

For VFM, we insert the measured displacement field $\bumeas^h$ into Eq. \eqref{eq:discrete_global_residual} with the BC specification shown in Figure \ref{fig:schematic_BC_FEMU_VFM} (c) and (d). It may seem odd to require BCs when the displacement field is known, but their specification has two notable effects. First, they define the space of admissible virtual fields through Eq. \eqref{eq:Vh_space}. Second, we assume a non-zero traction boundary $\Gamma^L_{h_2}$, which will allow us to account for the measured axial load.

The VFM objective is obtained by dotting the global residual with virtual field coefficients $\bvfvec \in \mathbb{R}^{n_{dof}}$ to yield

\begin{equation}
\begin{split}
   \underset{\bp}{\text{min}} \quad & V(\bp) \propto \left( \bvfvec \cdot \bs{R}(\bumeas^h, \bxi; \bp) \right)^2, \\
\text{s.t.} \quad & \bs{C}(\bu^h, \dot{\bu}^h, \bxi, \dot{\bxi}; \bp) = \bzero.
\end{split}
\label{eq:rough_VFM_objective}
\end{equation}

\noindent Load data is incorporated by choosing the virtual field coefficients to be equal to unity on one of the non-trivial traction boundaries and setting this integral equal to the experimentally measured axial load. As an example, consider the BCs in Figure \ref{fig:schematic_BC_FEMU_VFM} (d) where the boundary term gives

\begin{equation}
    \vfvec^a_2 \int_{\Gamma_{h_2}} h_2 N_a(\bx) \ d\Gamma \overset{set}{=} \Fmeas.
\end{equation}

\noindent Interestingly, under this logic the inclusion of the ``opposite'' traction boundary would cancel out the contribution of the load measurement, so $\Gamma^L_{h_2}$ only appears on one side of the specimen.

When the virtual field coefficients are chosen in this fashion, we have

\begin{equation}
    \bvfvec \cdot \bs{R}(\bumeas^h, \bxi; \bp) = \underbrace{\vfvec^a_i \int_{\Omega} \partial_j N_a(\bx)  \bSigma_{ij}(\bumeas^h,\bxi,\bp) \ d\Omega}_{\text{internal virtual work}} \quad -  \underbrace{\Fmeas}_{\text{external virtual work}}
    \label{eq:plugin_vfm},
\end{equation}

\noindent which will be equal to zero (for any set of virtual field coefficients) when $\bumeas^h$ is equal to the finite element solution $\bu^h$, and $\Fmeas$ is equal to the corresponding load prediction $F$.

Therefore, we informally conclude that VFM will work best when the measured experimental data is close to the finite element approximation for a set of constitutive model parameters $\bp$, which will depend on errors due to measurement noise, discretization, and model form error. With the understanding that in practice it is unlikely that Eq. \eqref{eq:plugin_vfm} will be exactly zero, we instead aim to make it small in a least-squares sense through minimization. We note that sums of virtual fields, i.e.

\begin{equation}
    V(\bp) \propto \sum_{k=1}^{N_{vf}} \left( \bvfvec_k \cdot \bs{R}(\bumeas^h, \bxi; \bp) \right)^2,
\end{equation}

\noindent may also be used.

The art of VFM manifests in the selection of coefficients for the virtual field(s). In this work, we make a traditional choice — a single virtual field based on analytical functions. However, recent developments include stress sensitivity-based virtual fields \cite{marek_sensitivity-based_2017,marek_extension_2019, marek_experimental_2020}, pseudo-real deformation virtual fields \cite{kim_finite_2021}, and the approach taken in the EUCLID method \cite{flaschel_discovering_2022} where the coefficients are unity over internal degrees of freedom and boundaries with known load data and zero otherwise.

\section{Inverse formulations}
\label{sec:inverse_formulations}

This section outlines the inverse problem framework, including the precise definitions of the objective functions and formal statements of the inverse problems for both FEMU and VFM. It also contains the novel presentation of forward and adjoint local sensitivity analyses for the VFM objective function gradient.

\subsection{FEMU objective function}

The FEMU objective function quantifies the mismatch between model predictions and corresponding experimental measurements subject to an FE model of the characterization experiment. The continuous form of this objective function is
\begin{equation}
\underset{\bp}{\text{min}} \quad 
\mathcal{J} := \frac{1}{2 \sigma^2_{u} T A } \int_0^T \int_{\Gamma_{\text{DIC}}}
\left\| \bu\left(\bx, t;\bp\right) - \bumeas\left(\bx, t\right)
\right\|^2 \, dA \, dt
+ \frac{\alpha}{2 \sigma^2_{F} T} \int_0^T
\left(F\left(t;\bp\right) - \Fmeas\left(t\right) \right)^2 \, dt,
\label{eq:femu_continuous_objective}
\end{equation}

\noindent where $\sigma^2_{u}$ and $\sigma^2_{F}$ are the variance of the assumed uncorrelated white noise in the displacement $\tilde{\bu}$ and axial load $\Fmeas$ measurements, respectively, $T$ is the total time duration, $\Gamma_{\text{DIC}}$ is the surface where DIC-measured displacement data is available with area $A$, and $\alpha$ is a balance factor for tuning the relative importance of the two pieces of the objective function. This objective function is unitless, however, in this work, we will absorb the noise variance constants into $\alpha$ for simplicity. In practice, modeling noise and selecting the balance factor $\alpha$ remain open avenues for research.

We discretize Eq. \eqref{eq:femu_continuous_objective} with the same FE basis functions described in the previous section and a backward Euler temporal discretization into $n = 0, \ldots, n_L,$ ``load steps'' to obtain

\begin{equation}
\underset{\bp}{\text{min}} \quad
J := \frac{1}{2 T A} \sum_{n=1}^{n_L}
\udiff \mass \udiff \left(\Delta t \right)^n
+ \frac{\alpha}{2 T}\sum_{n=1}^{n_L}
\left(F^n - \Fmeas^n \right)^2 \left(\Delta t\right)^n,
\label{eq:femu_discrete_objective}
\end{equation}

\noindent where $\mass$ is the FE mass matrix, $u^n_i \in \mathbb{R}^{n_{np}}$ are nodal displacement coefficients, and $\left(\Delta t\right)^n := t^n - t^{n-1}$ are the backward Euler time intervals. The predicted axial load $F^n$ is computed by summing ``reaction loads'' at load step $n$ as described in Ref. \cite{hughes_continuous_2000}. The initial load step $n=0$ is excluded from the objective, as the initial condition for the displacement field  is specified, and therefore both the displacement and the load do not depend on $\bp$.

The formal statement of the FEMU inverse problem, a PDE-constrained optimization problem, is

\begin{equation}
\begin{split}
\underset{\bp}{\text{min}} \quad &J(\bp) \\
\text{s.t.} \quad & \bs{R}^n\left( \bs{u}^n, \bs{\xi}^{n}; \bp \right) = \bzero, \quad n = 1, \ldots, n_L,  \\
& \bs{C}^n_e \left( \bs{u}^n_e,  \bs{u}^{n-1}_e, \bs{\xi}^{n}_e, \bs{\xi}^{n-1}_e; \bp \right) = \bzero, \quad e = 1, \ldots, n_{el}, \ n = 1, \ldots, n_L.
\end{split}
\label{eq:FEMU_problem}
\end{equation}

The computation of the FEMU objective gradient with finite differences, forward sensitivities, and adjoint sensitivities are described in \cite{seidl2022calibration} and will not be repeated here. We remark the main computational bottleneck in FEMU is the solution of the global residual for the equilibrium PDE.

\subsection{VFM objective function and its gradients}

In this section, we discuss the VFM objective function and methods for computing its gradient. In particular, we highlight how we can obtain the gradient using forward and adjoint sensitivities. The VFM inverse problem involves the minimization of an objective function that measures the discrepancy between the internal virtual work and the external virtual work with respect to the material parameters. We can therefore write the discrete VFM calibration problem as follows:

\begin{equation}
\begin{split}
\underset{\bp}{\text{min}} \quad &V(\bp) := \frac{1}{2 T} \sum_{n=1}^{n_L} \bigg[ \left( \bvfvec \cdot \bs{R}^n\left(\bs{\xi}^{n}; \bumeas^n, \bp \right) - \Fmeas^n \right) \Delta t_n \bigg]^2 \\
\text{s.t.} \quad & \bs{C}^n_e \left( \bs{\xi}^{n}_e, \bs{\xi}^{n-1}_e; \bumeas^n_e, \bumeas^{n-1}_e, \bp \right) = \bzero, \quad e = 1, \ldots, n_{el}, \ n = 1, \ldots, n_L,
\end{split}
\label{eq:vfm_objective}
\end{equation}

\noindent where we have used the same FE and temporal discretizations as in the discrete FEMU objective of Eq. \eqref{eq:femu_discrete_objective}, as well as a temporal averaging of the virtual work.

In what follows, we denote the scaled virtual work mismatch $\bar{V}^n$ at load step $n$, i.e.\ the difference between internal and external virtual work multiplied by a temporal scaling factor, as follows:
\begin{align}
V_{\text{int}}^n &:= \bvfvec \cdot \bs{R}^n\left(\bs{\xi}^{n}; \bumeas^n, \bp \right), \\
   \bar{V}^n &= (V_{\text{int}}^n - \Fmeas) \frac{\Delta t_n}{T}. 
\end{align}

\subsubsection{Forward sensitivity-VFM}

Forward sensitivity is often referred to as direct sensitivity analysis, where the gradient is computed by considering the objective function as an implicit function of the material parameters. In addition to the material parameters, the VFM objective function defined in Eq. \eqnref{eq:vfm_objective} depends on both local state variables and the virtual fields.  We  take the derivative of the objective function as follows:

\begin{equation}
    \frac{ d V}{d \bs{p}} =  \sum_{n=1}^{n_L} \sum_{e=1}^{n_{el}} \bigg( \frac{\partial V_e^n}{\partial \bs{p}} + \frac{\partial V^n_e}{\partial \bs{\xi}^n_e} \frac{\partial \bs{\xi}^n_e}{\partial \bs{p}} \bigg) = \sum_{n=1}^{n_L} \bar{V}^n \left( \bvfvec \cdot \sum_{e=1}^{n_{el}} \frac{d \bs{R}_e^n}{d \bs{p}}\right),
\end{equation}

\noindent where $\frac{d \bs{R}_e^n}{d \bs{p}}$ represents the total derivative of the global residual with respect to the material parameters $\bp$. The total derivative of the global residual is given as:

\begin{equation}
  \frac{d \bs{R}_e^n}{d \bs{p}} =  \frac{\partial \bs{R}_e^n}{\partial \bs{p}} + \frac{\partial \bs{R}^n_e}{\partial \bs{\xi}^n_e} \frac{\partial \bs{\xi}^n_e}{\partial \bs{p}}. 
  \label{eq:FS_VFM_dr_dp}
\end{equation}

One of the key challenges in the forward sensitivity approach lies in determining the derivative of the local state variables with respect to the material parameters $\frac{\partial \bs{\xi}^n_e}{\partial \bs{p}}$. This involves understanding how the state variables change in response to variations in the material parameters, which can be complex due to the implicit relationship between the objective function and the material parameters. 

We can compute $\frac{\partial \bs{\xi}^n_e}{\partial \bs{p}}$ at the element level by conducting a forward sensitivity analysis on the local residual $\bs{C}^n_e$. Note that the local residual is defined as a constraint equation in the VFM calibration problem, see Eq. \eqnref{eq:vfm_objective}. $\bs{C}^n_e$ is a function of the local state variable $\bs{\xi}^n_e$ at the current and previous load steps, so we can write the gradient of the local residual as:

\begin{equation}
 \frac{\partial \bs{C}^n_e}{\partial \bs{\xi}^n_e} \frac{\partial \bs{\xi}^n_e}{\partial \bs{p}} +  \frac{\partial \bs{C}^n_e}{\partial \bs{p}} + \frac{\partial \bs{C}^n_e}{\partial \bs{\xi}^{n-1}_e} \frac{\partial \bs{\xi}^{n-1}_e}{\partial \bs{p}} = \bs{0},  \hspace{0.2cm} e=1,\ldots,n_{el}, \;  n=1,\ldots, n_L\\
 \label{eq:FS_VFM}
\end{equation}

Eq. \eqnref{eq:FS_VFM} captures the dependencies of the local residual on both the current and previous state variables and their change with respect to the material parameters. By solving for $\frac{\partial \bs{\xi}^n_e}{\partial \bs{p}}$, we can substitute the result into the expression of the total derivative of the global residual and re-cast Eq. \eqnref{eq:FS_VFM_dr_dp} as follows:

\begin{equation}
  \frac{d \bs{R}_e^n}{d \bs{p}} =  \frac{\partial \bs{R}_e^n}{\partial \bs{p}} - \frac{\partial \bs{R}^n_e}{\partial \bs{\xi}^n_e} \left(\frac{\partial \bs{C}^n_e}{\partial \bs{\xi}^n_e}\right)^{-1}\bigg(\frac{\partial \bs{C}^n_e}{\partial \bs{p}} + \frac{\partial \bs{C}^n_e}{\partial \bs{\xi}^{n-1}_e} \frac{\partial \bs{\xi}^{n-1}_e}{\partial \bs{p}} \bigg)
\end{equation}

As noted earlier, automatic differentiation is utilized to compute these gradients effectively. AD enables us to compute the derivatives of the residuals with respect to the material parameters by “seeding” one of the variables (such as $\bs{p}, \bs{\xi}^n_e$ or $\bs{\xi}^{n-1}_e$), evaluating the corresponding derivative, and then “unseeding” it before moving on to the next variable. This procedure is repeated for each variable, allowing for efficient gradient computation. 

\subsubsection{Adjoint sensitivity-VFM}

This section presents an adjoint sensitivity analysis framework for computing gradients of VFM. We start by constructing a Lagrangian $(\mathcal{L})$ as follows:

\begin{equation}
    \mathcal{L} = \sum_{n=1}^{n_L} \bigg( {V}^n + \sum_{e=1}^{N_{el}} (\bs{\phi}_e^n)^T \bs{C}^n_e \left( \bs{\xi}^{n}_e, \bs{\xi}^{n-1}_e; \bumeas^n_e, \bumeas^{n-1}_e, \bp \right)  \bigg),
\end{equation}

\noindent where $\bs{\phi}^n_e$ represents the Lagrange multiplier that enforces the constraint $\bs{C}^n_e = \bs{0}$ defined in Eq. \eqnref{eq:vfm_objective}. We can then compute the gradient of the objective function with respect to the material parameters as:

\begin{equation}
    \frac{d \mathcal{L}}{d \bs{p}} =  \frac{d {V}}{d \bs{p}}  = \sum_{n=1}^{n_L} \sum_{e=1}^{n_{el}} \bigg( \frac{d V_e^n}{d \bs{p}}  + (\bs{\phi}_e^n)^T \frac{d \bs{C}_e^n}{d \bs{p}}\bigg).
\label{eq:dl_dp}
\end{equation}

Note that since $\bs{C}_e^n = \bs{0}$, the gradients of $\mathcal{L}$ and $V$ are identical. For computing the gradient we need to evaluate the adjoint variable $\bs{\phi}^n_e$ (a Lagrange multiplier). Using the chain rule,  we write

\begin{eqnarray}
      \frac{d V}{d \bs{p}} = \sum_{n=1}^{n_L} \sum_{e=1}^{n_{el}} \bigg( \frac{\partial V_e^n}{\partial \bs{p}} + \frac{\partial V_e^n}{\partial \bs{\xi}_e^n} \frac{\partial \bs{\xi}_e^n}{\partial \bs{p}}\bigg) + \sum_{n=1}^{n_L} \sum_{e=1}^{n_{el}} (\bs{\phi}_e^n)^T  \bigg( \frac{\partial \bs{C}_e^n}{\partial \bs{p}} + \frac{\partial \bs{C}_e^n}{\partial \bs{\xi}_e^n} \frac{\partial \bs{\xi}_e^n}{\partial \bs{p}} + \frac{\partial \bs{C}_e^n}{\partial \bs{\xi}_e^{n-1}} \frac{\partial \bs{\xi}_e^{n-1}}{\partial \bs{p}} \bigg)
    \label{eq:dl_dp_expand}
\end{eqnarray}

\noindent We can rearrange and express Eq. \eqnref{eq:dl_dp_expand} as

\begin{eqnarray}
    \begin{aligned}
      \frac{d V}{d \bs{p}} &= \sum_{n=1}^{n_L} \sum_{e=1}^{n_{el}} \bigg( \frac{\partial V_e^n}{\partial \bs{p}} +  (\bs{\phi}_e^n)^T  \frac{\partial \bs{C}_e^n}{\partial \bs{p}} \bigg)
       + \sum_{e=1}^{n_{el}} \Bigg[ \frac{\partial {V}_e^{n_L}}{\partial \bs{\xi}_e^{n_L}} + (\bs{\phi}_e^{n_L})^T  \frac{\partial \bs{C}_e^{n_L} }{\partial \bs{\xi}^{n_L}} \Bigg] \frac{\partial \bs{\xi}_e^{n_L}}{\partial \bs{p}}  \\
       &+ \sum_{n=1}^{n_L-1} \sum_{e=1}^{n_{el}}   \Bigg[ \frac{\partial V_e^n}{\partial \bs{\xi}_e^n} + (\bs{\phi}_e^n)^T\frac{\partial \bs{C}_e^n}{\partial \bs{\xi}_e^n} + (\bs{\phi}_e^{n+1})^T \frac{\partial \bs{C}_e^{n+1}}{\partial \bs{\xi}_e^{n}} \Bigg] \frac{\partial \bs{\xi}_e^n}{\partial \bs{p}} + \sum_{e=1}^{n_{el}} (\bs{\phi}_e^1)^T \cancel{\frac{\partial \bs{C}_e^n}{\partial \bs{\xi}_e^{0}} \frac{\partial \bs{\xi}_e^{0}}{\partial \bs{p}}}
    \end{aligned}
    \label{eq:dl_dp_rearranged}
\end{eqnarray}

The initial value for the sensitivity matrix $\frac{\partial \bs{\xi}_e^0}{\partial \bs{p}}$ is zero, as the initial values of the local state variables are given. We obtain the following system of adjoint equations for $\bs{\phi}^n_e$ by forcing the terms that multiply to the non-zero sensitivity matrices to vanish:

\begin{equation}
\begin{aligned}
 \frac{\partial {V}_e^{n_L}}{\partial \bs{\xi}_e^{n_L}}  + (\bs{\phi}_e^{n_L})^T \; \frac{\partial \bs{C}_e^{n_L} }{\partial \bs{\xi}^{n_L}} &= \bs{0}, \quad e=1,\ldots, n_{el} \\
\frac{\partial {V}_e^{n}}{\partial \bs{\xi}_e^{n}} + ( \bs{\phi}_e^{n})^T \;\frac{\partial \bs{C}_e^{n} }{\partial \bs{\xi}_e^{n}}  + (\bs{\phi}_e^{n+1})^T\; \frac{\partial \bs{C}_e^{n+1} }{\partial \bs{\xi}_e^{n}}  &= \bs{0}, \quad e=1,\ldots, n_{el}, \quad n=n_L-1,\ldots,1.
 \end{aligned}
  \label{eq:phi_n}  
\end{equation}

The solution of the adjoint problem of Eq. \eqref{eq:phi_n} yields the following simplified expression for the gradient:

\begin{equation}
      \frac{d V}{d \bs{p}} = \sum_{n=1}^{n_L} 
      \left(\bar{V}^n \left(\bvfvec \cdot  \sum_{e=1}^{n_{el}}  \frac{\partial\bs{R}^n_e}{\partial \bp}\right) +   \sum_{e=1}^{n_{el}}  (\bs{\phi}_e^n)^T  \frac{\partial \bs{C}_e^n}{\partial \bs{p}}\right)
\end{equation}

We conclude this section with a few remarks on forward and adjoint sensitivities for VFM. With FS, the gradient can be computed as the forward problem (i.e.\ the local residual) is being solved. It has minimal storage requirements, as only the local state variables and their sensitivities from neighboring load steps are needed at any given step. On the other hand, the adjoint approach requires the entire time history of the local state variables to be known before performing the adjoint solve, thus the forward problem must be completed first. However, the cost of the adjoint method is independent of the number of material model parameters, while that of FS scales linearly in this regard. Due to the lack of large-scale linear solves in VFM, the tradeoff between storage and computational cost is not as extreme as in FEMU, but there is still a crossover point between the two approaches.

\section{Numerical comparisons between FEMU and VFM}
\label{sec:results}
 \begin{figure}
\includegraphics[width=0.5\linewidth,center]{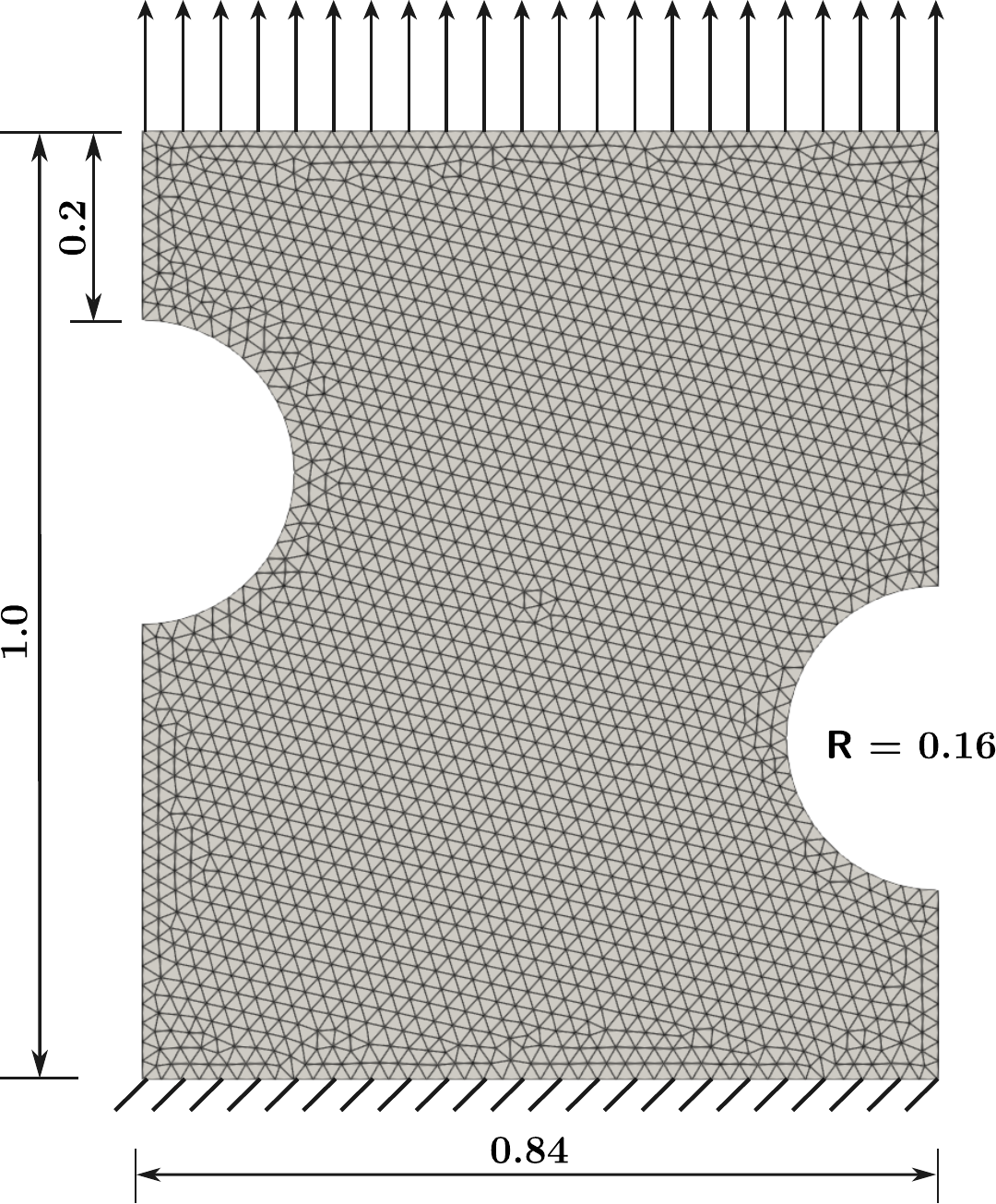}
 \caption{Schematic representation of the asymmetrically notched plate, including its geometry, mesh, and boundary conditions. Additionally, to replicate the effect of the rigid mechanical grips holding the specimen, the horizontal movement at the top edge of the specimen is constrained. All dimensions are in mm.}
 \label{fig:schematic}
 \end{figure}

 \begin{figure}
\includegraphics[width=0.5\linewidth,center]{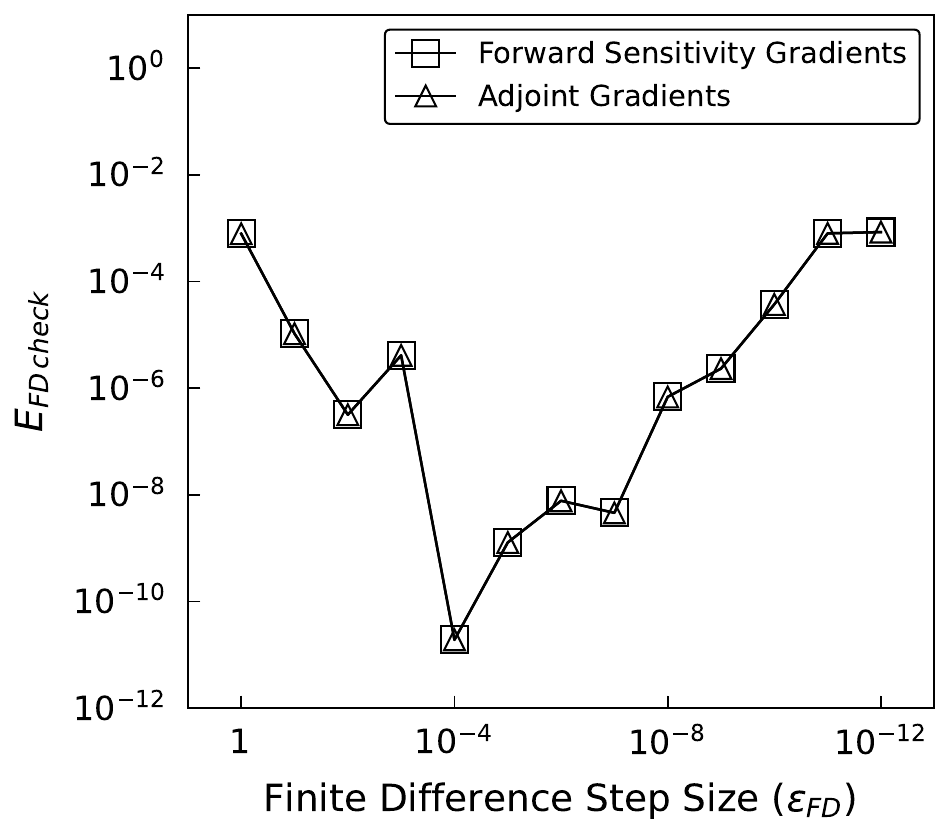}
 \caption{VFM finite difference gradient checks for the plate with asymmetric notches.}
 \label{fig:FD_gradCheck}
 \end{figure}

This section presents the results of numerical simulations to evaluate and compare the performance of FEMU and VFM in calibrating material parameters for finite strain elastoplastic constitutive models. 
Following the six numerical experiments (\textbf{E1} - \textbf{E6}) outlined in the introduction, we begin by assessing the accuracy of the calibrated material parameters, along with the computational efficiency (\textbf{E1}), sensitivity to the initial guess (\textbf{E2}), and robustness to noisy data (\textbf{E3}). Furthermore, we test the models under different conditions of model form errors, specifically examining how the methods perform when the functional form (here: hardening) is misspecified (\textbf{E4}). 
Additionally, we investigate the impact of spatial discretization mismatch (\textbf{E5}) between the synthetic calibration data and mesh employed in the inverse problem. Finally, we investigate the impact of the assumption of plane stress on the accuracy of FEMU and the VFM (\textbf{E6}). Before thoroughly studying these numerical experiments, we establish several assumptions.

\subsection{Assumptions and gradient verification}
The numerical simulations are conducted on a thin plate with an asymmetric notch, resembling the geometry described in \cite{kramer2014implementation}, and having dimensions \([0, 0] \times [0.84, 1] \times [0, T_0]\), where $T_0$ denotes the uniform thickness of the specimen, which we provide in the respective sections. We use millimeters (mm) as the unit of measurement for length. A schematic representation of the specimen, including its mesh and boundary conditions, is provided in Fig. \ref{fig:schematic}. The bottom surface of the specimen is fixed, while a displacement of magnitude 0.07 is applied incrementally in the upward direction at the top surface as $u_y = 0.01 \times t$. In all numerical experiments except for \textbf{E4}, the displacement is applied in 8 steps with $t = \{0.1, 0.15, 0.2, 0.5, 1, 3, 5, 7\}$. The smaller steps at the beginning help with the convergence of the nonlinear solver. All other faces of the specimen are assumed to be traction-free.

Plane stress conditions are assumed throughout the simulations, except in subsection \ref{sec:3D-2DSurfData}, where the effects of assuming plane stress for increasing specimen through-thicknesses are explored and a 3D model is utilized. The objective functions for FEMU and VFM, which guide the parameter identification processes, are defined in Eqs. \eqnref{eq:femu_discrete_objective} and \eqnref{eq:vfm_objective}, respectively.

We choose a virtual field with components $\vfvec_{\text{x}} = \cos\left(\pi \left(\text{y} - \frac{1}{2} \right) \right)$ and $\vfvec_{\text{y}} = \text{y}^2$, following \cite{kim2014determination, jones2018parameter}, where y represents the height along the loading direction, measured from the specimen’s bottom edge. The component $\vfvec_{\text{x}}$ is zero at both the top and bottom edges, while $\vfvec_{\text{y}}$ is one at the top (where the non-zero displacement is applied) and vanishes at the bottom (where the specimen is fixed). To evaluate the sensitivity of the calibrated material parameters to virtual field choice, Appendix \ref{appen:different_VF} compares results obtained using an alternative set of virtual field components (linear instead of quadratic for $\vfvec_{\text{y}}$). The calibrated material parameters for both virtual field choices are nearly identical for small levels of noise.

We employ the bound-constrained optimization algorithm L-BFGS-B \cite{zhu1997algorithm} to minimize the cost function, utilizing the implementation in \texttt{scipy.optimize} \cite{virtanen2020scipy}. In FEMU, the weighting between the displacement and force-matching components of the objective function is assigned on a trial basis, see Eq. \eqnref{eq:femu_discrete_objective}. We begin with an initial estimate and then update the balance factor based on the relative magnitude of the displacement and force objective function components so that they are comparable in size. After this optimization is completed, we again examine their relative sizes, adjust the balance factor in the same fashion, and optimize a second time. In each run, the algorithm is terminated based on a predefined stopping criterion that involves conditions such as the convergence of the objective function, gradient tolerance, and the maximum number of line search iterations.

With this setup, we attempt to ensure that the numerical simulations are both representative of experimental conditions and suitable for assessing the performance of FEMU and VFM in various scenarios.
The ground truth values of the material parameters denoted by $\bs{p}$ = [$E$ (GPa), $\nu$, $Y$ (MPa), $S$ (MPa), $D$] are taken as [200, 0.30, 330, 1000, 10], respectively.
The computation of the gradients in FEMU is restricted to adjoint and finite differences. We employ the adjoint method to compute gradients in FEMU for the numerical comparison in \textbf{E2}-\textbf{E6}.
 We report errors in the parameters using the normalized error defined as:
\begin{equation}
    \bp_{\text{normalized error}} = \frac{| \bp^{mean}_{calibrated} - \bp_{true} |}{\bp_{true}},
\end{equation}
where $ \bp^{mean}_{calibrated}$ represents the arithmetic mean of the calibrated parameters across the 10 sets of initial random guesses. $\bp_{true}$ denotes the corresponding truth value.

Before presenting the numerical experiments, we first verify that the presented gradients of the VFM objective obtained through forward sensitivity and adjoints are implemented correctly. In particular, we check the gradient by comparing it with its approximation computed using a finite difference approximation (FD) as follows:
\begin{equation}
    E_{\text{FD check}} = \Bigg | \left( \left[ \frac{d V}{d \bs{p}} \right]_{\text{FS/adjoint}}  \cdot \bs{\mathcal{D}}\right) - \left( \left[\frac{d V}{d \bs{p}}\right]_{\text{FD}} \cdot \bs{\mathcal{D}}\right) \Bigg |
\end{equation}
where $\bs{\mathcal{D}}$ denotes a direction vector in parameter space. We choose each component of this direction vector to be $\mathcal{D}_i = 0.1$. For computing the error, we use different step sizes as: $1, 10^{-1}, \cdots, 10^{-12}$. 
The finite difference error check for the gradients computed using forward sensitivity and adjoint approaches for VFM is shown in Fig \ref{fig:FD_gradCheck}. In finite difference methods, the error in the gradient approximation typically decreases as the step size is reduced, up to a critical point. In this study, we observe this trend until a step size of approximately $10^{-4}$, beyond which round-off errors dominate, leading to a loss of accuracy. This behavior is well-documented in computational finite difference techniques and strongly supports the accuracy of the gradients. Moreover, the gradients derived from forward and adjoint sensitivity methods exhibit numerical equivalence. This consistency reinforces the correctness of the gradients computed with the proposed forward and adjoint sensitivity schemes. 
Finally, we introduce the following naming conventions. FEMU-FD and FEMU-Adjoint refer to FEMU based on finite difference and adjoint gradients, respectively, while VFM-FD, VFM-Adjoint, and VFM-FS denote using VFM with finite differences, adjoints, and forward sensitivities methods for computing the gradient.

 \begin{figure}
\includegraphics[width=0.8\linewidth,center]{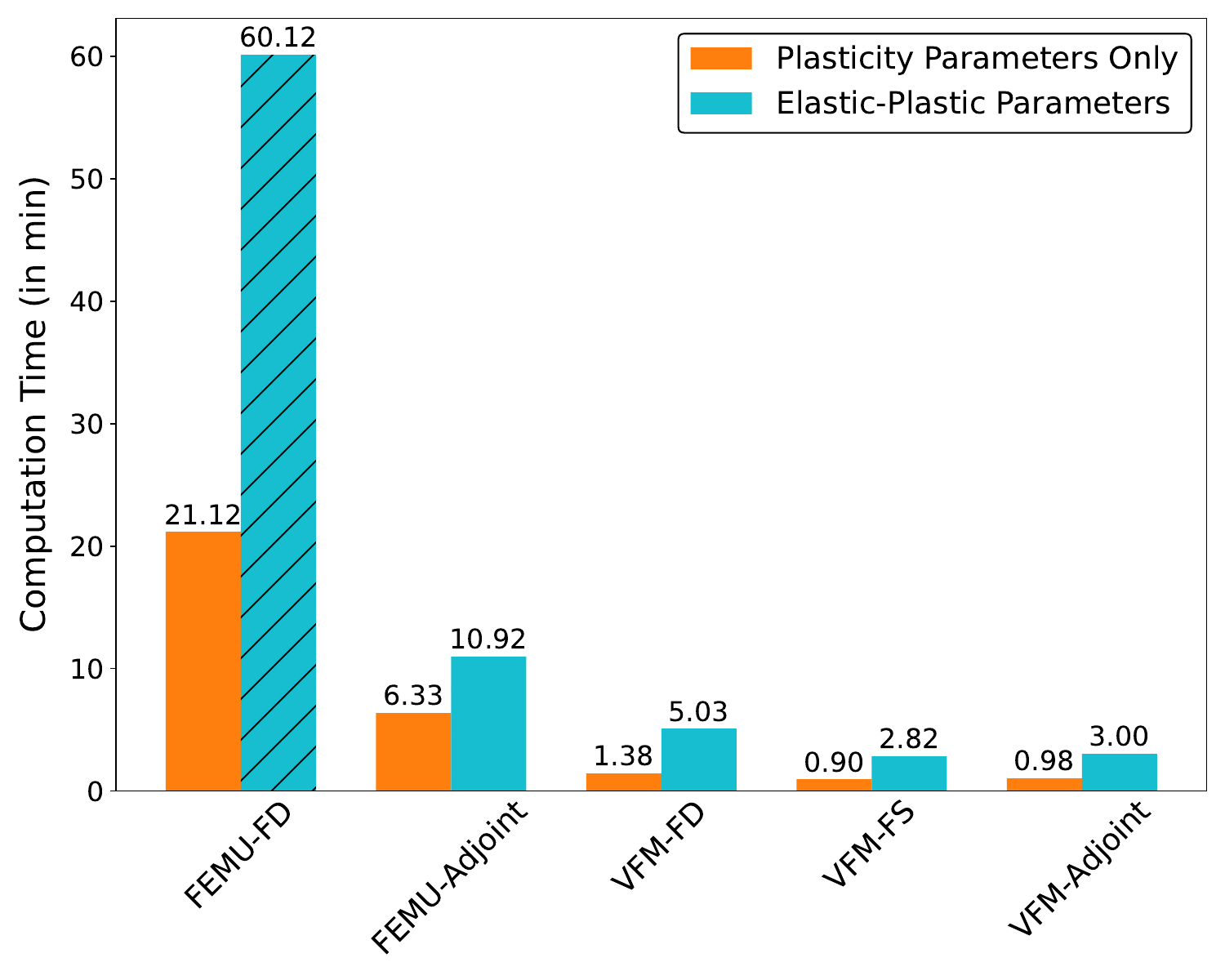}
 \caption{A comparison of computation times for different inverse methods in calibrating material parameters demonstrates that VFM outperforms FEMU in computational efficiency for both identifying plasticity parameters alone and identifying elastic-plastic parameters.}
 \label{fig:pr-1_computationTime}
 \end{figure}

\subsection{\textbf{E1:} Inverse solution accuracy and computation time}
\label{sec:pb-1}
\renewcommand{\arraystretch}{1.25}
\begin{table}[h!]
\centering
\caption{Comparison of inverse problem solutions for the plate with asymmetric notch in finite strain elasto-plastic deformation with synthetic data}
\resizebox{!}{5.5cm}{
\begin{tabular}{>{\raggedright\arraybackslash}m{2.9cm} c c c c c}
\toprule
 & $E$(GPa) & $\nu$ & $Y$ (MPa) & $S$(MPa) & $D$ \\
\midrule
\textbf{Truth} & 200  & 0.30 & 330 & 1000 & 10 \\
 & & & & & \\
\rowcolor{gray!15}
\multicolumn{6}{c}{\textbf{Plasticity parameters only}} \\  
\textbf{Initial} & - & - & 360 & 920 & 6 \\
\textbf{FEMU-FD} & - & - & $330$ & $999.9998 (0.0002\%)$ & $10$  \\ 
\textbf{FEMU-Adjoint} & - & - & $330$ & $1000$ & $10$  \\
\rowcolor{gray!5}
 & & & & & \\
\textbf{VFM-FD} & - & - & $330.0174 (0.0053\%)$ & $1000.2631 (0.0263\%)$ & $9.9957 (0.0430\%)$  \\ 
\textbf{VFM-FS} & - & - & $329.9994 (0.0002\%)$ & $999.9952 (0.0005\%)$ & $10.0001 (0.0010\%)$  \\ 
\textbf{VFM-Adjoint} & - & - & $329.9994 (0.0002\%)$ & $999.9952 (0.0005\%)$ & $10.0001 (0.0010\%)$  \\ 
 & & & & & \\
\rowcolor{gray!15}
\multicolumn{6}{c}{\textbf{Both elastic-plastic parameters}} \\
\textbf{Initial} & 220 & 0.24 & 360 & 920 & 6 \\
\textbf{FEMU-FD} & $200.0011 (0.001\%)$ & $0.2999 (0.033\%)$ & $330$ & $1000$ & $10$  \\ 
\textbf{FEMU-Adjoint} & 200 & 0.3 & 330 & 1000 & 10  \\ 
\rowcolor{gray!5}
 & & & & & \\
\textbf{VFM-FD} & $200.1033 (0.052\%)$ & $0.2988 (0.400\%)$ & $330.1016 (0.031\%)$ & $1000.3779 (0.038\%)$ & $9.9917 (0.083\%)$  \\ 
\textbf{VFM-FS}& $200.0043 (0.002\%)$ & $0.2999 (0.033\%)$ & $330.0030 (0.001\%)$ & 1000 & $9.9998 (0.002\%)$  \\ 
\textbf{VFM-Adjoint} & $200.0043 (0.002\%)$ & $0.2999 (0.033\%)$ & $330.0030 (0.001\%)$ & 1000 & $9.9998 (0.002\%)$  \\ 
\bottomrule
\end{tabular}
}
\caption*{\footnotesize\emph{Note}: Errors relative to the true solutions are reported next to the parameter values.}
\label{tb:accuracy_computationTime}
\end{table}

We start by assessing the accuracy and computational efficiency of both methods for a noiseless case. The accuracy assessment focuses on the discrepancy between the calibrated and true material parameters, while the computation time highlights the methods' feasibility for large-scale applications. We investigate two cases: one where we invert only plastic parameters and the second case where we invert both elastic and plastic parameters. For both cases, we present two levels of comparison, (i) comparison of adjoint and FS (for VFM only) with finite differences for computing the gradient, and (ii) comparing the impact of numerically-exact sensitivities on FEMU and VFM.
We consider noiseless synthetic data obtained from the forward finite element analysis. For a description of the plane stress forward model, see Appendix \ref{app:residuals}. 

The initial guess for the inversion of the plasticity parameters was chosen to be $\bp_{\text{p}} = [360, 920, 6]$. In contrast, we use $\bp_{\text{ep}} = [220, 0.24, 360, 920, 6]$ as the initial guess for inverting both elastic and plastic parameters and the lower and the upper bounds are specified as $\bp_{\text{lo}} = [ 100, 0.23, 250, 800, 2] $ and $\bp_{\text{hi}} = [300, 0.35, 400, 1150, 12]$, respectively. Note that, specifically for the inversion of only the plasticity parameters, it is not necessary to specify bounds on $E$ and $\nu$. For inverting plasticity parameters only, we use the lower and upper bounds for $Y$, $S$, and $D$ as specified in $\bp_{\text{lo}}$ and $\bp_{\text{hi}}$, respectively.

The results of the inverse problems are given in Table \ref{tb:accuracy_computationTime}. First, we focus on the prediction accuracy of FEMU and VFM in inverting only the plasticity parameters. Next, we present the calibration results for both elastic-plastic parameters together. We set the relevant accuracy of the parameters to four decimal places. The errors relative to the true values are reported in parenthesis next to the calibrated parameter values. We can see that the accuracy of the FEMU prediction is slightly greater than that of VFM, but both methods almost exactly recover the true parameter values.  

The results show that both FEMU and VFM can achieve excellent accuracy under ideal conditions viz., noise-free data and accurate model assumptions (i.e.\ the model that generated the data is exactly consistent with the one utilized for inversion). We can however also report higher errors when finite differences are used to approximate the sensitivities compared to the forward sensitivity and adjoint approach for both FEMU and VFM. As expected, adjoint or forward sensitivity-based approaches for a given method yield equivalent results. FEMU achieves exact or near-exact solutions for all parameters. VFM (with FS and Adjoint) shows negligible errors with VFM-FD being slightly less accurate.

Next, in Fig. \ref{fig:pr-1_computationTime}, we compare the computational efficiency of both FEMU and VFM in conjunction with the introduced gradient methods. We observe a stark contrast between FEMU and VFM methods, especially for FEMU-FD for which the computation time is $\approx 20 \times$ larger than the computation time needed for VFM-FS or VFM-Adjoint.
This high computational demand of FD hinders its scalability to practical problems and makes FEMU-FD impractical for large-scale computations. Notably, the FEMU-Adjoint has a considerably lesser computational cost than FEMU-FD. However, VFM methods are faster than all FEMU variants. In fact, VFM (FS, Adjoint) requires approximately one-sixth and one-third of the computation time compared to FEMU-Adjoint for calibrating plastic parameters only and elastic-plastic parameters, respectively. We note that the computation time for VFM-FD is also significantly smaller than FEMU-Adjoint. FEMU's reliance on forward FE simulations increases computational demand, making VFM the obvious choice for scenarios with ideal conditions (no noise, no model-form errors).

\subsection{ \textbf{E2:} Sensitivity of initial inputs in calibrating material parameters}

 \begin{figure}
\includegraphics[width=1.0\linewidth,center]{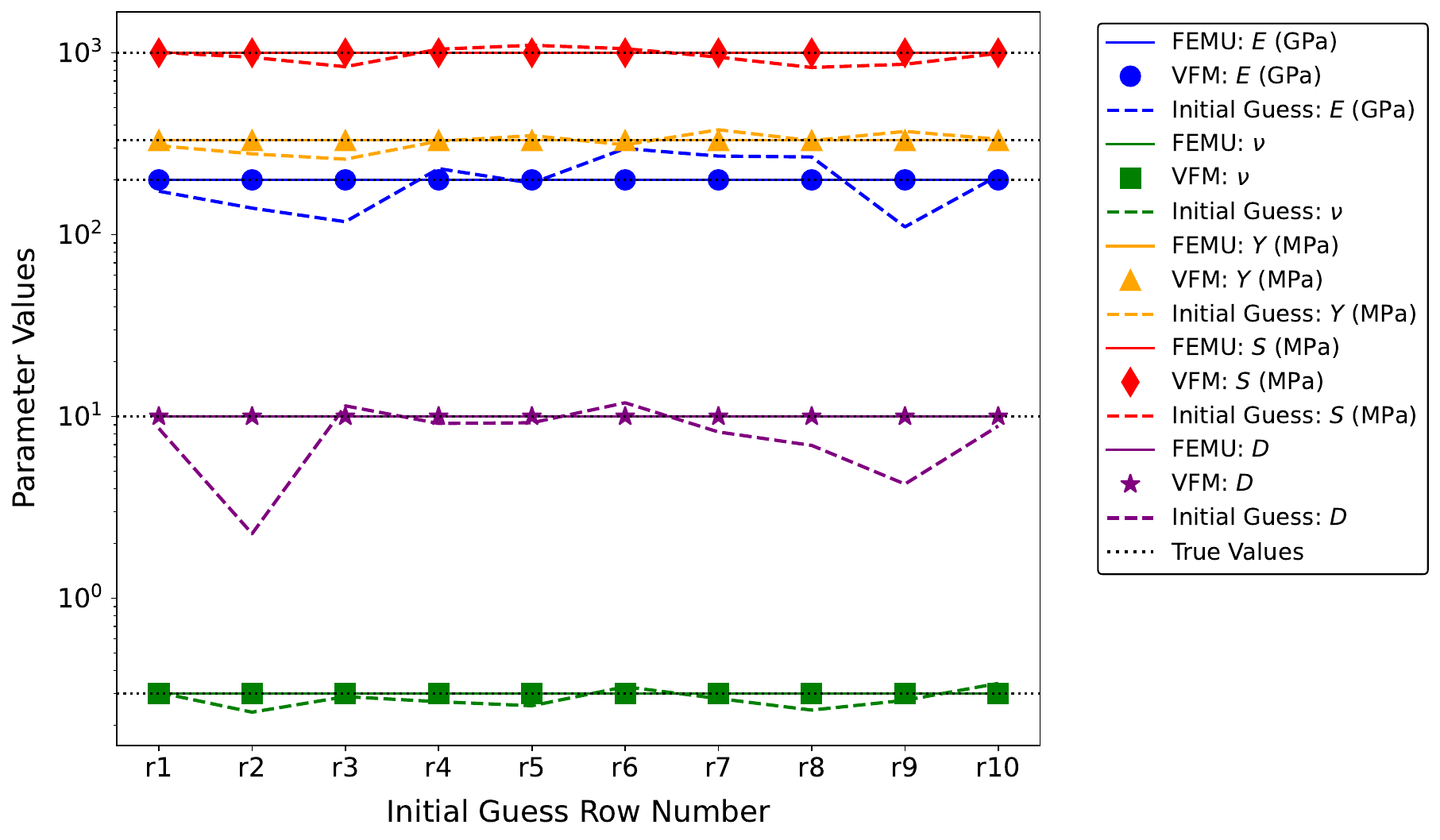}
 \caption{Comparison of inverse problem solutions with varying initial guesses on a logarithmic scale, illustrating the convergence behavior and accuracy of the calibrated parameters for different initial data sets. The variation in initial guesses is also depicted with dotted lines.}
 \label{fig:initialGuessSens}
 \end{figure}

\begin{table}[h!]
\centering
\caption{Comparison of inverse problem solutions for different initial guesses for the plate with asymmetric notch in finite strain elasto-plastic deformation using synthetic data}
\resizebox{\textwidth}{!}{%
\begin{tabular}{>{\raggedright\arraybackslash}m{1.7cm} | c c c c c | c c c c c}
\toprule
\rowcolor{gray!10}
 & \multicolumn{2}{c}{$E$(GPa)} & \multicolumn{2}{c}{$\nu$ } & \multicolumn{2}{c}{$Y$ (MPa)} & \multicolumn{2}{c}{$S$ (MPa)} & \multicolumn{2}{c}{$D$}\\
\midrule
\textbf{Truth} &  \multicolumn{2}{c}{200} & \multicolumn{2}{c}{0.30} & \multicolumn{2}{c}{330} & \multicolumn{2}{c}{1000} & \multicolumn{2}{c}{10} \\
\rowcolor{gray!10}
\multicolumn{11}{c}{\textbf{Random initial guess}} \\
\midrule
\rowcolor{gray!15}
&\multicolumn{2}{c}{173.27} & \multicolumn{2}{c}{0.30} & \multicolumn{2}{c}{308.44} & \multicolumn{2}{c}{1006.78} & \multicolumn{2}{c}{8.59} \\
&\multicolumn{2}{c}{139.86} & \multicolumn{2}{c}{0.24} & \multicolumn{2}{c}{278.10} & \multicolumn{2}{c}{944.37} & \multicolumn{2}{c}{2.26} \\
\rowcolor{gray!15}
&\multicolumn{2}{c}{117.71} & \multicolumn{2}{c}{0.29} & \multicolumn{2}{c}{260.01} & \multicolumn{2}{c}{838.40} & \multicolumn{2}{c}{11.43} \\
&\multicolumn{2}{c}{230.64} & \multicolumn{2}{c}{0.27} & \multicolumn{2}{c}{326.83} & \multicolumn{2}{c}{1046.61} & \multicolumn{2}{c}{9.15} \\
\rowcolor{gray!15}
&\multicolumn{2}{c}{191.87} & \multicolumn{2}{c}{0.26} & \multicolumn{2}{c}{350.13} & \multicolumn{2}{c}{1100.44} & \multicolumn{2}{c}{9.22} \\
&\multicolumn{2}{c}{297.54} & \multicolumn{2}{c}{0.33} & \multicolumn{2}{c}{311.92} & \multicolumn{2}{c}{1053.76} & \multicolumn{2}{c}{11.88} \\
\rowcolor{gray!15}
&\multicolumn{2}{c}{270.31} & \multicolumn{2}{c}{0.28} & \multicolumn{2}{c}{377.25} & \multicolumn{2}{c}{946.85} & \multicolumn{2}{c}{8.21} \\
&\multicolumn{2}{c}{267.39} & \multicolumn{2}{c}{0.24} & \multicolumn{2}{c}{328.67} & \multicolumn{2}{c}{830.49} & \multicolumn{2}{c}{6.92} \\
\rowcolor{gray!15}
&\multicolumn{2}{c}{110.29} & \multicolumn{2}{c}{0.27} & \multicolumn{2}{c}{369.62} & \multicolumn{2}{c}{865.03} & \multicolumn{2}{c}{4.24} \\
&\multicolumn{2}{c}{211.07} & \multicolumn{2}{c}{0.34} & \multicolumn{2}{c}{334.77} & \multicolumn{2}{c}{992.94} & \multicolumn{2}{c}{8.85} \\
\midrule
\rowcolor{gray!15}
\multicolumn{11}{c}{\textbf{Calibrated Parameters}} \\
& \multicolumn{5}{c|}{\textbf{FEMU-Adjoint}} & \multicolumn{5}{c}{\textbf{VFM-Adjoint}} \\
\cline{2-11}
&$E$ (GPa) & \bs{$\nu$} & $Y$ (MPa) & $S$ (MPa) & $D$ & $E$ (GPa) & $\nu$ & $Y$ (MPa) & $S$ (MPa) & $D$ \\
\midrule
\textbf{Mean} & 200 & 0.30 & 330 & 1000 & 10 & 200.0043 & 0.2999 & 330.0030 & 1000.0082 & 9.9998 \\
\midrule
\textbf{Error (\%)} & 0 & 0 & 0 & 0 & 0 & 0.0021 & 0.0154 & 0.0009 & 0.0008 & 0.0020 \\
\bottomrule
\end{tabular}%
}
\caption*{\footnotesize\emph{Note}: The mean errors are reported relative to the true solutions.}
\label{tb:initialGuessSens}
\end{table}
In this subsection, we examine the robustness of both the FEMU-Adjoint and VFM-Adjoint in yielding accurate material parameter estimates when starting from different sets of initial guesses. To this end, we analyze ten different initial guess sets, each chosen to significantly deviate from the true values, with the initial data generated from a uniform random distribution. The lower and upper bounds for generating the random data are set to match $\bp_{\text{lo}}$ and $\bp_{\text{hi}}$ (as given in Section \ref{sec:pb-1}), respectively. The results of this comparison are summarized in Table \ref{tb:initialGuessSens}.
FEMU-Adjoint provides an exact inversion of the material properties while VFM-Adjoint also yields almost perfect results. Figure \ref{fig:initialGuessSens} further illustrates the variations in initial guesses and the corresponding calibrated parameter values for both methods, in comparison to the true data, across different initial data sets on a logarithmic scale.
We can highlight the practical applicability of these methods, in the case of noiseless data with no model-form errors, and even when prior knowledge of material properties is limited and the initial guess is within reasonable bounds.

\subsection{\textbf{E3:}  Inverse solution accuracy with noisy data}

\begin{figure}
\includegraphics[width=0.8\linewidth,center]{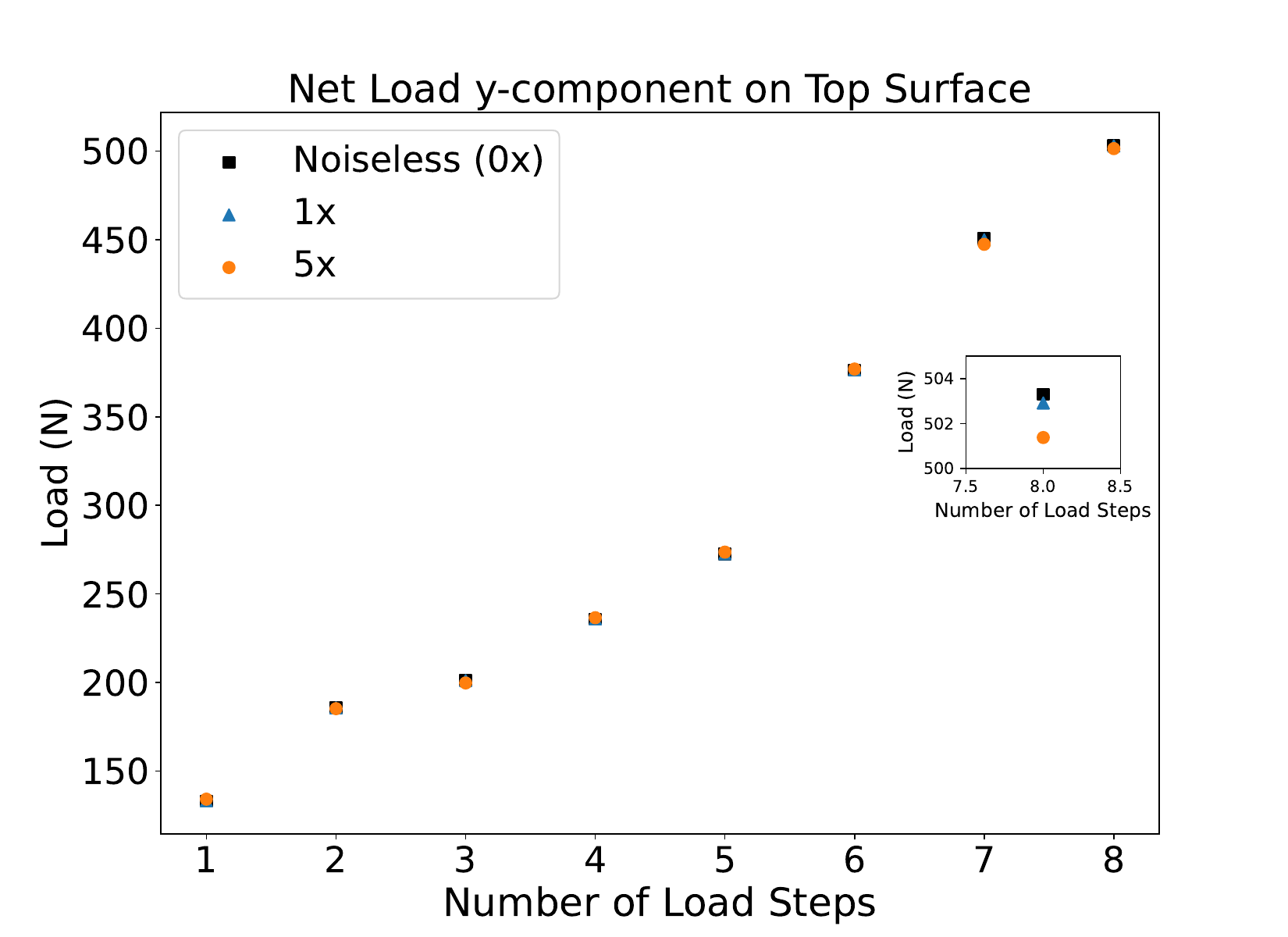}
 \caption{Scatter plot comparing the noiseless load data with its noisy counterparts at various load steps. Note that in the generated noisy synthetic data, 0x is equivalent to noiseless data.}
 \label{fig:noisy_load}
 \end{figure}

\begin{table}[h!]
\centering
\caption{Comparison of the calibrated material parameters for a plate with an asymmetric notch undergoing finite strain elasto-plastic deformation, utilizing noisy synthetic data}
\begin{tabular}{>{\raggedright\arraybackslash}m{1.7cm} |c c c | c c c}
\toprule
\rowcolor{gray!10}
& \multicolumn{2}{c}{$Y$ (MPa)} & \multicolumn{2}{c}{$S$ (MPa)} & \multicolumn{2}{c}{$D$}\\
\midrule
\textbf{Truth} &  \multicolumn{2}{c}{330} & \multicolumn{2}{c}{1000} & \multicolumn{2}{c}{10} \\
\midrule
\rowcolor{gray!15}
\multicolumn{7}{c}{\textbf{calibrated parameters using unfiltered noisy synthetic data }} \\
 & \multicolumn{3}{c|}{\textbf{FEMU-Adjoint}} & \multicolumn{3}{c}{\textbf{VFM-Adjoint}} \\
\cline{2-7}
\textbf{DNSF} & $Y$ (MPa) & $S$ (MPa) & $D$ & $Y$ (MPa) & $S$ (MPa) & $D$ \\
\midrule
0x & 329.8700 & 999.5013 & 9.9939 &  331.1416 & 1001.9545 & 9.9066 \\
\rowcolor{gray!7}
0.5x & 329.8797 & 999.5129 & 9.9933 &  335.0300 & 1037.2422 & 9.3195 \\
1x & 329.8894 & 999.5243 & 9.9928 &  350.0212 & 1216.3747 & 7.1817 \\
\rowcolor{gray!7}
2x & 329.9090 & 999.5474 & 9.9917 &  394.0667 & \underline{2000} & 3.4152 \\
5x & 329.9675 & 999.6165 & 9.9883 &  478.6309 & \underline{2000} & 2.3789 \\

\midrule
\rowcolor{gray!15}
\multicolumn{7}{c}{\textbf{calibrated parameters using filtered noisy synthetic data }} \\
\midrule
0x & 330.1098 & 1000.5099 & 9.9704 &  332.2053 & 1002.2356 & 9.8980 \\
\rowcolor{gray!7}
0.5x & 330.1198 & 1000.5221 & 9.9698 &  332.2978 & 1007.9728 & 9.8173 \\
1x & 330.1300 & 1000.5346 & 9.9692 &  332.6146 & 1015.6207 & 9.7053 \\
\rowcolor{gray!7}
2x & 330.1499 & 1000.5589 & 9.9680 &  334.4073 & 1041.4420 & 9.3205 \\
5x & 330.2097 & 1000.6321 & 9.9646 &  351.9797 & 1314.0205 & 6.4964 \\
\bottomrule
\end{tabular}%
\caption*{\footnotesize\emph{Note}: The mean values of the calibrated parameters are reported pertaining to the ten sets of random initial guesses. Load noise scale factor is 1x viz. noise in the load is at the base level ${\epsilon}^{\text{F}}_{\text{noise}} = 0.25$. DNSF: Displacement noise scale factor. For the displacement base level noise see Eq. \eqnref{eq:disp_noise_base}. The calibrated parameters that are hitting the bound are indicated by $\underline{(\cdot)}$.}
\label{tb:noisy_data}
\end{table}

\begin{figure}
\begin{subfigure}{0.32\textwidth}
\centering
\includegraphics[width=0.99\linewidth]{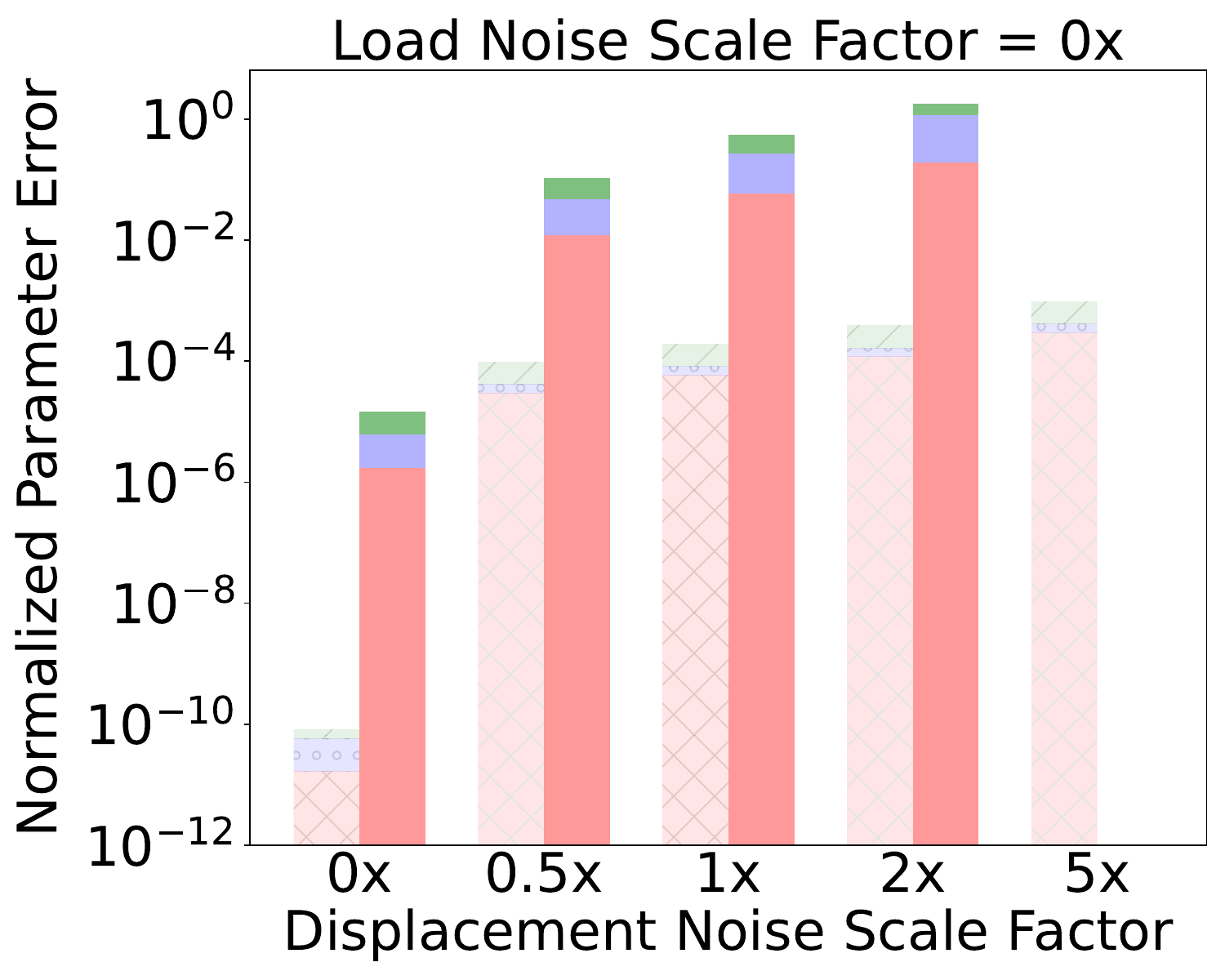}
\caption{}
\label{fig:pb-3_LNSF_0}
\end{subfigure}
\begin{subfigure}{0.32\textwidth}
\centering
\includegraphics[width=0.99\linewidth]{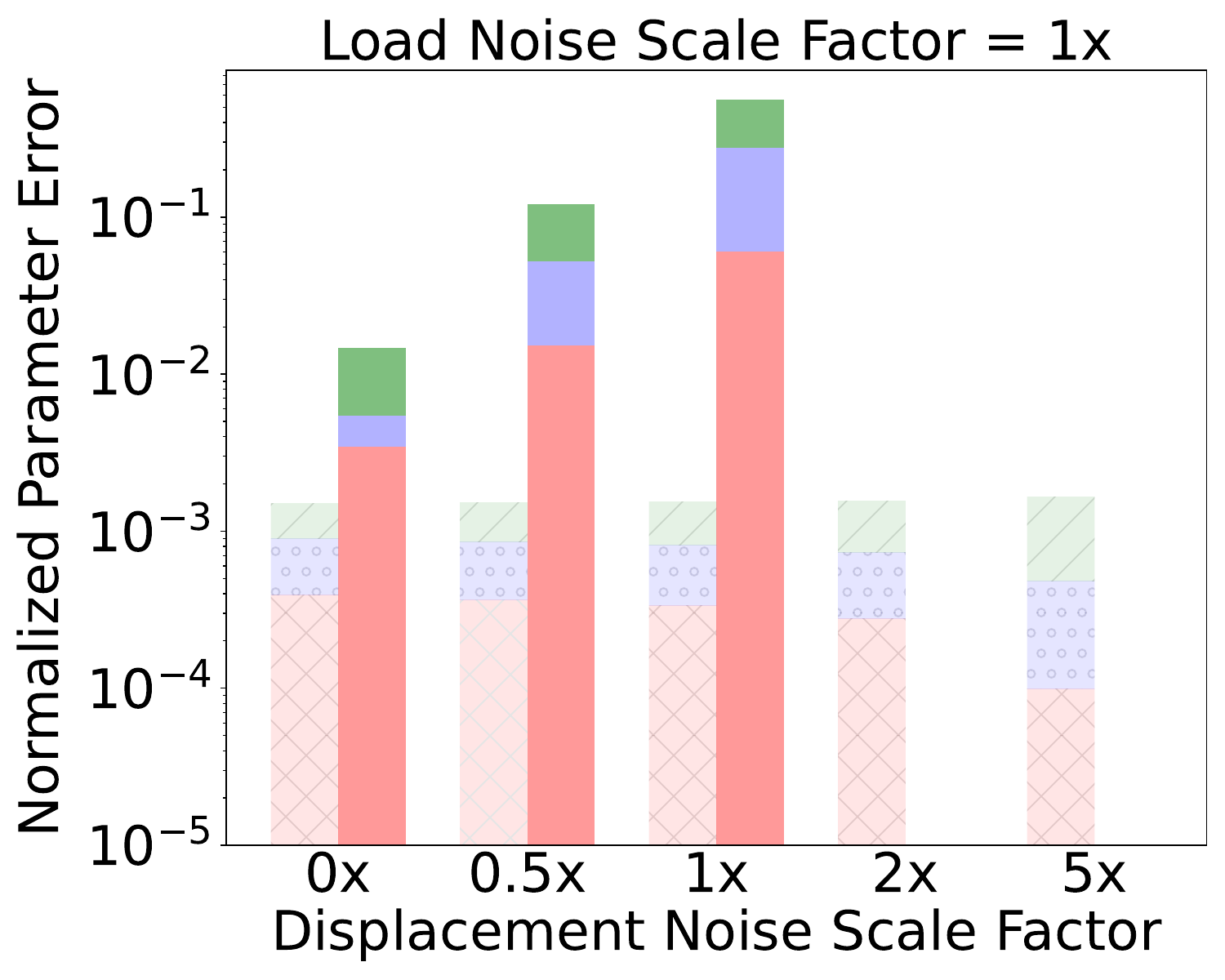}
\caption{}
\label{fig:pb-3_LNSF_1}
\end{subfigure}
\begin{subfigure}{0.32\textwidth}
\centering
\includegraphics[width=0.99\linewidth]{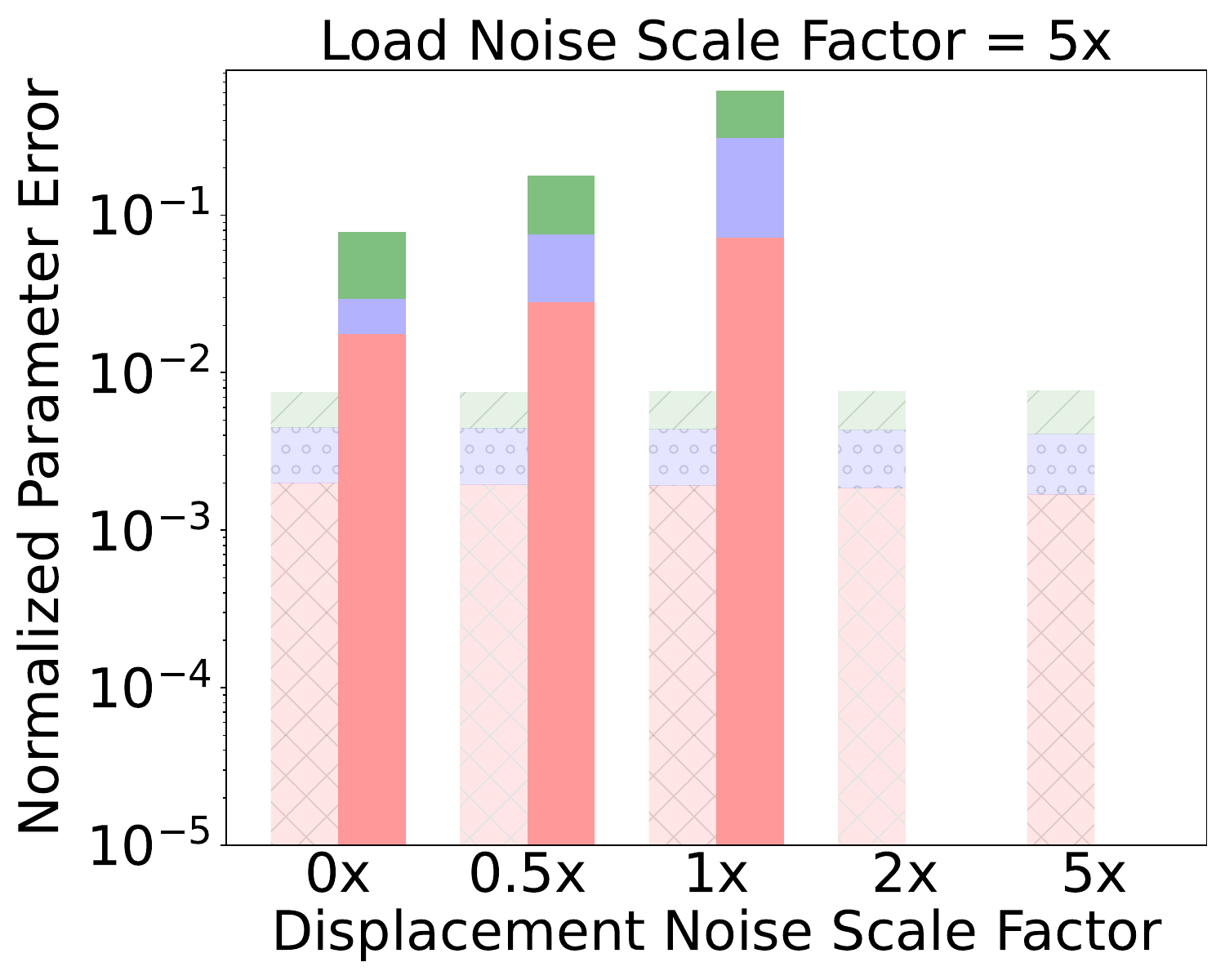}
\caption{}
\label{fig:pb-3_LNSF_10}
\end{subfigure}

\vspace{0.5em}
\begin{subfigure}{0.32\textwidth}
\centering
\scalebox{1.0}{\includegraphics[width=0.99\linewidth]{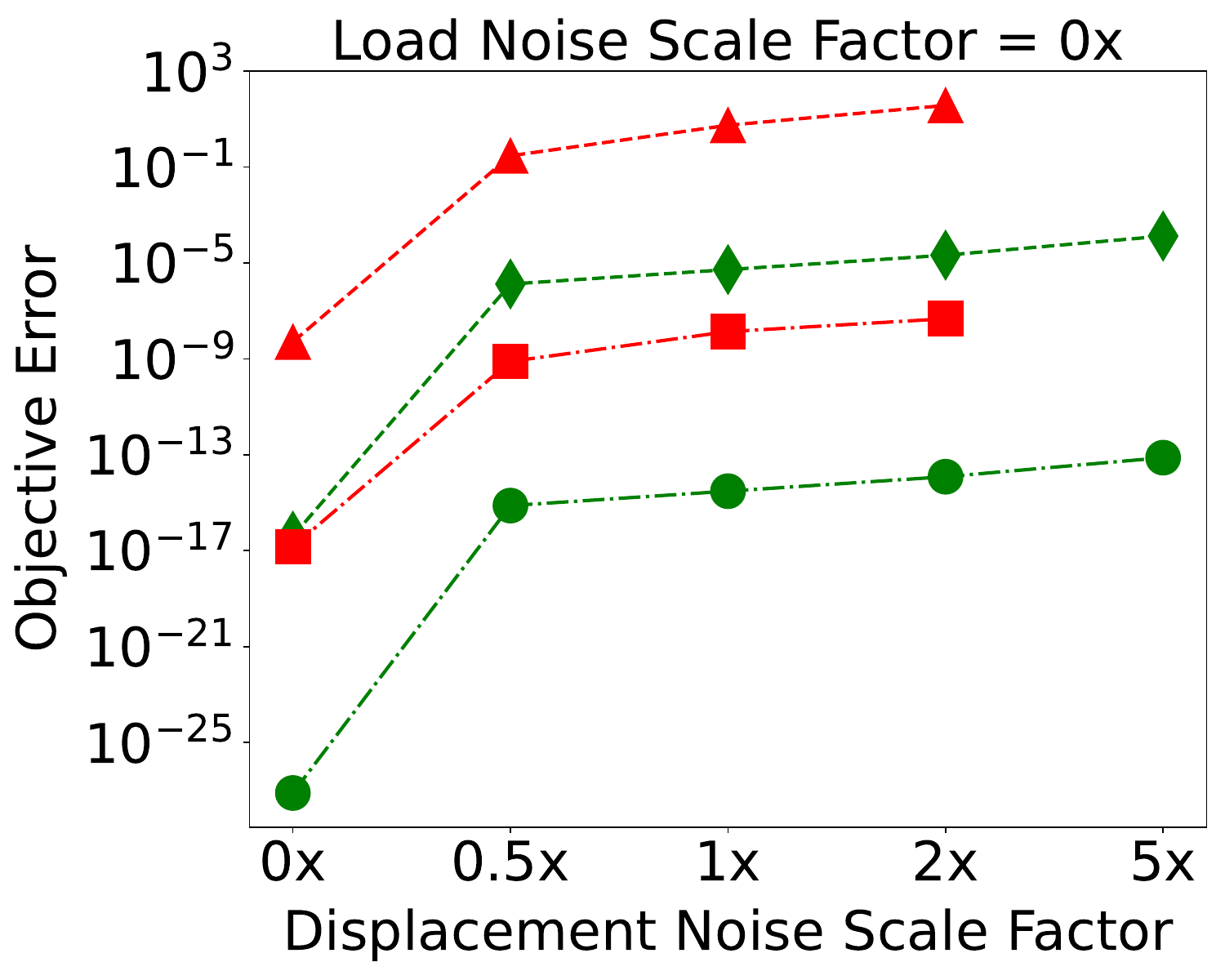}}
\caption{}
\label{fig:pb-3_load_disp_LNSF_0}
\end{subfigure}
\begin{subfigure}{0.32\textwidth}
\centering
\includegraphics[width=0.99\linewidth]{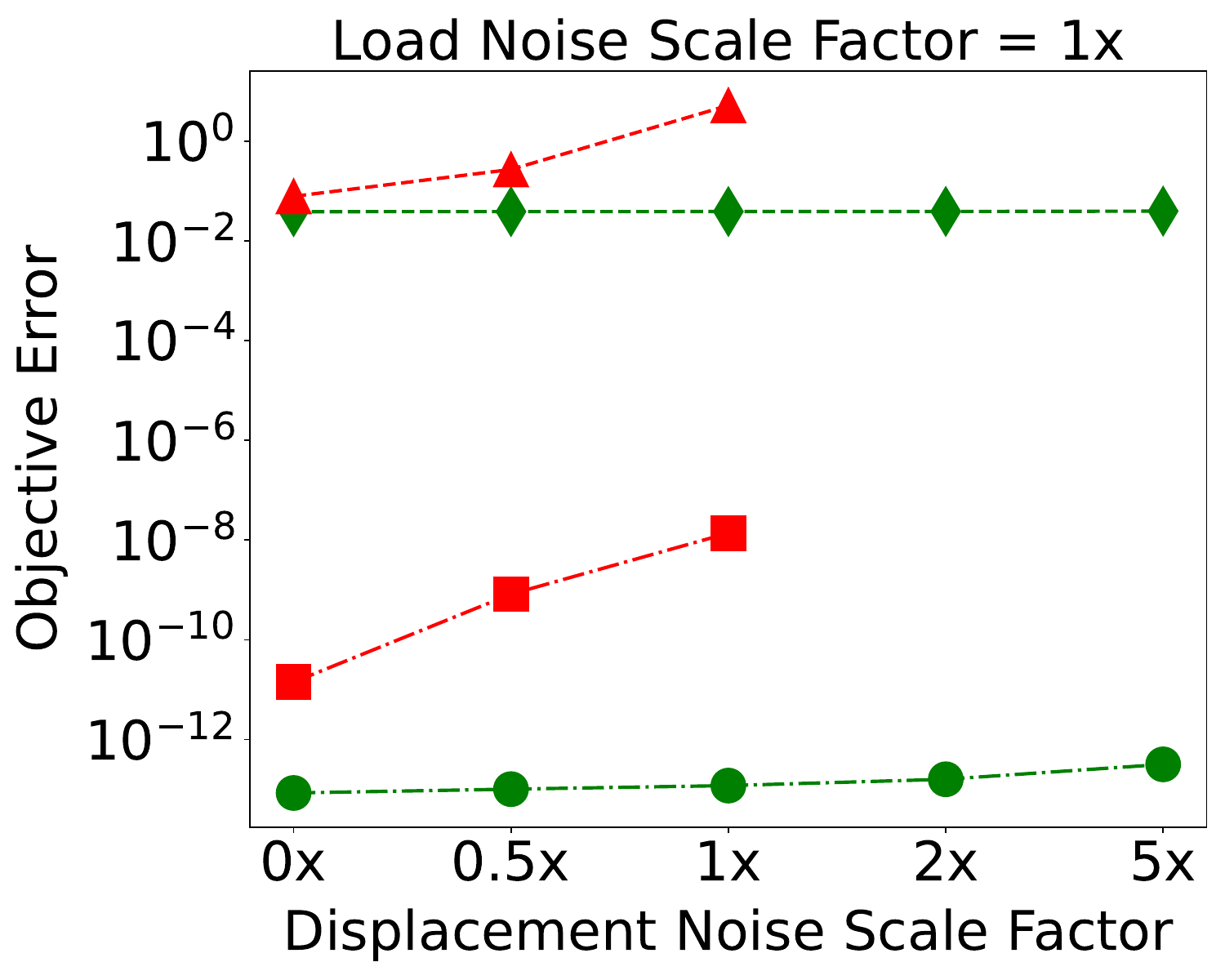}
\caption{}
\label{fig:pb-3_load_disp_LNSF_1}
\end{subfigure}
\begin{subfigure}{0.32\textwidth}
\centering
\includegraphics[width=0.99\linewidth]{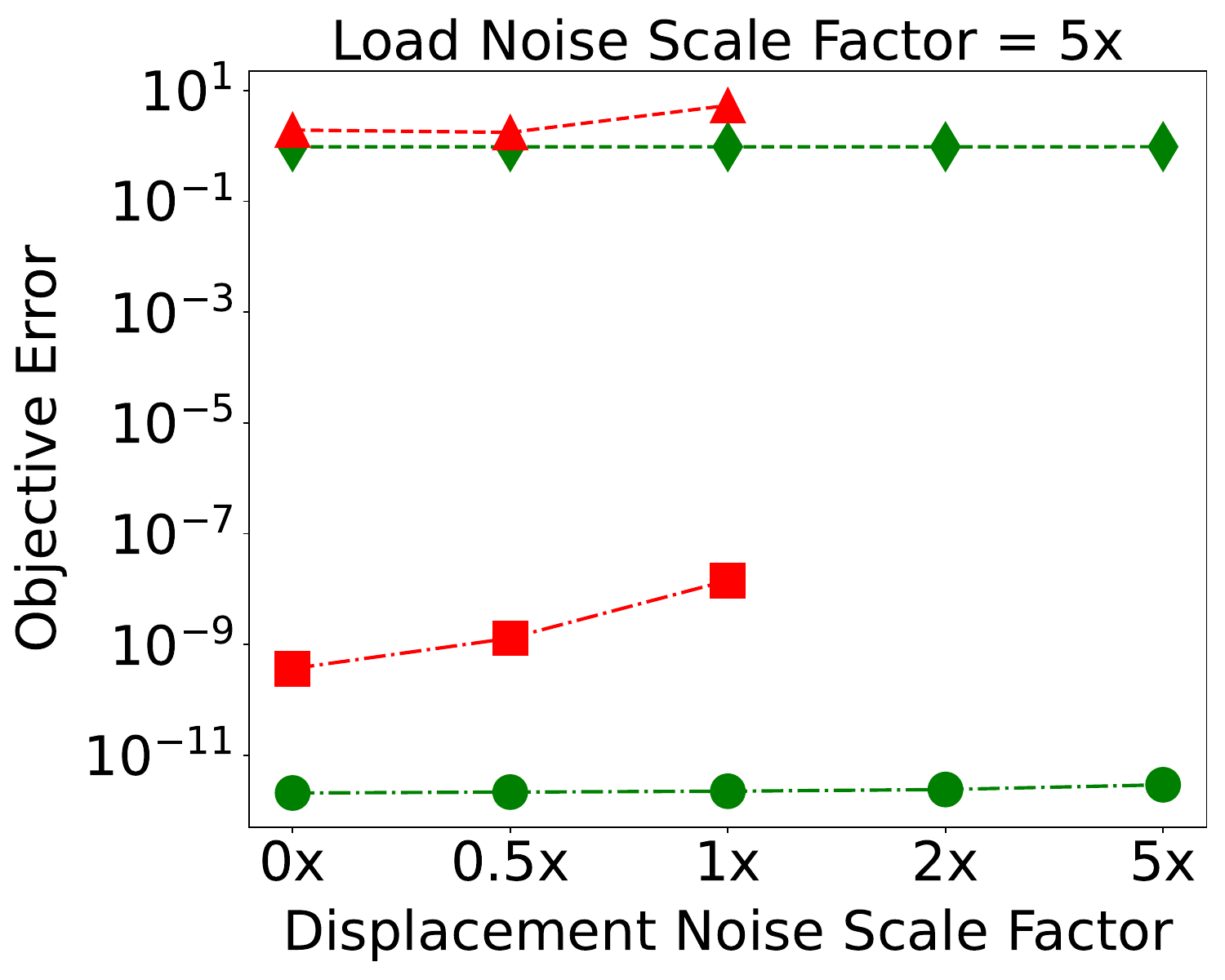}
\caption{}
\label{fig:pb-3_load_disp_LNSF_10}
\end{subfigure}

\vspace{0.2em}
\begin{subfigure}{0.45\textwidth}
\centering
\includegraphics[width=0.95\linewidth]{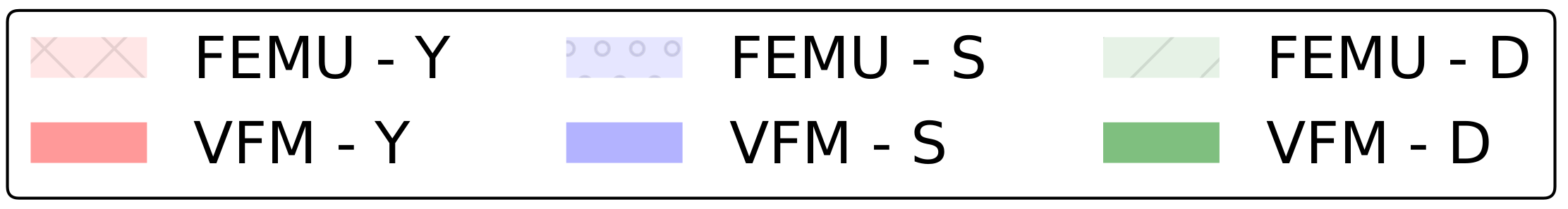}
\caption*{}
\end{subfigure}
\begin{subfigure}{0.46\textwidth}
\centering
\includegraphics[width=0.99\linewidth]{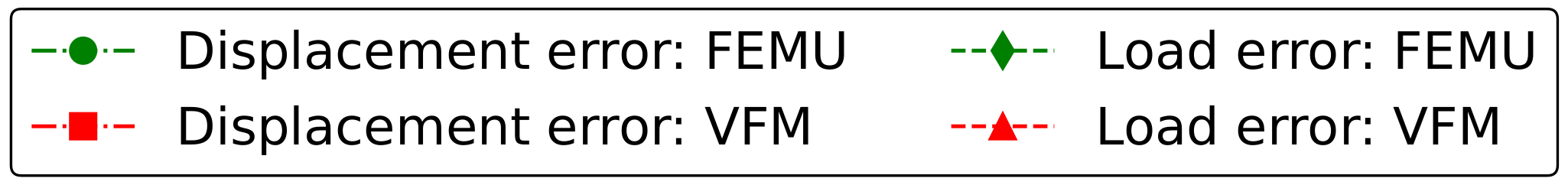}
\caption*{}
\end{subfigure}
\caption{Comparison of FEMU-Adjoint and VFM-Adjoint in calibrating the material parameters with unfiltered noisy synthetic data at varying noise levels, (a)-(c) normalized error in the calibrated material parameters (d)-(f) error in load and displacement objectives computed using the calibrated material parameters. For some noise levels, VFM-Adjoint fails to invert the material parameters because it hits the parameter bounds, which is why the VFM-Adjoint error data is not plotted. Error in load and displacement objective are in N$^2$ and mm$^2$, respectively.} 
\label{fig:pb-3_noisy_data}
\end{figure}

\begin{figure}
\begin{subfigure}{0.32\textwidth}
\centering
\includegraphics[width=0.99\linewidth]{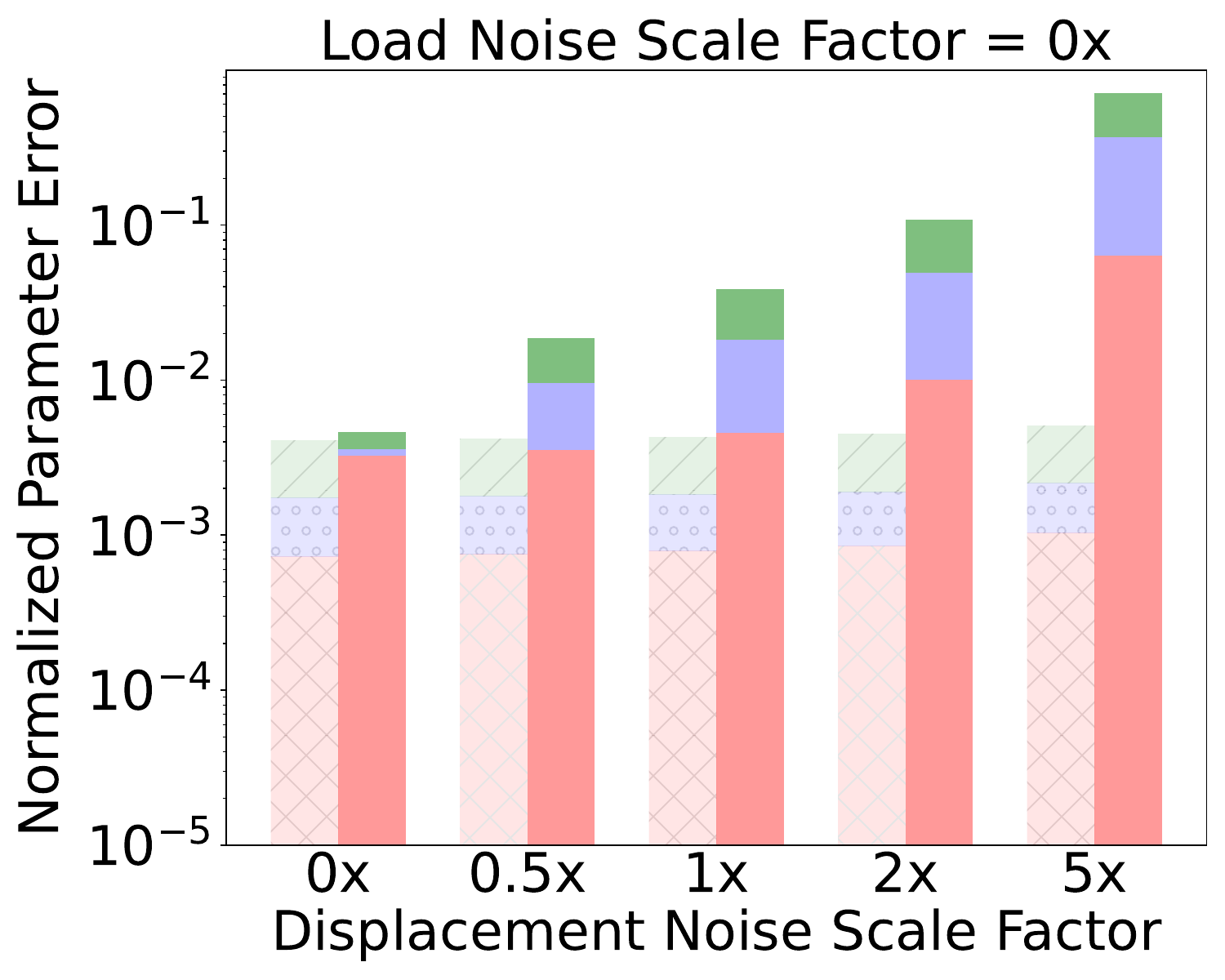}
\caption{}
\end{subfigure}
\begin{subfigure}{0.32\textwidth}
\centering
\includegraphics[width=0.99\linewidth]{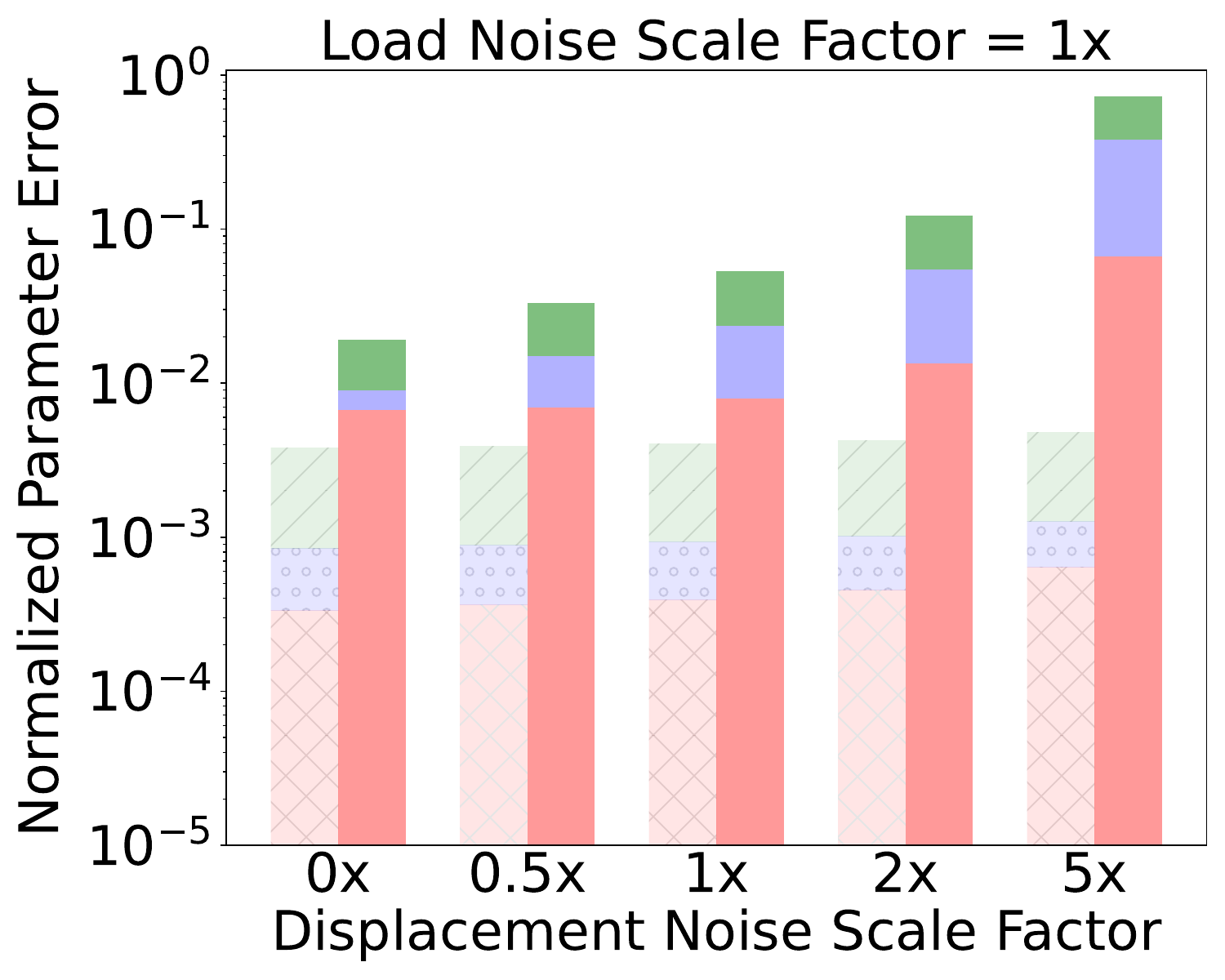}
\caption{}
\end{subfigure}
\begin{subfigure}{0.32\textwidth}
\centering
\includegraphics[width=0.99\linewidth]{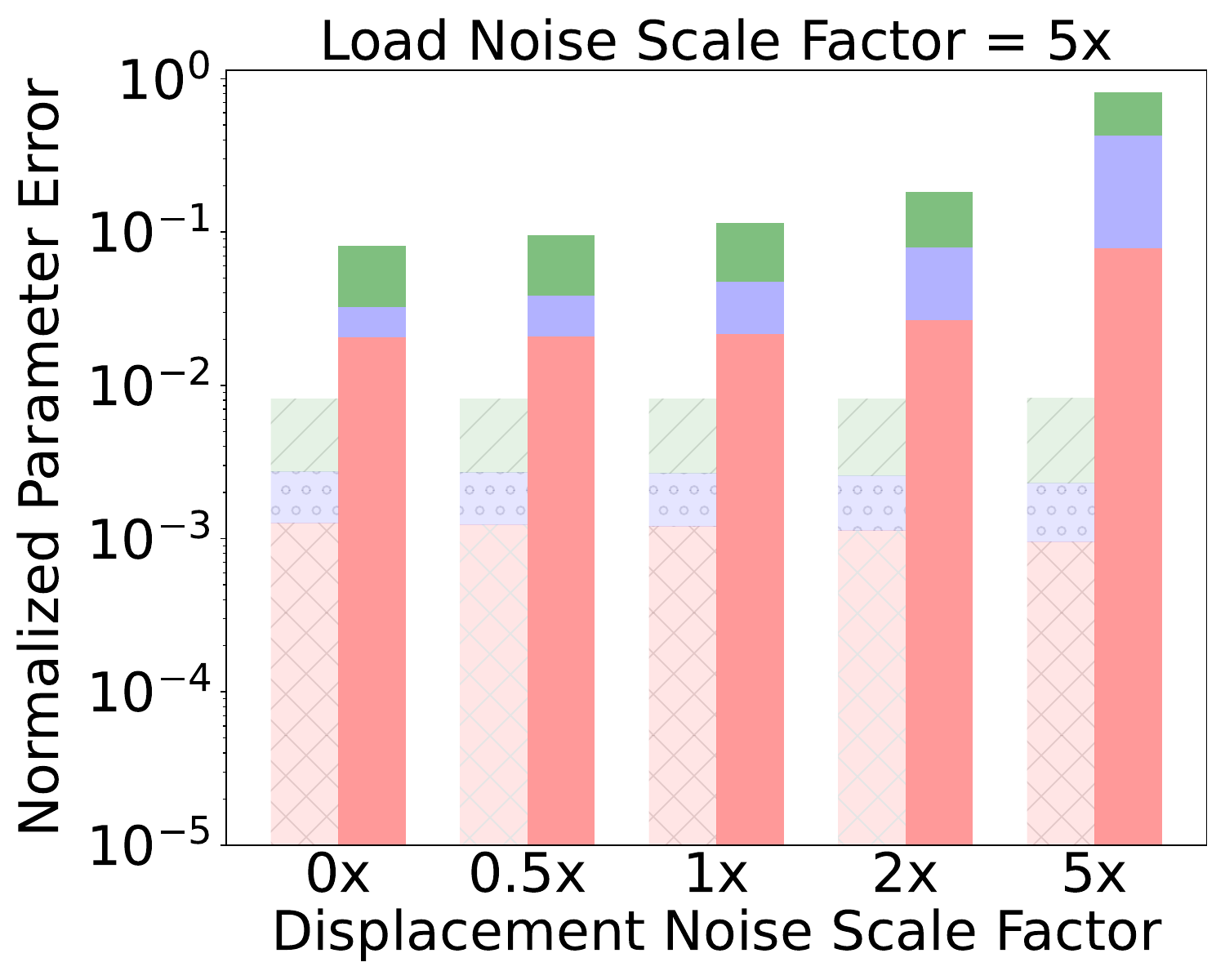}
\caption{}
\end{subfigure}

\vspace{0.5em}
\begin{subfigure}{0.32\textwidth}
\centering
\scalebox{1.0}{\includegraphics[width=0.99\linewidth]{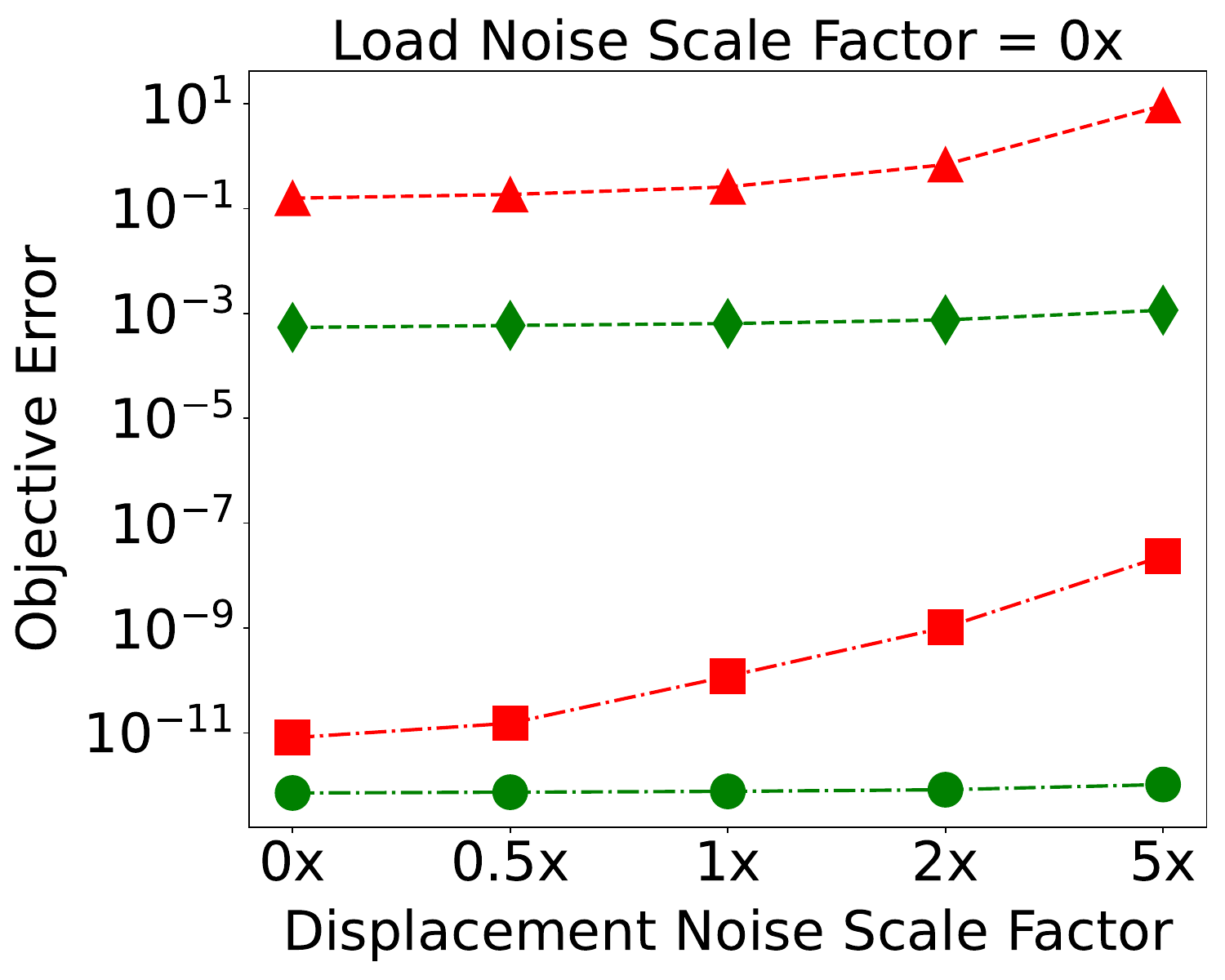}}
\caption{}
\end{subfigure}
\begin{subfigure}{0.32\textwidth}
\centering
\includegraphics[width=0.99\linewidth]{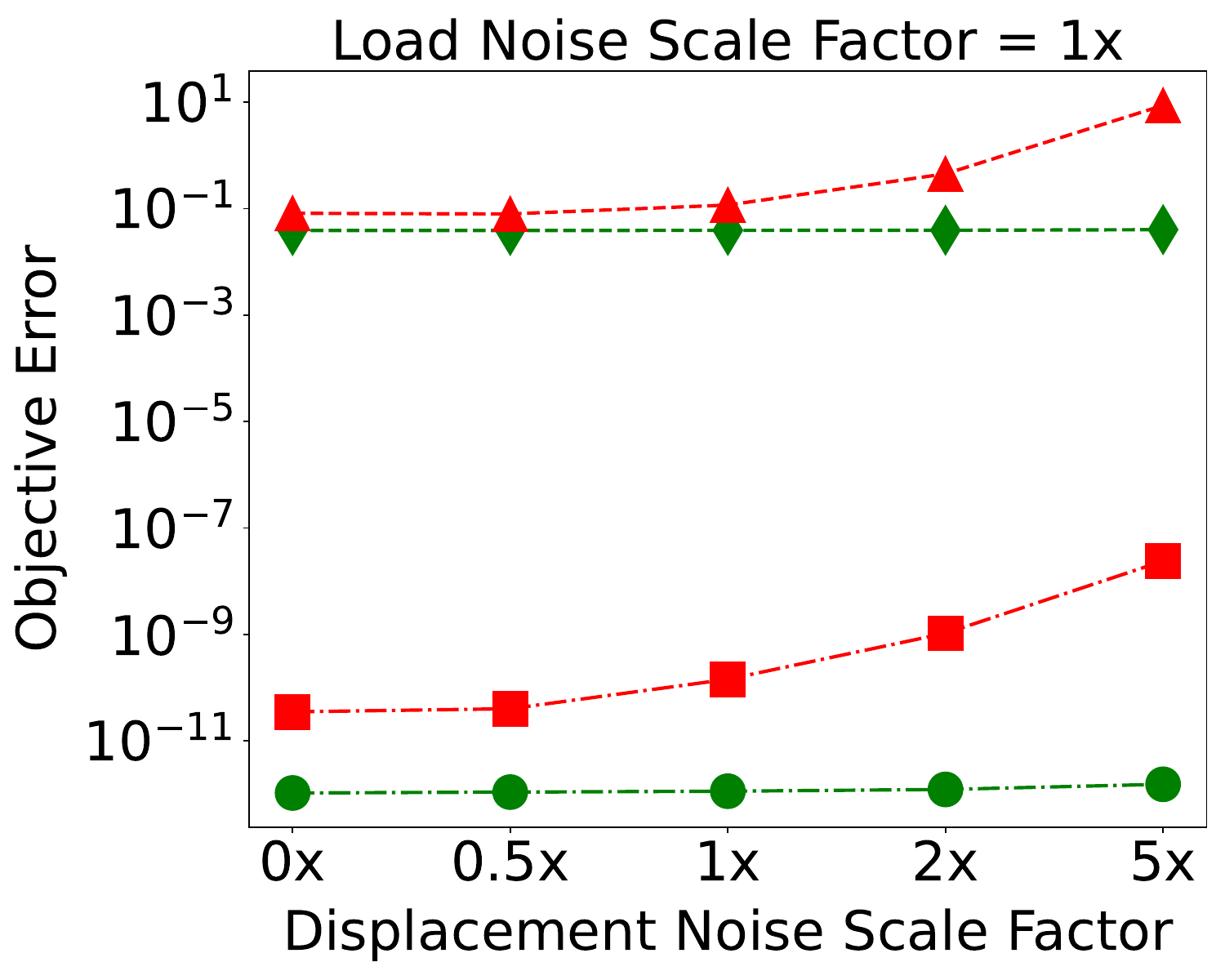}
\caption{}
\end{subfigure}
\begin{subfigure}{0.32\textwidth}
\centering
\includegraphics[width=0.99\linewidth]{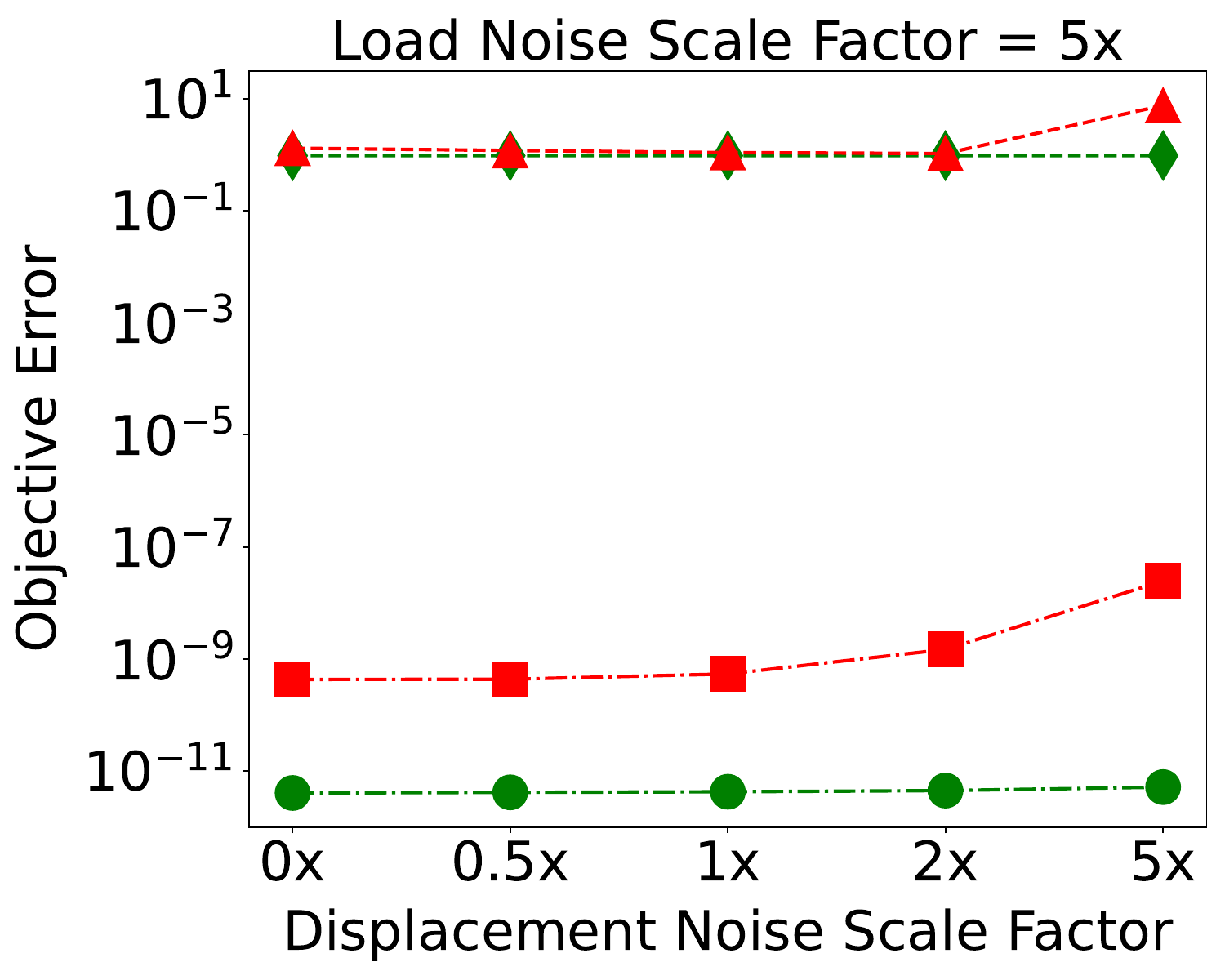}
\caption{}
\end{subfigure}

\vspace{0.2em}
\begin{subfigure}{0.45\textwidth}
\centering
\includegraphics[width=0.95\linewidth]{./figures/pb-3/pb-3_legend_error_bar.png}
\caption*{}
\end{subfigure}
\begin{subfigure}{0.46\textwidth}
\centering
\includegraphics[width=0.99\linewidth]{./figures/pb-3/pb-3_legend_load_disp.png}
\caption*{}
\end{subfigure}
\caption{Comparison of FEMU-Adjoint and VFM-Adjoint in calibrating the material parameters with filtered noisy synthetic data at varying noise levels, (a)-(c) normalized error in the calibrated material parameters (d)-(f) error in load and displacement computed using the calibrated material parameters. FEMU-Adjoint exhibits lower error compared to VFM-Adjoint at all noise levels. Error in load and displacement objective are in N$^2$ and mm$^2$, respectively.} 
\label{fig:pb-3_noisy_data_filtered}
\end{figure}

\begin{figure}
\begin{subfigure}{0.32\textwidth}
\centering
\includegraphics[width=0.99\linewidth]{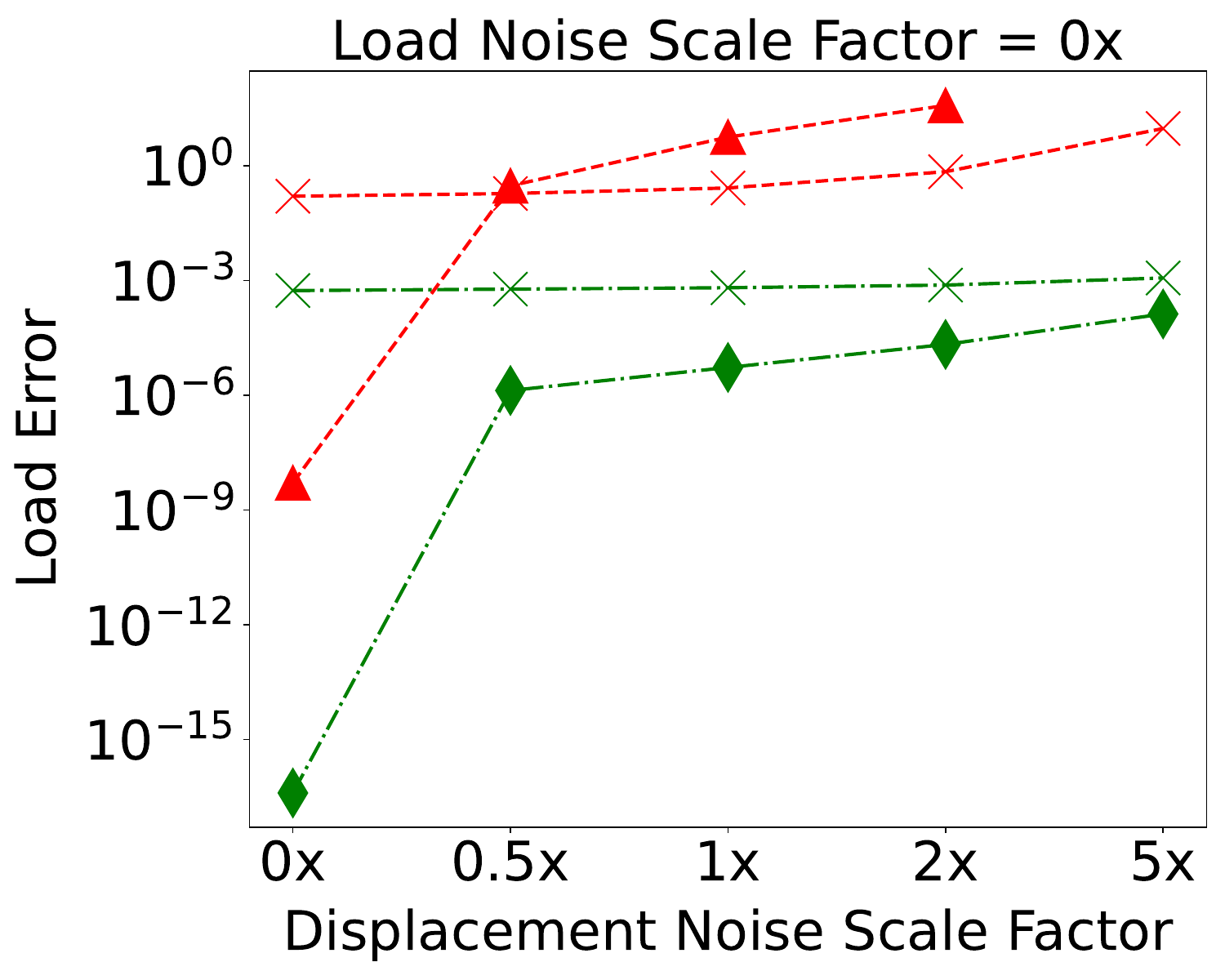}
\caption{}
\end{subfigure}
\begin{subfigure}{0.32\textwidth}
\centering
\includegraphics[width=0.99\linewidth]{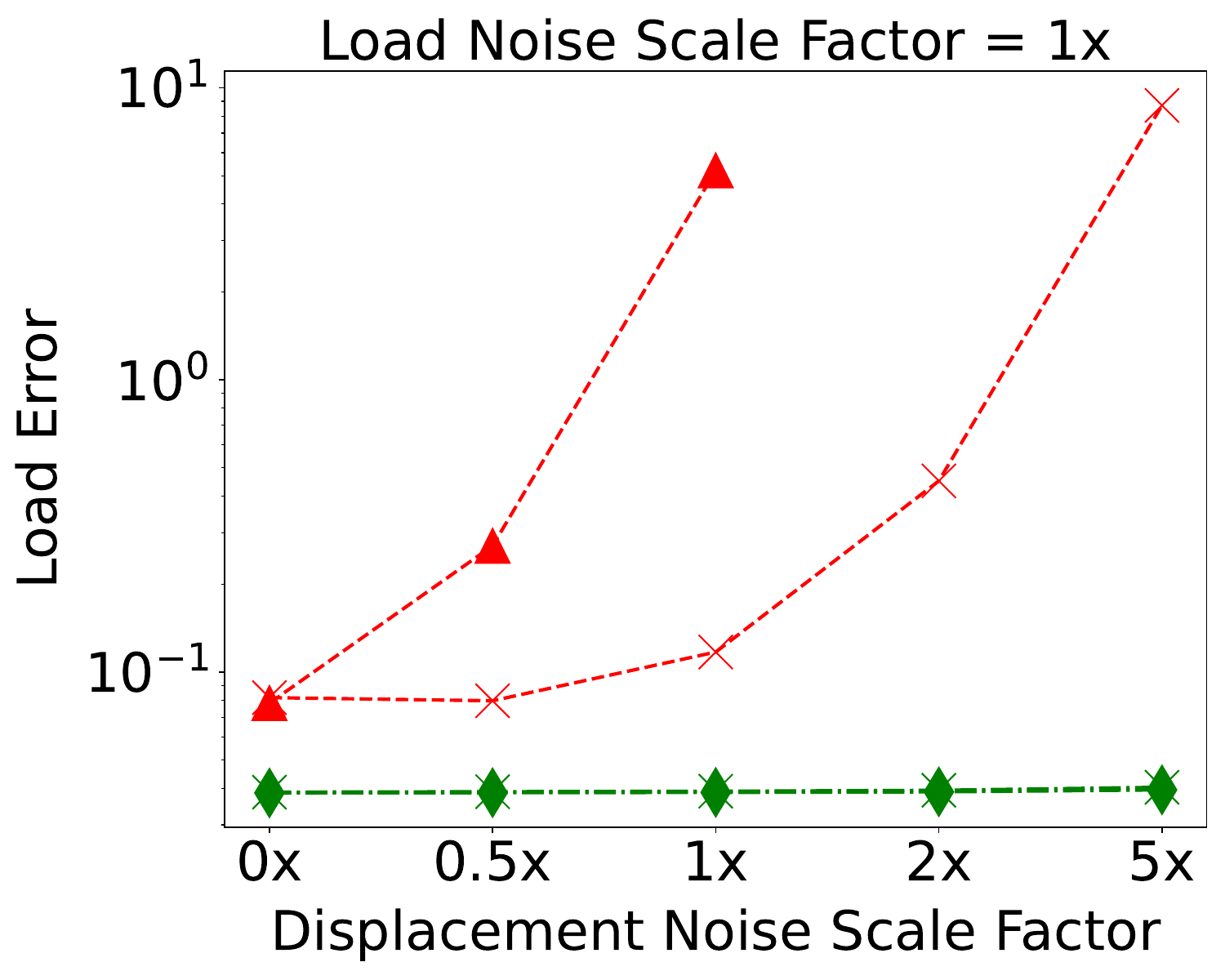}
\caption{}
\end{subfigure}
\begin{subfigure}{0.32\textwidth}
\centering
\includegraphics[width=0.99\linewidth]{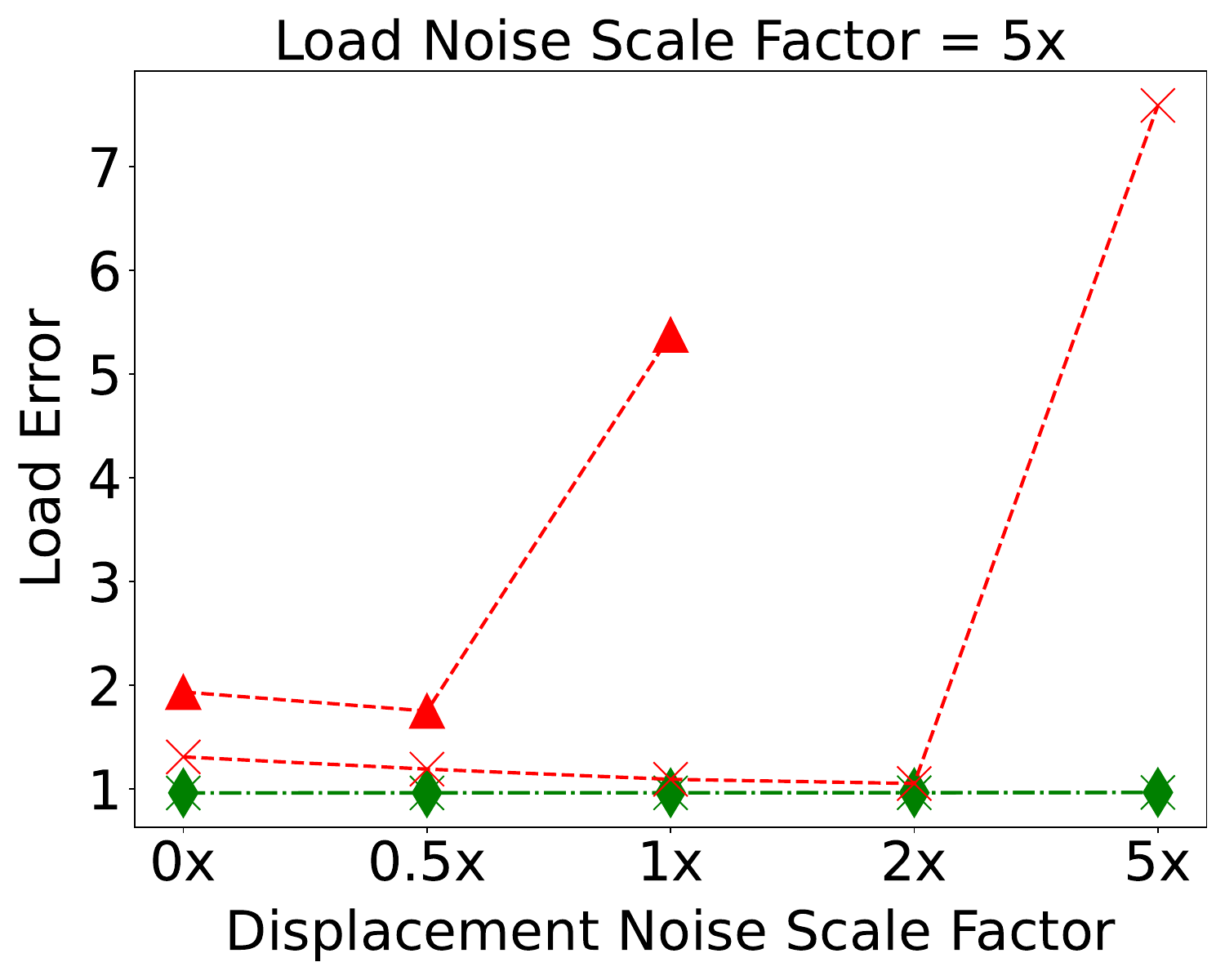}
\caption{}
\end{subfigure}

\vspace{0.5em}
\begin{subfigure}{0.32\textwidth}
\centering
\scalebox{1.0}{\includegraphics[width=0.99\linewidth]{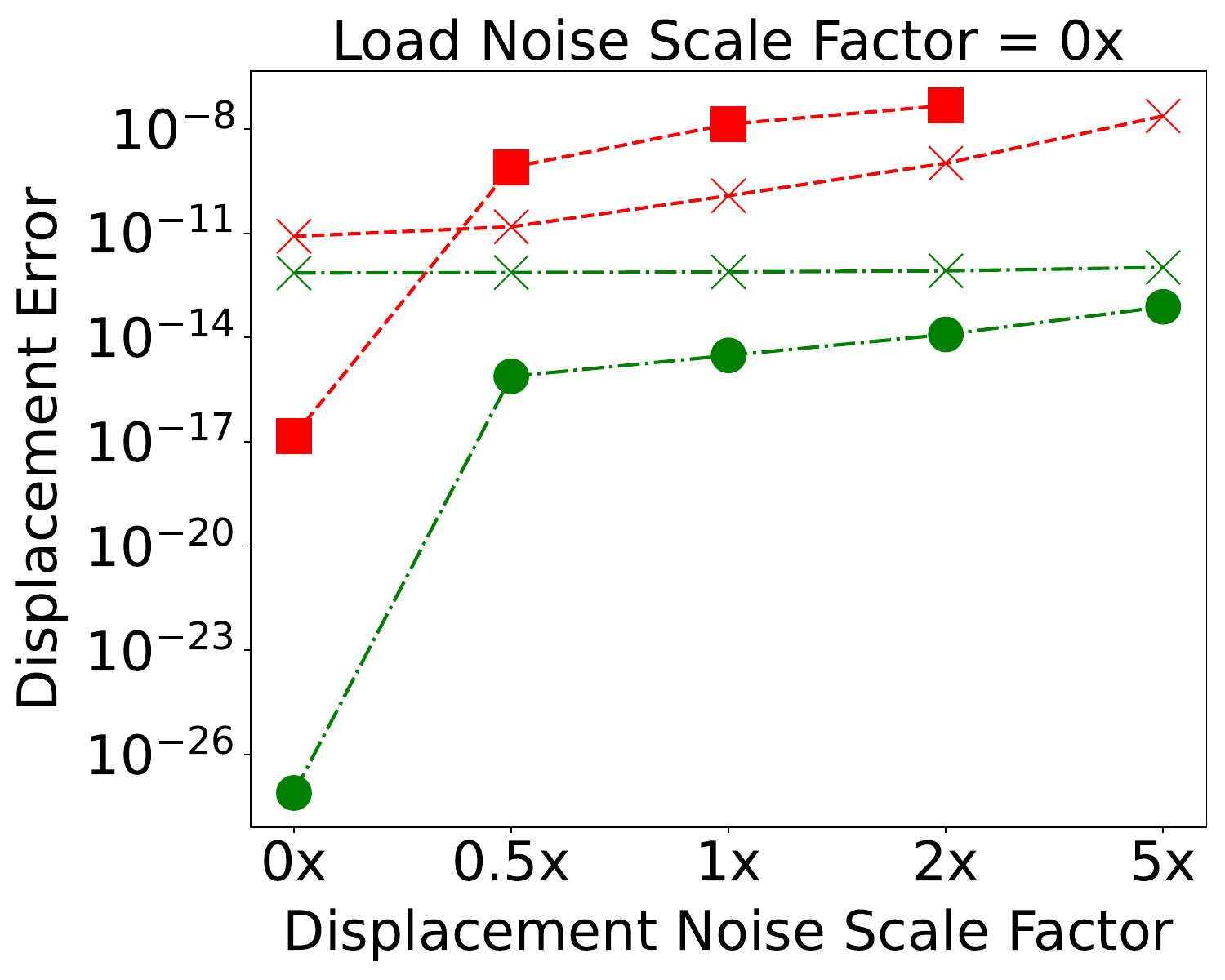}}
\caption{}
\end{subfigure}
\begin{subfigure}{0.32\textwidth}
\centering
\includegraphics[width=0.99\linewidth]{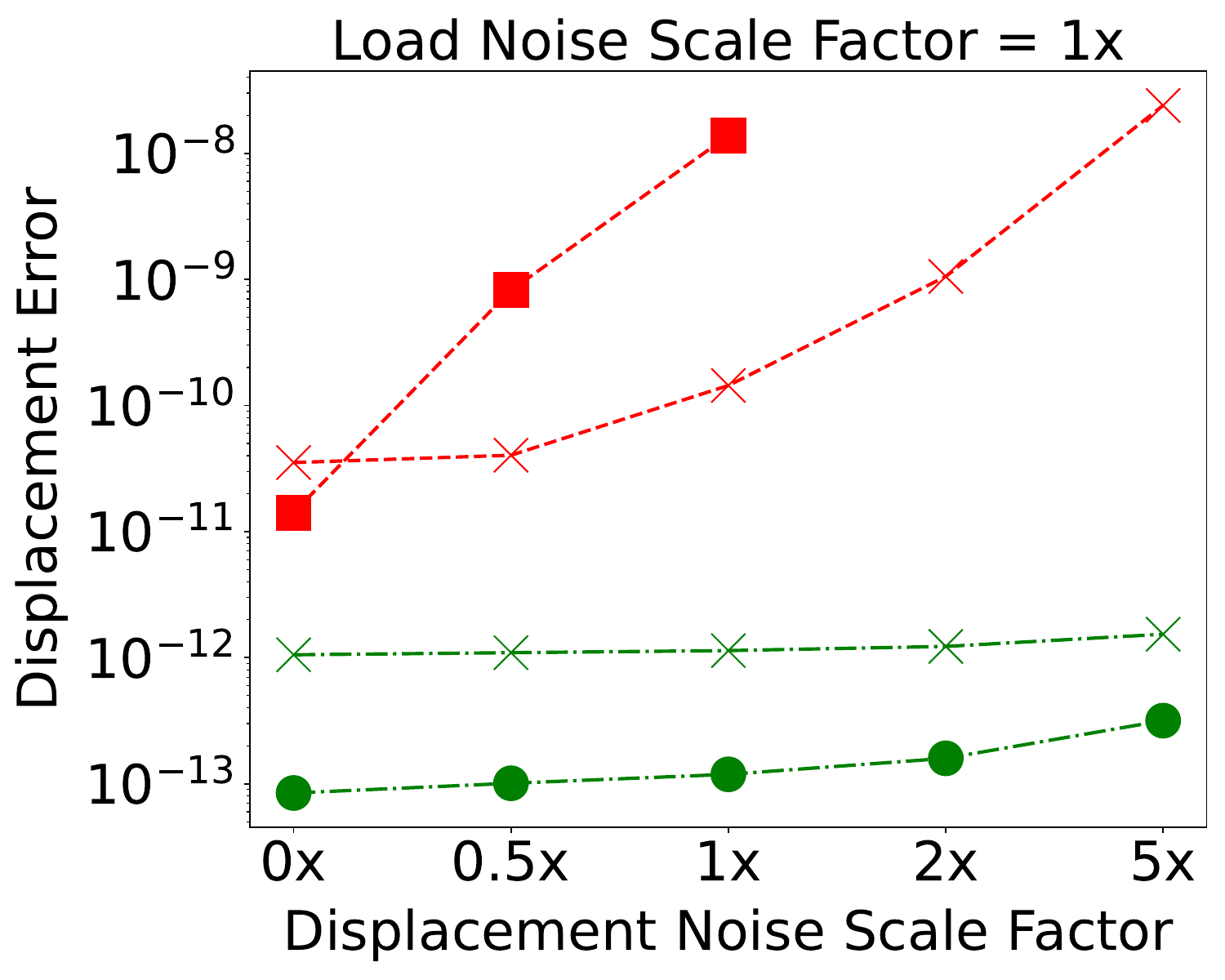}
\caption{}
\end{subfigure}
\begin{subfigure}{0.32\textwidth}
\centering
\includegraphics[width=0.99\linewidth]{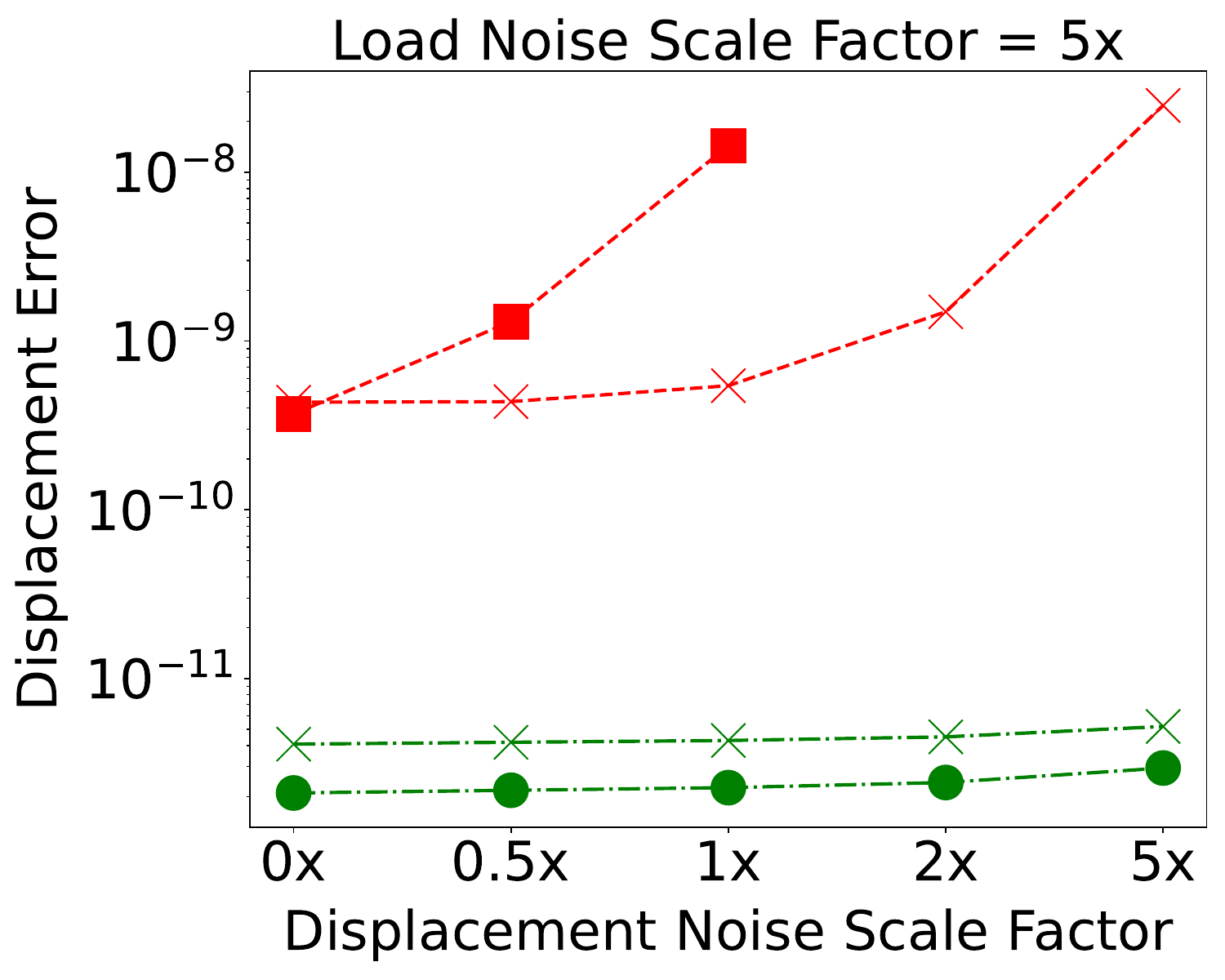}
\caption{}
\end{subfigure}

\vspace{0.2em}
\begin{subfigure}{0.54\textwidth}
\centering
\includegraphics[width=0.95\linewidth]{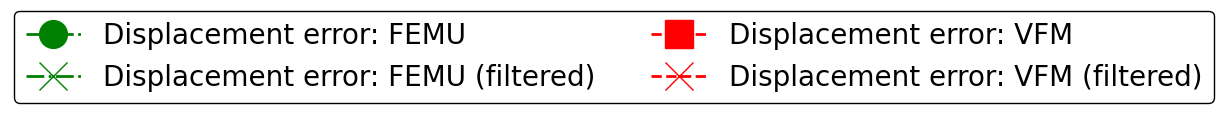}
\caption*{}
\end{subfigure}
\begin{subfigure}{0.42\textwidth}
\centering
\includegraphics[width=0.99\linewidth]{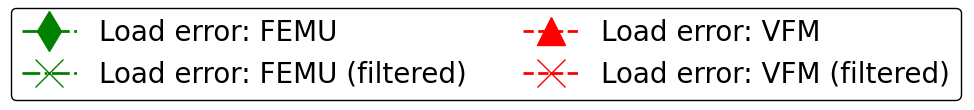}
\caption*{}
\end{subfigure}
\caption{Comparison of load and displacement errors calculated using calibrated parameters for FEMU-Adjoint and VFM-Adjoint, based on filtered and unfiltered noisy synthetic data: (a)-(c) load error, (d)-(f) displacement error. FEMU-Adjoint successfully inverts material parameters regardless of noise filtering, even at higher noise levels. In contrast, VFM-Adjoint fails to invert material parameters and hits parameter bounds when noise is unfiltered. Error in load and displacement objective are in N$^2$ and mm$^2$, respectively.} 
\label{fig:pb-3_filtered_vs_unfiltered}
\end{figure}


In this subsection, we evaluate the performance of the FEMU and VFM methods in calibrating material parameters when noise is introduced into both the displacement and load calibration data. We generate synthetic noisy measurements for both of these quantities by adding a random value sampled from a normal distribution with zero mean and unit variance, scaled by a noise factor, as follows:

\begin{equation}
\begin{aligned}
    \bumeas_{\text{noisy}} &= \bumeas +  n \, {\epsilon}^{\text{u}}_{\text{noise}}\\
    \Fmeas_{\text{noisy}} &= \Fmeas +  n \, {\epsilon}^{\text{F}}_{\text{noise}},
\end{aligned}
\end{equation}

\noindent where $n \sim \mathcal{N}(0,1)$. ${\epsilon}^{\text{u}}_{\text{noise}}$ and ${\epsilon}^{\text{F}}_{\text{noise}}$ are given as: $ {\epsilon}^{\text{u}}_{\text{noise}} = \epsilon^{*\text{u}}_{\text{noise}} \times \alpha^{\text{u}}_{\text{noise}}$ and ${\epsilon}^{\text{F}}_{\text{noise}} = \epsilon^{*\text{F}}_{\text{noise}} \times \alpha^{\text{F}}_{\text{noise}}$. Here, $\epsilon^{*\text{u}}_{\text{noise}}$ and $\epsilon^{*\text{F}}_{\text{noise}}$ denote base level noise in the displacement and load, respectively. The base level noise is meant to be close to the noise encountered in actual DIC measurements. $\alpha^{\text{u}}_{\text{noise}}$ and $\alpha^{\text{F}}_{\text{noise}}$ respectively represent the displacement noise scale factor (DNSF) and load noise scale factor. To estimate $\epsilon^{*\text{u}}_{\text{noise}}$, we consider typical parameters for experimental DIC measurements. Specifically, we assume that the $y$-extent of the undeformed sample, $L_y$, fills $80\%$ of the field of view, the standard deviation of the image noise is 0.01 pixels, and the camera resolution is 100 MP ($2048 \times 2048$ pixels). This leads to the following equation:

\begin{equation}
    \epsilon^{*\text{u}}_{\text{noise}} = \frac{0.01 L_y}{0.8 \times 2048}.
    \label{eq:disp_noise_base}
\end{equation}

For the load we assume ${\epsilon}^{\text{F}}_{\text{noise}} = 0.25$. 
We visualize the effect of the load noise in Fig. \ref{fig:noisy_load}, which shows the net load on the top surface at different steps for various noise levels. In the following, we call a 1x noise scale factor the base-level noise for both displacement and load.

The results for FEMU-Adjoint and VFM-Adjoint under base-level load noise and with varying DNSF data are summarized in Table \ref{tb:noisy_data} for both unfiltered and filtered. The DNSF ranges from 0x to 5x. We refer to Figure \ref{fig:smoothing} for a sense of the impact of the noise and the filtering process on the displacement data for a DNSF of 5x.
For each DNSF, the parameters are inverted for ten sets of random initial guesses to mitigate bias. We then report the mean values for each DNSF used in the analysis.

For the unfiltered data, FEMU-Adjoint inverts all parameters accurately for all DNSFs. In contrast, for VFM-Adjoint, the calibrated values start to deviate significantly from the true values as the DNSF increases. On top of that, for DNSF = 2x and 5x, the inverted hardening saturation modulus $(S)$ with VFM-Adjoint hits the parameter bounds for the unfiltered noisy data, which we indicate by underlined values in the Table \ref{tb:noisy_data}. 
This highlights that VFM is more sensitive to noise in the data. 
The calibrated material parameters based on the smoothed data are reported in Table \ref{tb:noisy_data} for both FEMU-Adjoint and VFM-Adjoint. In a secondary study, we also apply a moving least squares (MLS) filter to the noisy synthetic displacement data \cite{wendland2004scattered, compadre_toolkit}. A comparison between unfiltered vs. filtered noisy displacement data is shown in Appendix \ref{app:filtered_unfiltered_noisy_data}. Notably, with the filtered data, VFM successfully inverts the material parameters even at high noise levels (DNSF = 2x and 5x). However, FEMU-Adjoint still inverts the values more accurately.

 Fig. \ref{fig:pb-3_noisy_data} visualizes the parameter errors of FEMU-Adjoint and VFM-Adjoint for the unfiltered noisy synthetic data at varying noise levels. Figs. (\ref{fig:pb-3_LNSF_0})-(\ref{fig:pb-3_LNSF_10}) depict a comparison of the normalized parameter errors, while Figs. (\ref{fig:pb-3_load_disp_LNSF_0})-(\ref{fig:pb-3_load_disp_LNSF_10}) depict the squared objective error in load and displacement using the calibrated material parameters.
  The corresponding error plots for the noise-filtered case are shown in Fig. \ref{fig:pb-3_noisy_data_filtered}. We also report a direct comparison of filtered and unfiltered cases in Fig. \ref{fig:pb-3_filtered_vs_unfiltered}.
Overall, these results allow us to highlight the importance of the quality of the measured data, i.e., level of noise, on the accuracy of the inverse solutions. While both FEMU and VFM are obtaining acceptable solutions under low- and base noise, FEMU, in particular, seems more robust for higher noise levels.

\subsection{ \textbf{E4:}  Model form error due to hardening misspecification}

\begin{figure}
\begin{subfigure}{0.49\textwidth}
\centering
\includegraphics[width=0.99\linewidth]{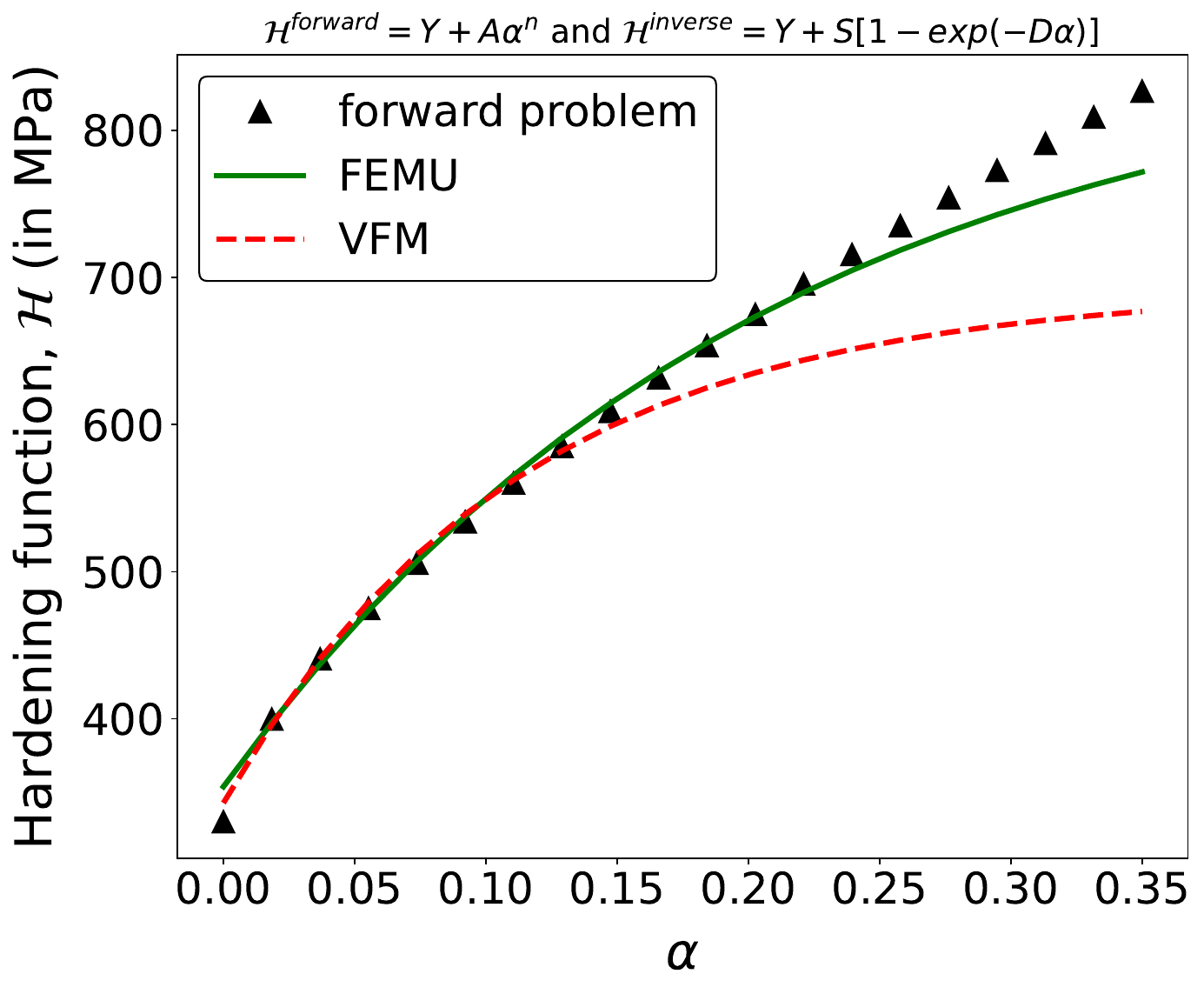}
\caption{$K = 0$}
\label{fig:compre_hardening_K_0}
\end{subfigure}
\begin{subfigure}{0.49\textwidth}
\centering
\includegraphics[width=0.99\linewidth]{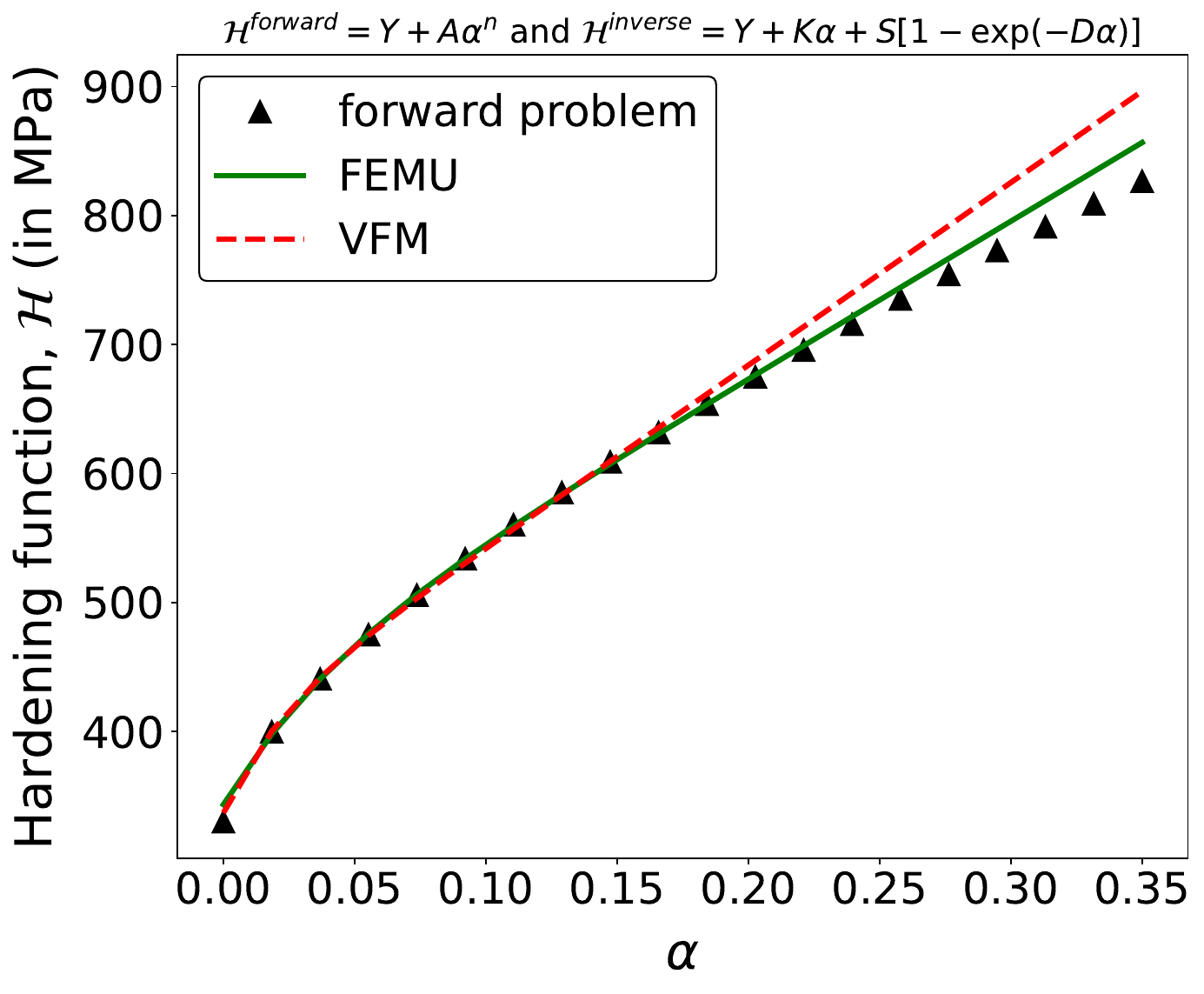}
\caption{$K \neq 0$}
\label{fig:compre_hardening_K}
\end{subfigure}
\caption{Comparison of the hardening functions in the forward model $( \mathcal{H}^{forward})$ and the misspecified inverted model $( \mathcal{H}^{inverse})$ problems. For the cases (a) without, and (b) with linear hardening included in the functional form of $ \mathcal{H}^{inverse}$. The curves are plotted to $\alpha = 0.35$, which is the maximum value of the isotropic hardening variable at the final applied displacement $u_y = 0.07$ in the forward problem.}
\label{fig:compre_hardening}
\end{figure}
\begin{table}[h!]
\centering
\caption{Comparison of displacement and load error of FEMU and VFM predictions for model form error due to hardening misspecification.}
\resizebox{\textwidth}{!}{
\begin{tabular}{>{\raggedright\arraybackslash}m{1.5cm} c c c c c c | c c}
\toprule
 & \multirow{3}{*}{$Y$ (MPa)} & \multirow{3}{*}{$K$ (MPa)} & \multirow{3}{*}{$S$ (MPa)} & \multirow{3}{*}{$D$} & \multirow{3}{*}{$A$ (MPa)} & \multirow{3}{*}{$n$} & \multicolumn{2}{c}{\textbf{Error}} \\
 \cline{8-9}
 &  &  & & & & & \textbf{Displacement} & \textbf{Load} \\
 &  &  & & & & & \textbf{(mm)$^2$} & \textbf{(N)$^2$} \\
\midrule
\textbf{Forward} &  330 & - & - & - & 1000 & 0.667  &  & \\
(\textbf{Truth})& & & & & & & &\\
 \rowcolor{gray!15}
\multicolumn{9}{c}{$\mathcal{H}^{inverse} = Y  + S [1 - \text{exp} (-D\alpha)]$} \\  
\textbf{FEMU} &  353.831 & - & 513.193 & 4.811 & - & -  & 3.718$\times 10^{-9}$ & 4.839$\times 10^{-1}$  \\
\textbf{VFM} &  342.765  & - & 349.310 & 8.961 & - & - & 6.902$\times 10^{-8}$ & 4.818$\times 10^{0}$  \\
& & & & & & &\\
 \rowcolor{gray!15}
\multicolumn{9}{c}{$\mathcal{H}^{inverse} = Y + K\alpha + S [1 - \text{exp} (-D\alpha)]$} \\  
\textbf{FEMU} &  343.307 & 1219.217 & 86.518 & 25.310 & - & -  & 2.845$\times 10^{-10}$ & 4.730$\times 10^{-2}$  \\
\textbf{VFM} &  336.870 & 1415.826 & 63.912 & 47.493 & - & - & 2.430$\times 10^{-9}$ & 5.041$\times 10^{-2}$  \\
\bottomrule
\end{tabular}%
}
\caption*{\footnotesize\emph{Note}: The error in load and displacement for FEMU is approximately one order of magnitude smaller compared to VFM.}
\label{tb:compteDispLoad_error}
\end{table}

In this section, we investigate the impact of model form errors arising from mismatched constitutive model specifications. Practically, all models are approximations of reality, so this issue will invariably be encountered with real data. To isolate their influence, we limit ourselves to noiseless data.
As a representative example, we restrict ourselves here to misspecification hardening behavior. 
In the present study, the underlying material behavior was assumed to follow a non-linear power law $\mathcal{H}^{forward}(\alpha)$ expressed as
\begin{equation}
     \mathcal{H}^{forward} = Y  + A \alpha^{n}.
\end{equation}
To evaluate the effect of misspecification, we perform the inverse analysis using the following hardening function:
\begin{equation}
     \mathcal{H}^{inverse} = Y  + K \alpha + S [1 - \text{exp} (-D\alpha)],
\end{equation}
We solve the inverse problem with FEMU-Adjoint and VFM-Adjoint, respectively, based on synthetic data generated using $\mathcal{H}^{forward}$. In the inverse problem, we investigate two cases of hardening $\mathcal{H}^{inverse}$: one without linear hardening $(K = 0)$ and one with linear hardening $(K \neq 0)$. To mitigate any bias in the calibrated material parameters that could arise from the initial guess, we generate 10 sets of random initial guesses for the parameters in the inverse problem. We then report the mean value of the calibrated parameters for both methods in Table \ref{tb:compteDispLoad_error}.

The hardening functions for the averaged inverted parameters with FEMU-Adjoint and VFM-Adjoint are shown in Fig. \ref{fig:compre_hardening} alongside the true hardening behavior. Figs. \ref{fig:compre_hardening_K_0} and \ref{fig:compre_hardening_K} depict the comparison of material hardening for $K = 0$ and $K \neq 0$, respectively. 

Both FEMU and VFM approximate the hardening behavior well for early plastic deformation quantified in terms of $\alpha$. The $\mathcal{H}^{inverse}$ with linear hardening $(K \neq 0)$ provides a closer match to the true (forward) hardening compared to the without linear hardening case. FEMU-Adjoint captures the material hardening behavior more closely than VFM-Adjoint in both cases, i.e., whether or not linear hardening is considered. In Table \ref{tb:compteDispLoad_error} we also report the squared errors of both the load and displacement objectives. It is important to note that we compute these errors using the mean value of the inverted parameters. As anticipated from the hardening plots, the error in load and displacement objectives for FEMU is considerably smaller (approximately one order of magnitude) compared to VFM.

\subsection{\textbf{E5:}  Model form error due to mesh density mismatch}

\begin{figure}
\begin{subfigure}{0.32\textwidth}
\centering
\includegraphics[width=0.99\linewidth]{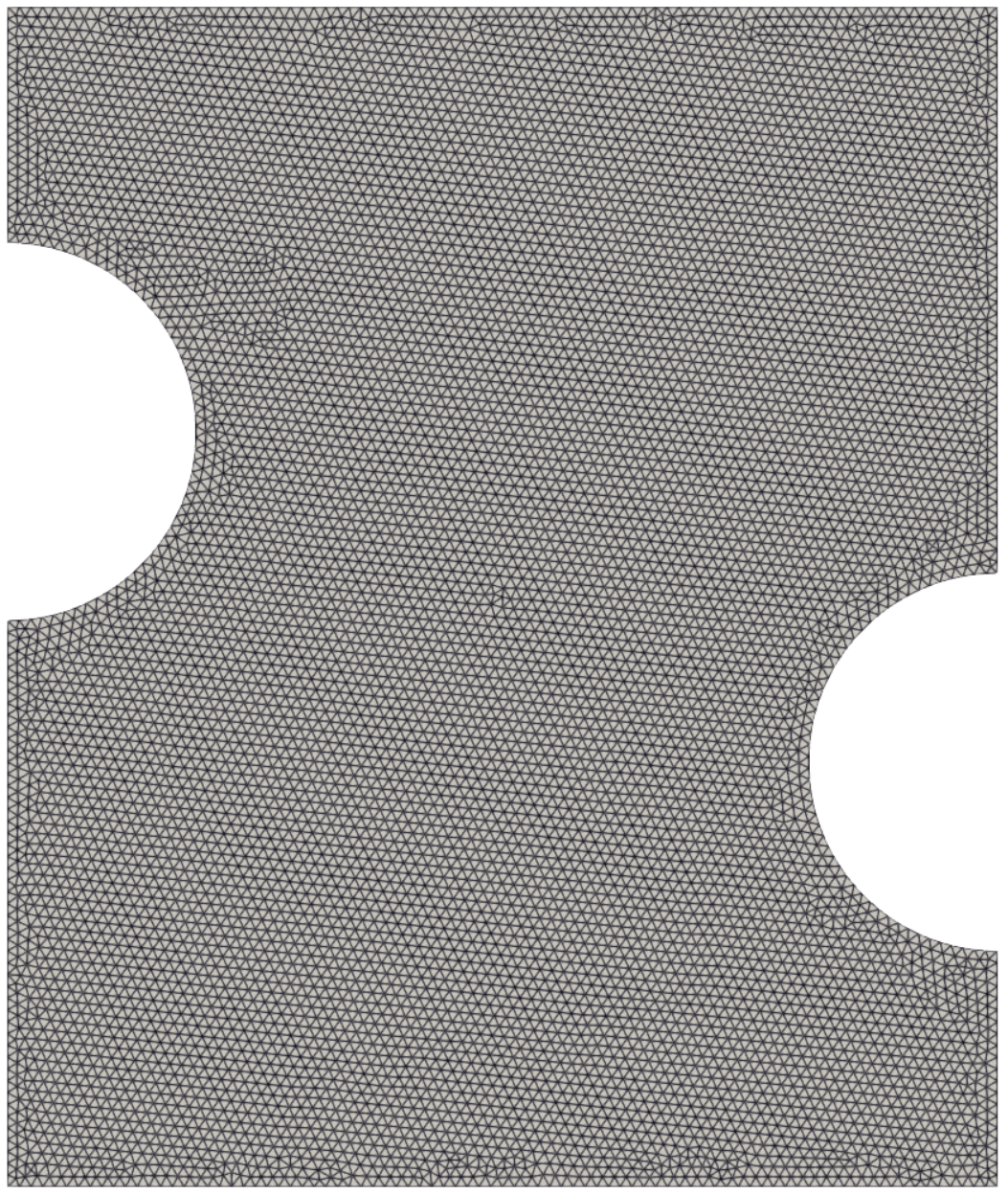}
\caption{}
\label{fig:compre_hardening_K_0}
\end{subfigure}
\hfill
\begin{subfigure}{0.32\textwidth}
\centering
\includegraphics[width=0.99\linewidth]{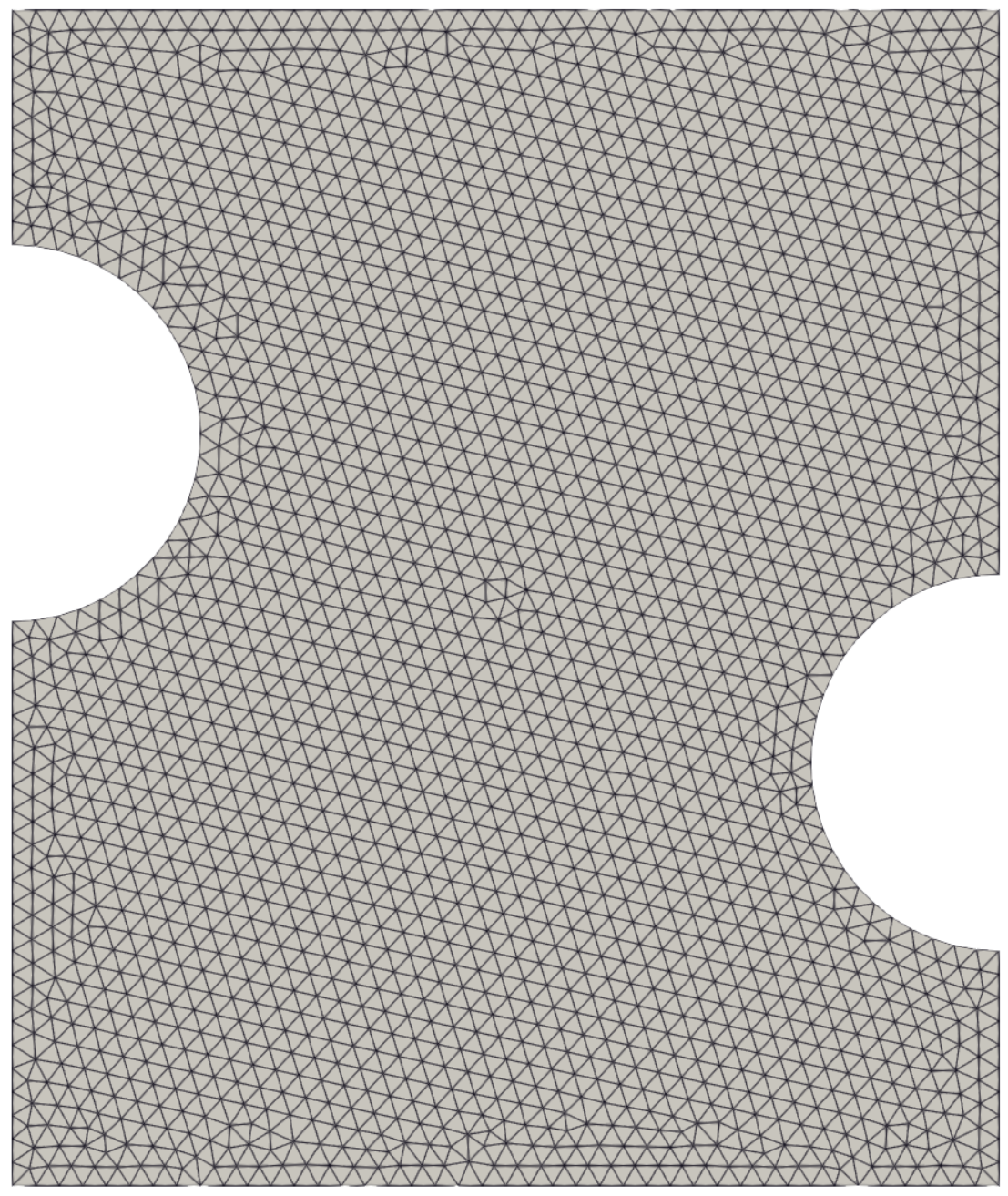}
\caption{}
\label{fig:compre_hardening_K}
\end{subfigure}
\hfill
\begin{subfigure}{0.32\textwidth}
\centering
\includegraphics[width=0.99\linewidth]{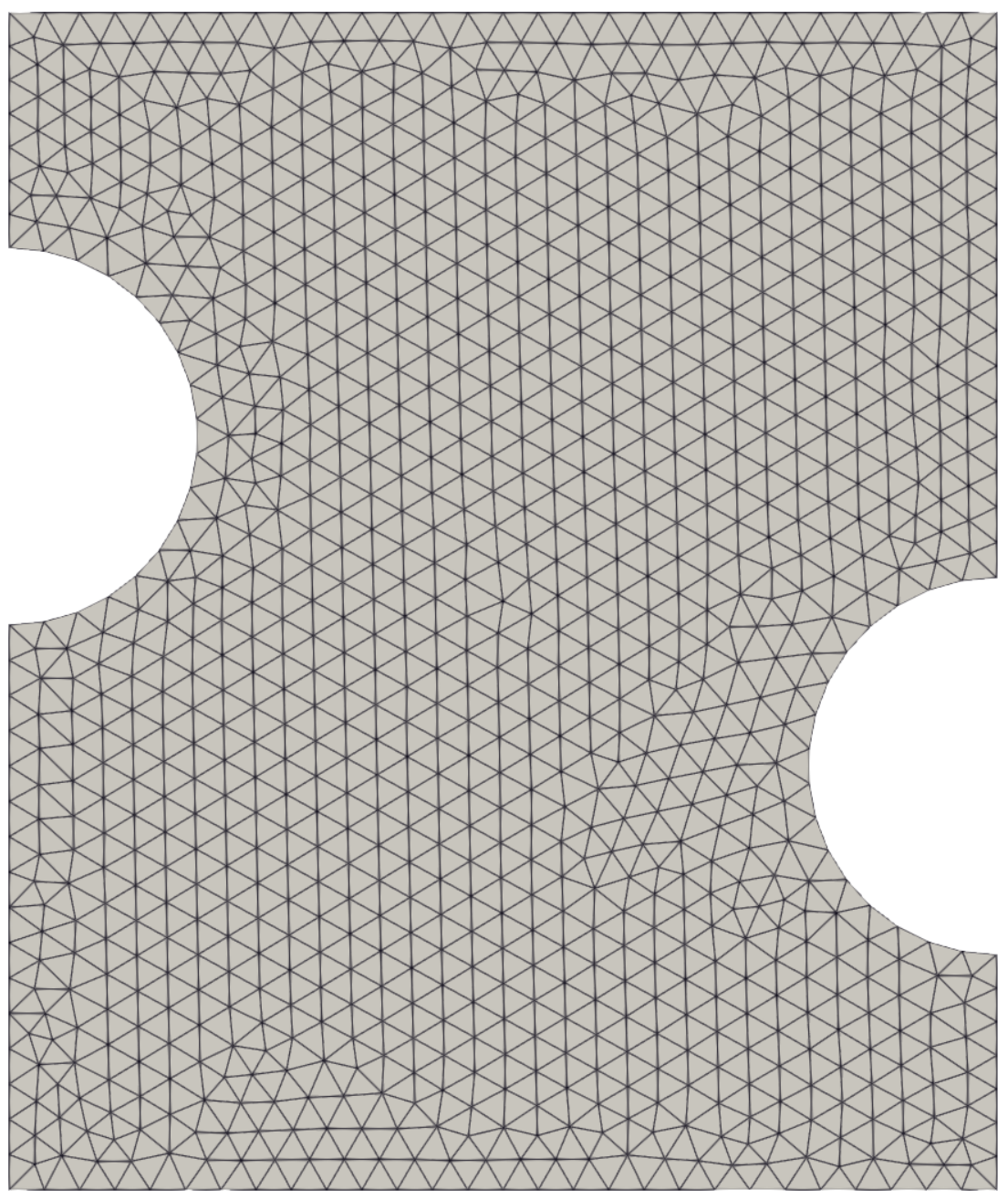}
\caption{}
\label{fig:compre_hardening_K}
\end{subfigure}
\caption{(a) Fine mesh used in the forward problem, element edge length x = 0.01 mm, to mimic high data resolution obtained from full-field measurements (b) First mesh used in the inverse problem with element edge length $\approx $2x larger than fine mesh (c) Second mesh used in the inverse problem with element edge length $\approx$ 3x larger.}
\label{fig:fine_coarse_mesh}
\end{figure}

\begin{figure}
\begin{subfigure}{0.49\textwidth}
\centering
\includegraphics[width=0.99\linewidth]{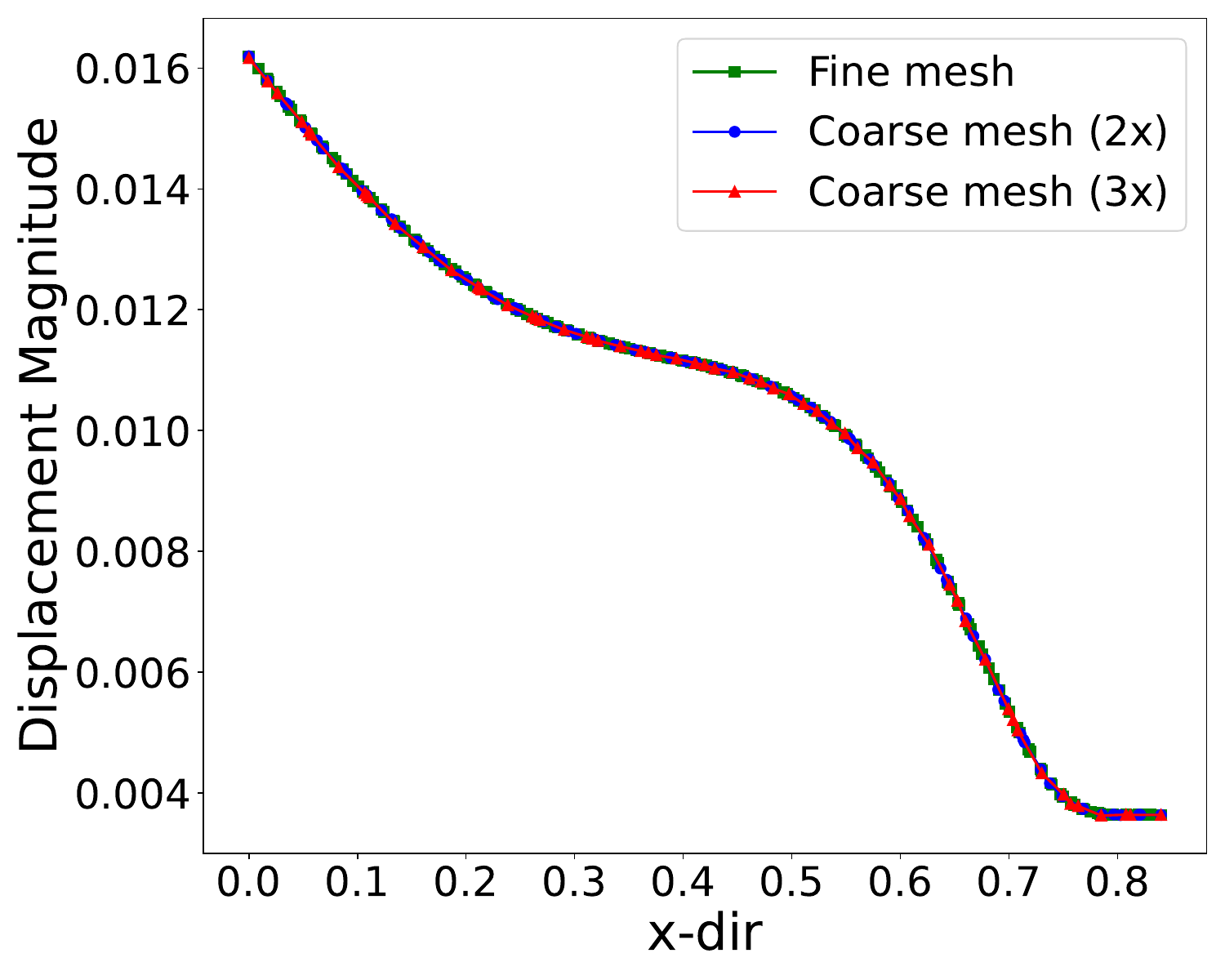}
\caption{y = 0.2}
\end{subfigure}
\begin{subfigure}{0.49\textwidth}
\centering
\includegraphics[width=0.99\linewidth]{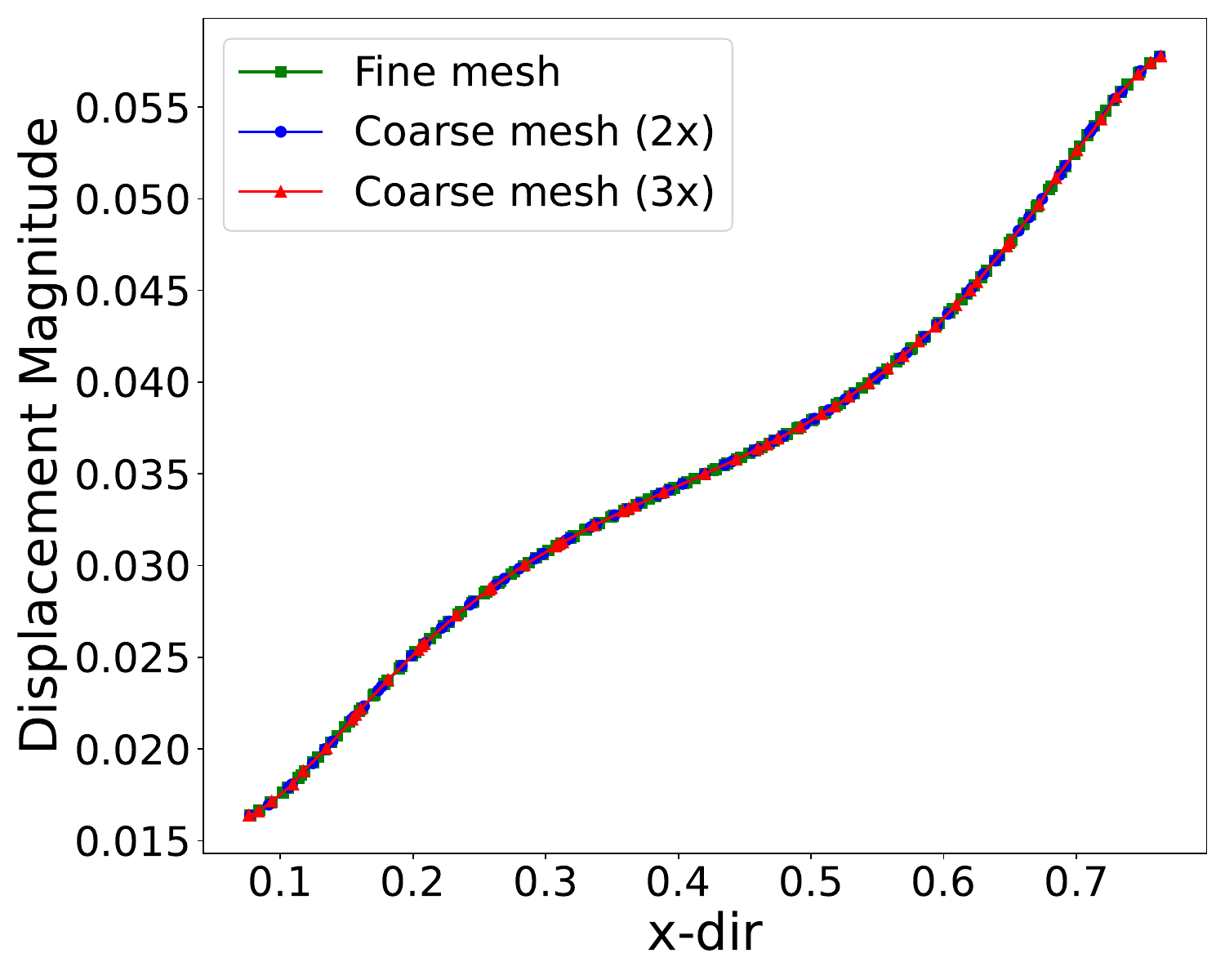}
\caption{y = 0.5}
\end{subfigure}

\vspace{0.5em}
\begin{subfigure}{0.49\textwidth}
\centering
\scalebox{1.0}{\includegraphics[width=0.99\linewidth]{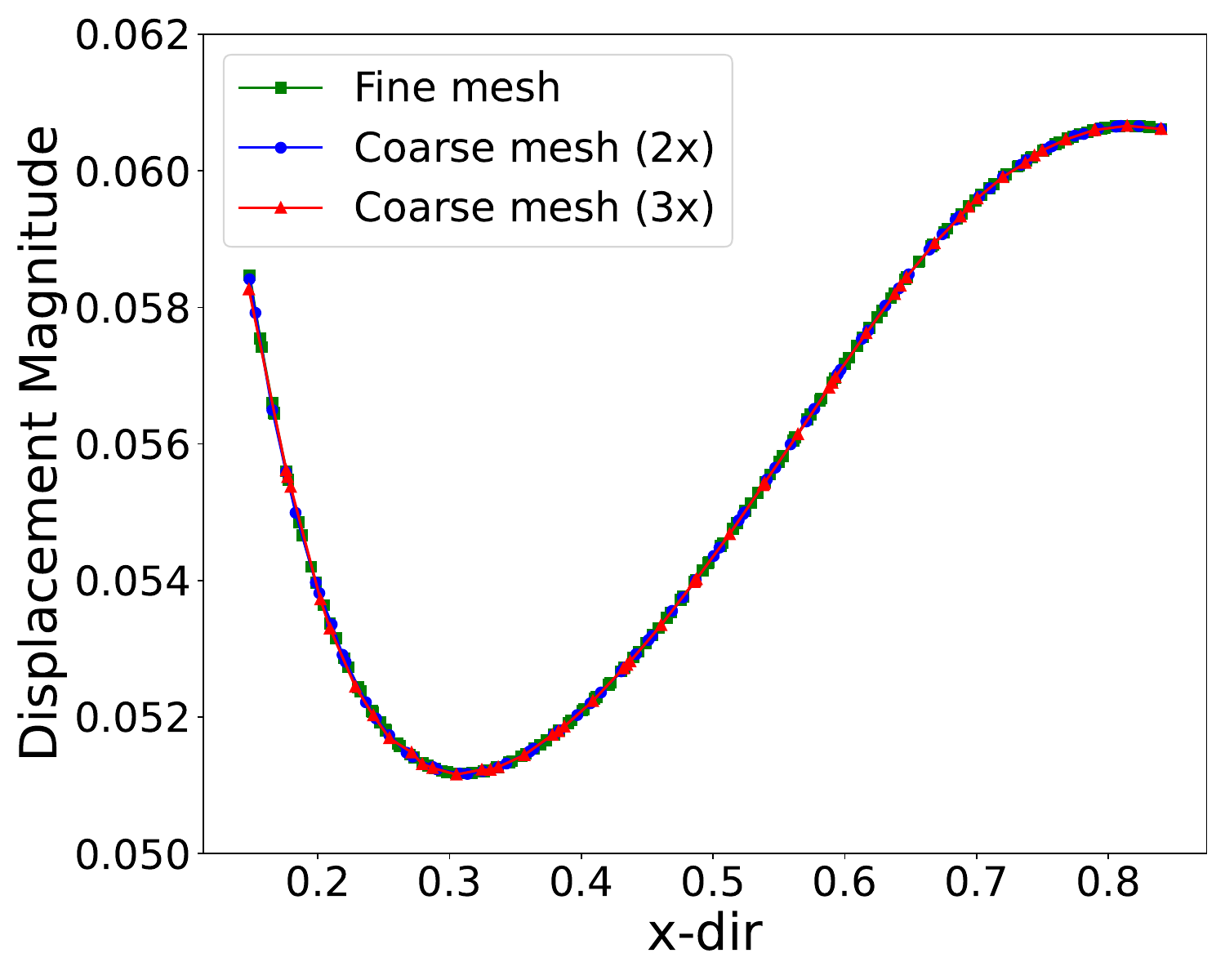}}
\caption{y = 0.7}
\end{subfigure}
\begin{subfigure}{0.49\textwidth}
\centering
\includegraphics[width=0.99\linewidth]{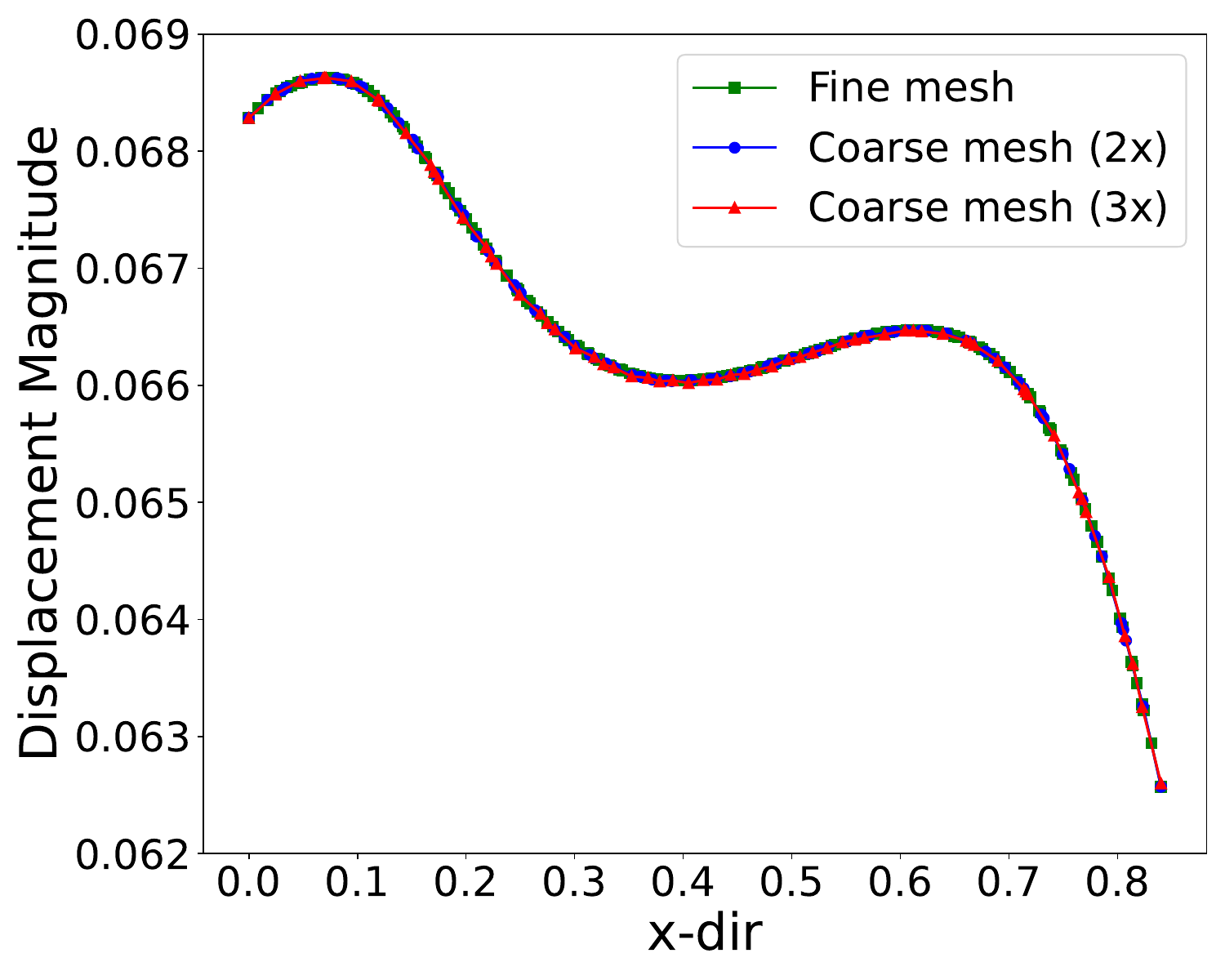}
\caption{y = 0.9}
\end{subfigure}

\caption{Mapping of synthetic displacement data from fine mesh used in the forward problem to coarse meshes used in the inverse problem for inverting the material parameters. Note `2x' and `3x' denote approximately two and three times the element edge length of the fine mesh, respectively.} 
\label{fig:mapping_fine_coarse_mesh}
\end{figure}

\begin{table}[h!]
\centering
\caption{Comparison of inverted parameters and error in load and displacement for FEMU-Adjoint and VFM-Adjoint, highlighting the impact of model form error caused by mapping high-resolution full-field data onto coarser meshes using discretization differences as a proxy for our synthetic study.}
\resizebox{13cm}{!}{
\begin{tabular}{>{\raggedright\arraybackslash}m{1.5cm} c c c | c c}
\toprule
 & \multirow{3}{*}{$Y$ (MPa)}  & \multirow{3}{*}{$S$ (MPa)} & \multirow{3}{*}{$D$} & \multicolumn{2}{c}{\textbf{Error}} \\
 \cline{5-6}
 &  &  & &  \textbf{Displacement} & \textbf{Load} \\
 &  &  & &   \textbf{(mm)$^2$} & \textbf{(N)$^2$} \\
\midrule
\textbf{Forward} &  330  &  1000 & 10  &  & \\
(\textbf{Truth})& & &  &&\\
 \rowcolor{gray!15}
\multicolumn{6}{c}{Nominal element edge length used in the inverse problem = 0.02 mm ($\approx$ 2x)} \\  
\textbf{FEMU-Adjoint} &  330.2722  & 997.2464 & 9.9664 & 1.7534$\times 10^{-10}$ & 3.6547$\times 10^{-5}$  \\
\textbf{VFM-Adjoint} &  329.8630   & 996.8598 & 10.0688 &  1.9205$\times 10^{-10}$ & 1.5000$\times 10^{-4}$  \\
& & & & \\
 \rowcolor{gray!15}
\multicolumn{6}{c}{Nominal element edge length used in the inverse problem = 0.03 mm ($\approx$ 3x)} \\   
\textbf{FEMU-Adjoint} &  331.7501 & 996.3535 & 9.8468   & 9.3938$\times 10^{-10}$ & 1.6903$\times 10^{-4}$  \\
\textbf{VFM-Adjoint} &  329.4042  & 988.5500 & 10.2358  & 1.1313$\times 10^{-9}$ & 9.6852$\times 10^{-4}$  \\
\bottomrule
\end{tabular}%
}
\caption*{\footnotesize\emph{Note}: Nominal element edge length used in the forward problem, $x = 0.01$ mm. Both FEMU-Adjoint and VFM-Adjoint successfully invert the material parameters with reasonable accuracy.}
\label{tb:fine_coarse_mesh}
\end{table}

Full-field deformation from DIC typically comes in the form of a dense point cloud, which is mapped onto a coarser point cloud of FE mesh nodes for inversion. Setting aside the issue of measuring a physical displacement field and comparing it to an FE model's predictions (which would be difficult to faithfully emulate in a synthetic study), we generate displacement data on a finer FE mesh and invert using a coarser one to emulate issues that arise from disparate spatial discretizations employed in the data-generating process and the computational model. More specifically, the inverse analysis is performed on the two distinct meshes with element edge lengths that are approximately two and three times larger than the mesh for the data-generation problem. The FE meshes used in the forward and inverse problems are shown in Fig \ref{fig:fine_coarse_mesh}. We use moving least squares to map the synthetic fine mesh data to the coarse FE meshes for the inverse problems. The results of this mapping are visualized in Fig. \ref{fig:mapping_fine_coarse_mesh} along different heights of the specimen. The inverted parameters and squared error in load and displacement objectives for both coarser meshed are reported in Table \ref{tb:fine_coarse_mesh}. Both FEMU-Adjoint and VFM-Adjoint successfully inverted the material parameters with reasonable accuracy for both coarse meshes, with FEMU performing slightly better, especially in the coarsest case. Therefore, we argue that both FEMU and VFM do not appear to be dramatically sensitive to mismatches between the data generation and inversion mesh sizes in this particular problem.

\subsection{\textbf{E6:} Model form error due to plane stress assumption}
\label{sec:3D-2DSurfData}

\begin{figure}
\includegraphics[width=0.98\linewidth,center]{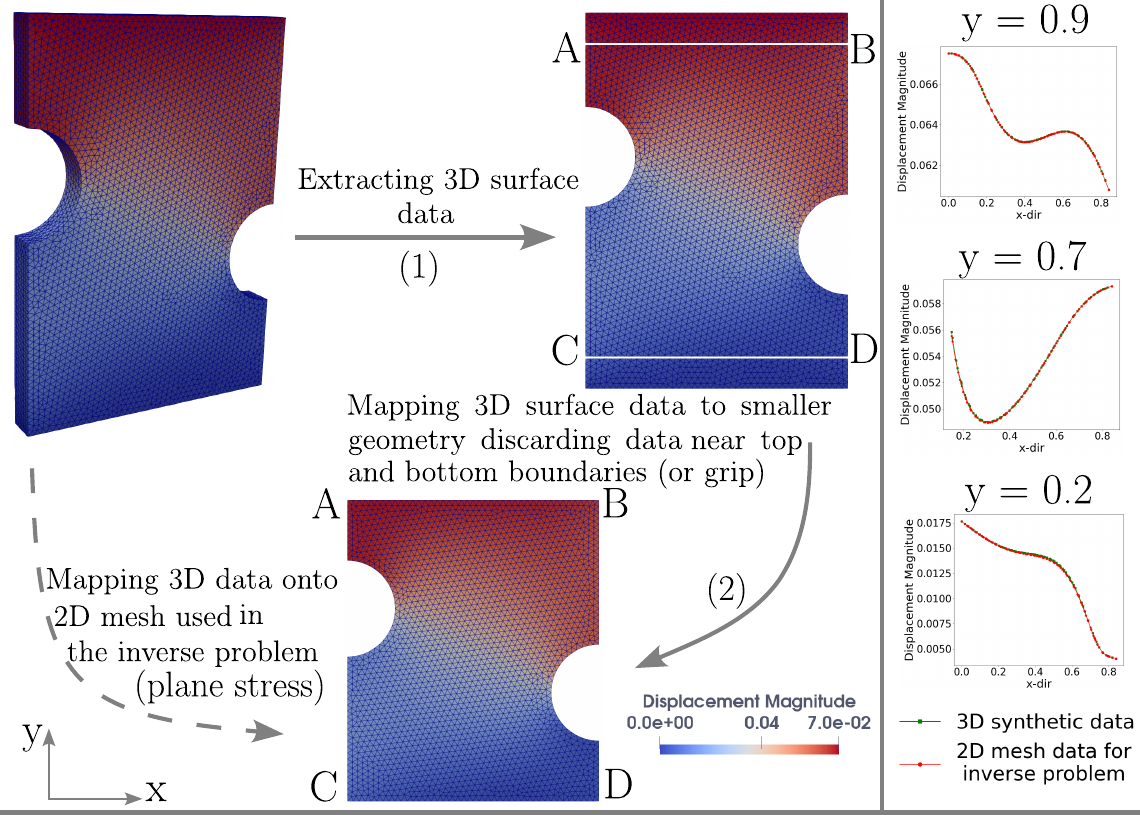}
 \caption{The schematic illustrates the projection of 3D synthetic data onto a 2D mesh, with data near the top and bottom edges removed to reduce the potential boundary effects caused by the mechanical grips holding the specimen during the experiment. The variable y denotes the height of the specimen from CD.}
 \label{fig:3D_2D_mesh_data_mapping}
 \end{figure}

\begin{table}[h!]
\centering
\caption{Comparison of inverted parameters and errors in load and displacement objectives for FEMU-Adjoint and VFM-Adjoint, highlighting model form error caused by plane stress assumptions for specimens with increasing thicknesses.}
\resizebox{\textwidth}{!}{
\begin{tabular}{>{\raggedright\arraybackslash}m{1.9cm} |c c c | c c c | c c}
\toprule
\rowcolor{gray!10}
& \multicolumn{2}{c}{$Y$ (MPa)} & \multicolumn{2}{c}{$S$ (MPa)} & \multicolumn{2}{c|}{$D$} & \multicolumn{2}{c}{\textbf{Error}} \\
\midrule
\textbf{Truth} & \multicolumn{2}{c}{330} & \multicolumn{2}{c}{1000} & \multicolumn{2}{c|}{10} & \textbf{Displacement} & \textbf{Load} \\
\cline{1-7}
\rowcolor{gray!15}
\textbf{Thickness} & \multicolumn{6}{c|}{\textbf{FEMU}} & (mm$^2$) & (N$^2$) \\
\midrule
$0.02$ & \multicolumn{2}{c}{330.1094} & \multicolumn{2}{c}{999.2902} & \multicolumn{2}{c|}{10.0083}  & 5.9528$\times 10^{-10}$ & 1.8474$\times 10^{-8}$ \\
\rowcolor{gray!7}
$0.04$ & \multicolumn{2}{c}{330.0133} & \multicolumn{2}{c}{999.4619} & \multicolumn{2}{c|}{10.0084} &  2.3390$\times 10^{-10}$ & 8.6563$\times 10^{-8}$ \\
$0.06$ & \multicolumn{2}{c}{330.0479} & \multicolumn{2}{c}{1000.0737} & \multicolumn{2}{c|}{9.9980}  & 1.6078$\times 10^{-10}$ & 3.9233$\times 10^{-8}$ \\
\rowcolor{gray!7}
$0.08$ & \multicolumn{2}{c}{330.0756} & \multicolumn{2}{c}{1000.0509} & \multicolumn{2}{c|}{9.9977} &  1.8799$\times 10^{-10}$ & 1.1790$\times 10^{-7}$ \\
$0.10$ & \multicolumn{2}{c}{330.0858} & \multicolumn{2}{c}{1000.6235} & \multicolumn{2}{c|}{9.9882}  & 1.5318$\times 10^{-10}$ & 2.1596$\times 10^{-7}$ \\
\rowcolor{gray!7}
$0.15$ & \multicolumn{2}{c}{330.0321} & \multicolumn{2}{c}{1000.0513} & \multicolumn{2}{c|}{9.9971} &  1.7082$\times 10^{-10}$ & 2.5201$\times 10^{-7}$ \\
\rowcolor{gray!7}
$0.20$ & \multicolumn{2}{c}{330.0484} & \multicolumn{2}{c}{1000.1525} & \multicolumn{2}{c|}{9.9961} &  1.6299$\times 10^{-10}$ & 1.3671$\times 10^{-6}$ \\
\midrule
\rowcolor{gray!15}
& \multicolumn{6}{c}{\textbf{Plane Stress}} & &\\
 & \multicolumn{3}{c|}{\textbf{FEMU}} & \multicolumn{3}{c|}{\textbf{VFM}} & \multicolumn{2}{c}{} \\
\cline{2-7}
\textbf{Thickness} & $Y$ (MPa) & $S$ (MPa) & $D$ & $Y$ (MPa) & $S$ (MPa) & $D$ &  &  \\
\midrule
$0.02$ & 330.0060 & 1002.2137 & 9.9679 &  330.9204 & 993.0635 & 10.1042  & 1.1917$\times 10^{-10}$ & 3.7399$\times 10^{-8}$ \\
\rowcolor{gray!5}
 \multicolumn{7}{c|}{} & (2.2645$\times 10^{-10}$) & (9.7693$\times 10^{-5}$) \\
$0.04$ & 330.0323 & 1001.6629 & 9.9833 &  330.7619 & 987.5222 & 10.1746 & 2.2681$\times 10^{-10}$ & 1.5034$\times 10^{-7}$ \\
\rowcolor{gray!5}
 \multicolumn{7}{c|}{} & (5.5985$\times 10^{-10}$) & (1.7902$\times 10^{-4}$) \\
$0.06$ & 330.1136 & 1003.0767 & 9.9627 &  329.9893 & 991.9646 & 10.0667 & 7.5208$\times 10^{-10}$ & 6.7612$\times 10^{-7}$ \\
\rowcolor{gray!5}
 \multicolumn{7}{c|}{} & (1.1123$\times 10^{-9}$) & (1.3024$\times 10^{-3}$) \\
$0.08$ & 330.1591 & 1004.0024 & 9.9544 &  328.6129 & 985.5031 & 10.1600 & 2.2478$\times 10^{-9}$ & 2.1042$\times 10^{-6}$ \\
\rowcolor{gray!5}
 \multicolumn{7}{c|}{} & (3.0108$\times 10^{-9}$) & (9.3614$\times 10^{-3}$) \\
$0.10$ & 330.1772 & 1005.3713 & 9.9409 &  327.8218 & 981.3063 & 10.2163 & 5.2322$\times 10^{-9}$ & 5.7854$\times 10^{-6}$ \\
\rowcolor{gray!5}
 \multicolumn{7}{c|}{} & (6.5646$\times 10^{-9}$) & (2.7087$\times 10^{-2}$) \\
$0.15$ & 330.3137 & 1007.5637 & 9.9439 &  325.4692 & 969.9175 & 10.4632 & 2.3166$\times 10^{-8}$ & 6.5427$\times 10^{-5}$ \\
\rowcolor{gray!5}
 \multicolumn{7}{c|}{} & (2.6352$\times 10^{-8}$) & (1.1584$\times 10^{-1}$) \\
$0.20$ & 330.8049 & 1010.3071 & 9.9685 & 323.7537 & 941.3201 & 11.1232 & 6.3410$\times 10^{-8}$ & 4.4206$\times 10^{-4}$ \\
\rowcolor{gray!5}
 \multicolumn{7}{c|}{} & (7.4366$\times 10^{-8}$) & (1.7525$\times 10^{-1}$) \\
\bottomrule
\end{tabular}%
}
\caption*{\footnotesize\emph{Note}: The errors in load and displacement for VFM are reported in parenthesis.}
\label{tb:3D-2D_surface_data}
\end{table}

In this section, we aim to study the impact of the plane stress assumption on the accuracy of the inverse problems with the aim of highlighting differences between FEMU and VFM. Full-field measurement data is usually acquired from thin three-dimensional (3D) mechanical characterization test specimens. If the thickness is small enough it is often assumed that the specimen can be treated as a structure under plane stress. This is generally done to speed up the necessary computational time. However, in the context of VFM, this assumption is furthermore necessary because the method requires full-field experimental data across the entire domain, i.e., VFM can not be used on partially observed data. On the other hand, FEMU is able to handle partially observed data and is therefore also able to invert material parameters in specimens where plane stress (or plane strain) conditions are not applicable. DIC provides data on a surface, so these considerations are frequently encountered in practice.

To isolate the effect, we again investigate the problem synthetically and assume a perfect model form (minus the 3D-plane stress difference studied here) and no noise. In particular, 
we create a 3D geometry featuring asymmetric notches and introduce a 0.1 mm offset at both the top and bottom edges, as in real experiments DIC measurements may not be available at the edge of the domain. Synthetic load and displacement data are then generated using a 3D finite element model that uses a mesh of this extended geometry with varying thicknesses. In particular, the longer side of the specimen is 1 unit long, see Figure \ref{fig:schematic}. By varying the thickness of the specimen between $0.02$ and $0.2$ we argue that we observe the range within which the plane stress assumption begins to become questionable.

Next, we extract the 3D surface data from the longer geometry, as shown in Fig. \ref{fig:3D_2D_mesh_data_mapping}. The geometry enclosed by the A-B-D-C-A path represents the geometry used in the inverse problem. Subsequently, we map the displacement data of the top surface of the 3D geometry onto a 2D mesh of the shorter geometry using generalized moving least squares. We thereby essentially discard the data near the top and bottom boundaries. We then use this 2D displacement data in the inverse problem for the case where we assume plane stress.

We furthermore also map the displacement data of the top surface of the 3D extended geometry to a 3D mesh associated with the shortened geometry. This data will be used for the 3D inverse problem where the data generation and inverse FE models are consistent. 

The inverted parameters, along with the squared errors in the load and displacement components of the objective function, are summarized in Table \ref{tb:3D-2D_surface_data}. Firstly, we can see that 3D FEMU-Adjoint provides material parameter estimates that closely match the true values, regardless of specimen thickness, demonstrating the robustness of FEMU in cases where the plane stress assumption becomes less appropriate (thickness $\approx  >0.1$). 

Next, we also report the inverted parameters under the assumption of plane stress case for both FEMU-Adjoint and VFM-Adjoint. We see that for specimens with smaller thicknesses ($\leq0.06$), both plane stress FEMU-Adjoint and VFM-Adjoint invert materials parameters close to the true value. For thicker specimens, however, the inverted parameters of VFM-Adjoint, in particular, seem to deviate more and more from the true parameter values. In contrast, for thicker specimens, the predictions of FEMU-Adjoint, even under the assumption of plane stress remain relatively close to the true values. They are, however, not as accurate as the 3D FEMU-Adjoint results.


\section{Discussion}
\label{sec:discussion}

Though we use synthetic FE data in place of full-field data obtained from experiments via DIC, we attempt to compare FEMU and VFM by mimicking practical scenarios through a series of numerical experiments that include solving the inverse problem by adding randomly generated Gaussian noise to load and displacement data, testing the sensitivity to initial guesses, constitutive model misspecifications, and model-form errors due to data resolution mismatch and plane stress assumptions. 
We can draw the following conclusions. Under ideal conditions, both FEMU and VFM successfully invert the true material parameters. However, in cases where model-form errors are present, FEMU appears to be more accurate at the expense of a considerably greater computational cost relative to VFM. Additionally, FEMU seems more robust in the case of noisy data. VFM benefits from a preprocessing step of filtering or smoothing when noise is present. 
 
Furthermore, VFM is also more affected by constitutive law misspecification, which we have here restricted to the hardening law. 
Resolution differences between the data generation and inverse spatial discretizations did not appear to have a sizable impact on the performance of either method. 

We highlight the versatility of FEMU by showing that the method is able to accurately invert material parameters from surface measurements in cases where the plane stress assumption is no longer reasonable due to increasing sample thickness. This might be important in some use cases, as VFM requires deformation data at all points in the computational domain to compute the virtual work, and therefore cannot be used in problems where the plane stress assumption is invalid.

Finally, a notable limitation of our study is its restriction to numerical results with synthetic data. Several issues not present in our study become critically important when performing constitutive model calibration with real experimental data (e.g. boundary conditions issues, various flavors of model form errors). This choice, however, allowed us to compare inversion results against a ground truth in a controlled manner for both FEMU and VFM inverse methods.

\section{Conclusion}
\label{sec:conclusion}

In this article, we have compared two widely used inverse methods developed by the experimental mechanics community, Finite Element Model Updating (FEMU) and the Virtual Fields Method (VFM), to explore their accuracy, sensitivity, and viability under multiple scenarios motivated by issues that arise in real calibration studies. Existing reviews and comparisons in the literature employ finite difference schemes to approximate objective function gradients. However, both methods can leverage forward and adjoint sensitivities to compute gradients at significantly lower computational costs compared to finite difference approximations. To our knowledge, for the first time, we developed forward and adjoint local sensitivity analyses to compute the gradient in VFM and verified their correctness. However, we have also shown that the computational cost of FEMU is considerably higher than that of VFM. Therefore, it is up to the engineer to decide if the generally greater accuracy of FEMU warrants the considerably more time-consuming inversion process for their particular calibration problem.

\section*{Acknowledgments}
D.T.\ Seidl and B.N.\ Granzow were supported by the Advanced Simulation and Computing program at Sandia National Laboratories, a multimission laboratory managed and operated by National Technology and Engineering Solutions of Sandia LLC (NTESS), a wholly owned subsidiary of Honeywell International Inc. for the U.S.  Department of Energy’s National Nuclear Security Administration under contract DE-NA0003525.  This written work is authored by an employee of NTESS. The employee, not NTESS, owns the right, title, and interest in and to the written work and is responsible for its contents. Any subjective views or opinions that might be expressed in the written work do not necessarily represent the views of the U.S. Government. The publisher acknowledges that the U.S. Government retains a non-exclusive, paid-up, irrevocable, world-wide license to publish or reproduce the published form of this written work or allow others to do so, for U.S. Government purposes. The DOE will provide public access to results of federally sponsored research in accordance with the DOE Public Access Plan.

\section*{Code availability}

The source code and files needed to generate our numerical results are available at \url{https://github.com/sanjais37/calibr8}.

\begin{appendices}
\label{appen_formulation}

\section{Global and local residuals}
\label{app:residuals}

In this appendix, we describe the finite deformation plane stress model plasticity model utilized in this work. The model is completely specified through the definition of global and local residuals, which come from the equilibrium PDE and constitutive model evolution equations. This model is the plane stress specialization of the 3D model presented in \cite{seidl2022calibration}, which was based on \cite{simo2006computational,borden2016phase}. 

The formulation assumes a multiplicative decomposition of the deformation gradient $\bs{F}$ into elastic and plastic components such that $\bs{F} = \bs{F}^e \bs{F}^p$ and the Jacobian determinant of deformation gradient $J := \text{det} [ \bs{F} ] = J^eJ^p$. The local residual explicitly enforces the isochoric assumption for plastic flow such that $J^p = 1$ and $J^e = J$.

We recall the constitutive model that forms the basis for the local residual:

\begin{align}
\bs{\tau} &= \bs{s} - J p \bs{I} , \label{eq:kirchhoff_stress} \\
-p &:= \frac{\kappa}{2} (J^2 - 1) / J ,\\
\bs{s} &:= \text{dev} \left[ \bs{\tau} \right] = \mu \ \text{dev} \left[ \bar{\bs{b}}^e \right] = \mu \bzeta,
\label{eq:dev_kirchhoff}
\end{align}

\noindent where $\mu$ is the shear modulus, $\kappa$ in the bulk modulus, and $\bar{\bs{b}}^e := J^{-\frac{2}{3}} \bs{b}^e$ is the volume-preserving part of the elastic left Cauchy-Green strain tensor $\bs{b}^e := \bs{F}^e \left({\bs{F}^e}\right)^T$. We decompose $\bar{\bs{b}}^e$ into spherical and deviatoric parts $\Ie$ and $\bzeta$, respectively, such that $\bar{\bs{b}}^e = \bzeta + \Ie \bs{I}$ by means of the operators $\text{tr} \left[\cdot\right]$ (trace) and $\text{dev} \left[ \cdot \right] := \left[\cdot\right] - \frac13 \text{tr} \left[\cdot \right] \bs{I}$ (deviator). We perform a similar decomposition for the Kirchhoff stress tensor $\bs{\tau} := J \bs{\sigma}$ in Eq. \eqref{eq:kirchhoff_stress}, where
$p$ is the pressure and $\bs{s}$ is the deviatoric part of $\bs{\tau}$.

For the plasticity model, we utilize a $J_2$ effective stress $\phi (\bs{s}) = \sqrt{\frac{3}{2}} \| \bs{s} \|$,  isotropic hardening with state variable $\alpha$ and generic hardening function $H(\alpha)$, yield stress $Y$, and an associative flow rule.

The plane stress constraint requires $\sigma_{13} = \sigma_{23} = \sigma_{33} = 0$. One way to ensure that the out-of-plane shear stresses vanish is to impose $F_{13} = F_{23} = F_{32} = F_{31} = 0$, which results in

\begin{equation}
    [\bs{F}]_{ij} = 
\begin{bmatrix}
F_{11} & F_{12} & 0 \\
F_{21} & F_{22} & 0 \\
0 & 0 & F_{33}
\end{bmatrix} \quad \text{and} \quad
J = \text{det}[\F_{\text{2D}}] F_{33}.
\end{equation}

\noindent We use Eqs. \eqref{eq:kirchhoff_stress}-\eqref{eq:dev_kirchhoff} with the remaining constraint to obtain

\begin{equation}
    \begin{aligned}
        \tau_{33} &= J \sigma_{33} = 0 = \mu \bar{\zeta}_{33}^e + \frac{\kappa}{2} \left(J^2 - 1\right),\\
        &= \mu \bar{\zeta}_{33}^e + \frac{\kappa}{2} \left[\left(F_{33} J_{2D}\right)^2 - 1\right],\\
        \implies F_{33} &= \left( \frac{1 - \frac{2\mu \bar{\zeta}_{33}^e}{\kappa}}{J_{2D}^2} \right)^{\frac{1}{2}} = \left( \frac{1 + \frac{2\mu \left(\nbzeta_{11} + \nbzeta_{22}\right)}{\kappa}}{J^2_{2D}}   \right)^{\frac{1}{2}},
    \end{aligned}
\end{equation}

\noindent and note that by construction $\nbzeta_{33} = - \left( \nbzeta_{11} + \nbzeta_{22}\right).$

In section \ref{sec:variational_problem}, we stated the a generic global residual for a single load step. We now specialize this to finite deformation plane stress. First, we make the generic stress measure $\bSigma$ the first Piola-Kirchhoff stress $\bs{P} = J \bs{\sigma} \bs{F}^{-T}$ and obtain

\begin{equation}
R^a_i := \int_{A} \partial_j N_a(\bx)  \bs{P}_{ij}(\bu^h,\bxi;\bp) \ d\Omega - \sum_{i=1}^\nsd \int_{\Gamma_{H_i}} H_i N_a(\bx) \ d\Gamma = 0,
\label{eq:full_plane_stress_discrete_global_residual}
\end{equation}

\noindent where all integrals are taken over the reference configuration and $\bs{H}$ is the reference configuration traction. Moving forward, we will drop the $h$ superscript for the finite element displacement field and suppress arguments for brevity.

Next, we only consider the in-plane components of displacement global state variable such that $\bu := \{ u_1, u_2\}$ and $i, j, k = 1, n_{sd}$, where $n_{sd} = 2$. The integral over the volume in the reference configuration may be written as an internal over the area if we assume a uniform initial thickness in the out-of-plane direction $T_0$. The plane stress global residual is

\begin{equation}
R^a_i := \int_{A} \partial_j N_a  J_{\text{2D}} F_{33} \sigma_{ik} [\bs{F}^{-T}_{\text{2D}}]_{kj} T_0 \ dA - \sum_{i=1}^\nsd \int_{\Gamma_{H_i}} H_i N_a \ d\Gamma = 0.
\label{eq:simplified_plane_stress_discrete_global_residual}
\end{equation}

As in our previous work, we employ a backward Euler time discretization for load steps $n=1, \ldots n_L$. The global residual (c.f. Eq. \eqref{eq:simplified_plane_stress_discrete_global_residual}) is satisfied at each load step, i.e. 

\begin{equation}
\bs{R}^n\left(\bu^n, \bxi^n, \bp\right) = 0, \ n = 1, \ldots, n_L. 
\end{equation}

The local residual describes the time evolution of six local state variables $\bxi := \{\nbzeta_{11}, \nbzeta_{12}, \nbzeta_{22}, \Ie, \alpha, F_{33} \}$. An independent local residual is present at each integration point in the mesh, but we employ a single integration point per element and thus use the index $e = 1, \ldots, n_{el}$ for discrete local residuals $\bC^n_e$, local state variables $\bxi^n_e$, and global state variables interpolated to integration points $\bu^n_e$ at load step $n$.

The plane stress local residual is a minor modification from the 3D version in \cite{seidl2022calibration}. Two important differences include the focus on the in-plane components of $\bzeta$, and we must solve for $F_{33}$. At each step $n$, we compute the relative deformation gradient $\bs{f}^n$
and its isochoric version $\bar{\bs{f}}^n$ according to

\begin{align}
\f^n &:= \F^n  \left(\F^{n-1} \right)^{-1},
\label{eq:relative_f}\\
\bar{\f}^n &:= \text{det} \left( \f^n \right)^{-\frac{1}{3}} \f^n.
\end{align}

\noindent Next, we evaluate trial quantities where it is assumed that the step is elastic, such that

\begin{align}
\alpha^n_{\text{trial}} &= \alpha^{n-1}, \\
\left(\bebar \right)^n_{\text{trial}} &= \bar{\f}^n \left( \left( \bzeta \right)^{n-1} + (\Ie)^{n-1} \bs{I} \right) \left( \bar{\f}^n \right)^T, \\
\left(\bzeta \right)^n_{\text{trial}} &= \text{dev} \left[ \left(\bebar\right)^n_{\text{trial}}  \right], \\
\left(\Ie \right)^n_{\text{trial}} &= \frac{1}{3} \text{tr} \left[ \left(\bebar\right)^n_{\text{trial}}  \right], \\
\bs{s}^n_{\text{trial}}  &= \mu \left(\bzeta \right)^n_{\text{trial}}, \label{eq:sn_trial}
\end{align} 

\noindent where all tensor quantities in Eqs. \eqref{eq:relative_f}-\eqref{eq:sn_trial} are 3D.

We then evaluate the yield surface $f(\bs{s}, \alpha) = \|\bs{s}\| - \sqrt{\frac{2}{3}} H(\alpha)$ at the trial quantities for step $n$ to determine which branch to take. The plastic branch is chosen when the trial value for the yield surface is positive. Otherwise, the elastic branch is taken and the local state variables for step $n$ are the trial values.

The discrete local residual is

\begin{equation}
\bs{C}_e^n :=
\begin{cases} 
f^n_{\text{trial}} \leq 0: 
\left\{
\begin{aligned}
&(\bzeta)^n - (\bzeta)^n_{\text{trial}}, \\ 
&({\bar{I}}^e)^n - ({\bar{I}}^e)_{\text{trial}}^n, \\ 
&\alpha^n - \alpha^n_{\text{trial}}.
\end{aligned}
\right. \\[25pt]
f^n_{\text{trial}} > 0: 
\left\{
\begin{aligned}
&(\bzeta)^n - (\bzeta)^n_{\text{trial}} + 2 \sqrt{\frac{3}{2}} (\alpha^n - \alpha^{n-1}) ({\bar{I}^e})^n \frac{\bs{s}^n}{\|\bs{s}^n\|}, \\ 
&\det \bigg[ (\bs{\bar{\zeta}}^e)^n + ({\bar{I}}^e)^n \bs{I} \bigg] - 1 \quad \text{(full 3D tensors)}, \\ 
&f(\bs{s}^n, \alpha^n).
\end{aligned}
\right.\\[45pt]
\begin{aligned}
    F_{33} - \left( \frac{1 + \frac{2\mu \left(\nbzeta_{11} + \nbzeta_{22}\right)}{\kappa}}{J^2_{2D}}   \right)^{\frac{1}{2}} \text{(always holds)}.
\end{aligned}
\end{cases}  
\end{equation}

The local residuals are satisfied at every integration point and for all times, such that
\begin{equation}
\bs{C}^n_e\left(\bu^n, \bu^{n-1}_e, \bxi^n_e, \bxi^{n-1}_e, \bp\right) = 0, \ e=1, \ldots, n_{el}, n = 1, \ldots, n_L. 
\end{equation}

\clearpage
\section{Comparison of material parameter calibration using different virtual fields}
\label{appen:different_VF}

\begin{table}[h!]
\centering
\caption{Comparison of calibrated material parameters in VFM for a plate with an asymmetric notch subjected to finite strain elasto-plastic deformation, using noisy synthetic data and different virtual fields.}
\begin{tabular}{>{\raggedright\arraybackslash}m{1.7cm} |c c c | c c c}
\toprule
\rowcolor{gray!10}
& \multicolumn{2}{c}{$Y$ (MPa)} & \multicolumn{2}{c}{$S$ (MPa)} & \multicolumn{2}{c}{$D$}\\
\midrule
\textbf{Truth} &  \multicolumn{2}{c}{330} & \multicolumn{2}{c}{1000} & \multicolumn{2}{c}{10} \\
\midrule
\rowcolor{gray!15}
\multicolumn{7}{c}{\textbf{calibrated parameters using unfiltered noisy synthetic data }} \\
\textbf{Virtual Fields} & \multicolumn{3}{c|}{$\vfvec_{\text{x}} = \cos(\pi (\text{y}-\frac{1}{2})), \vfvec_{\text{y}} = \text{y}$} & \multicolumn{3}{c}{{$\vfvec_{\text{x}} = \cos(\pi (\text{y}-\frac{1}{2})), \vfvec_{\text{y}} = \text{y}^2$}} \\
\cline{1-7}
\textbf{DNSF} & $Y$ (MPa) & $S$ (MPa) & $D$ & $Y$ (MPa) & $S$ (MPa) & $D$ \\
\midrule
0x & 331.1426 & 1001.9660 & 9.9064 &  331.1416 & 1001.9545 & 9.9066 \\
\rowcolor{gray!7}
0.5x & 334.8144 & 1032.7531 & 9.3811 &  335.0300 & 1037.2422 & 9.3195 \\
1x & 349.5128 & 1199.6416 & 7.3218 &  350.0212 & 1216.3747 & 7.1817 \\
\rowcolor{gray!7}
2x & 394.2154 & \underline{2000} & 3.4151 &  394.0667 & \underline{2000} & 3.4152 \\
5x & 479.8452 & \underline{2000} & 2.3719 &  478.6309 & \underline{2000} & 2.3789 \\

\bottomrule
\end{tabular}%
\caption*{\footnotesize\emph{Note}: The mean values of the calibrated parameters are reported pertaining to the ten sets of random initial guesses. Load noise scale factor is 1x viz. noise in the load is at the base level ${\epsilon}^{\text{F}}_{\text{noise}} = 0.25$. DNSF: Displacement noise scale factor. For the displacement base level noise see Eq. \eqnref{eq:disp_noise_base}. The calibrated parameters that are hitting the bound are indicated by $\underline{(\cdot)}$.}
\label{tb:different_choices_VF}
\end{table}

\clearpage
\section{Comparison of unfiltered vs. filtered noisy displacement data}
\label{app:filtered_unfiltered_noisy_data}

\begin{figure}[!bh]
\begin{subfigure}{0.49\textwidth}
\centering
\includegraphics[width=0.99\linewidth]{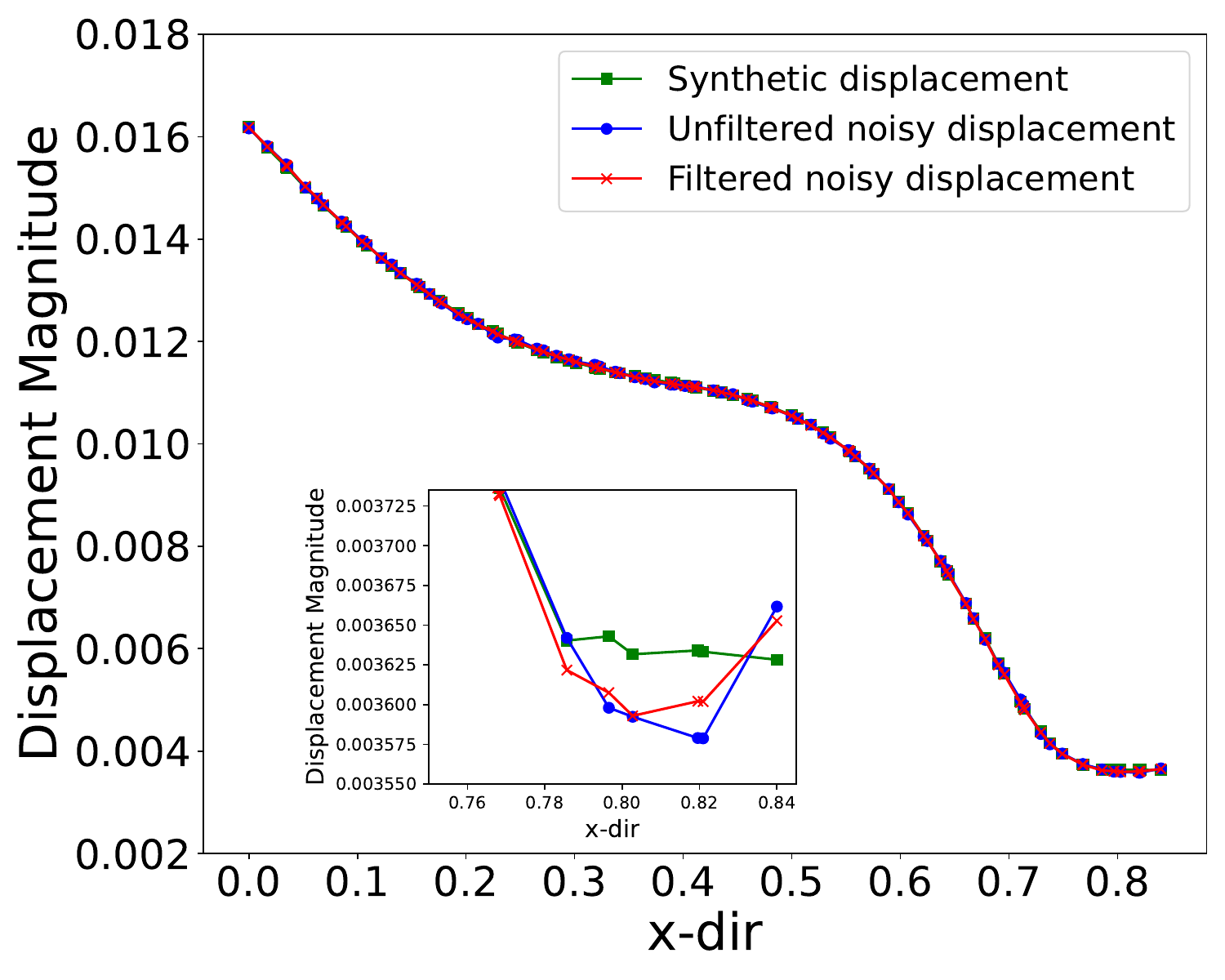}
\caption{y = 0.2}
\end{subfigure}
\begin{subfigure}{0.49\textwidth}
\centering
\includegraphics[width=0.99\linewidth]{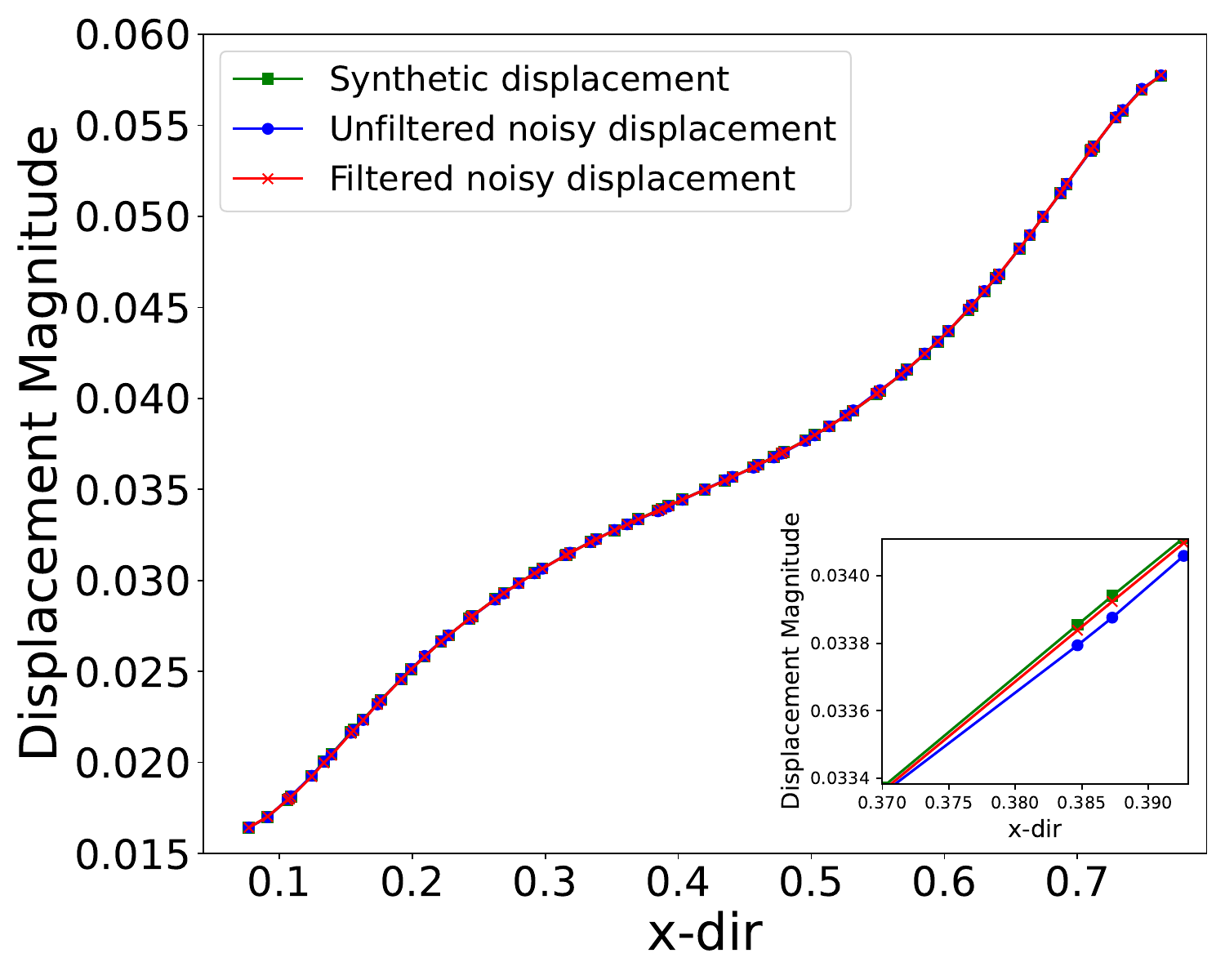}
\caption{y = 0.5}
\end{subfigure}

\vspace{0.5em}
\begin{subfigure}{0.49\textwidth}
\centering
\scalebox{1.0}{\includegraphics[width=0.99\linewidth]{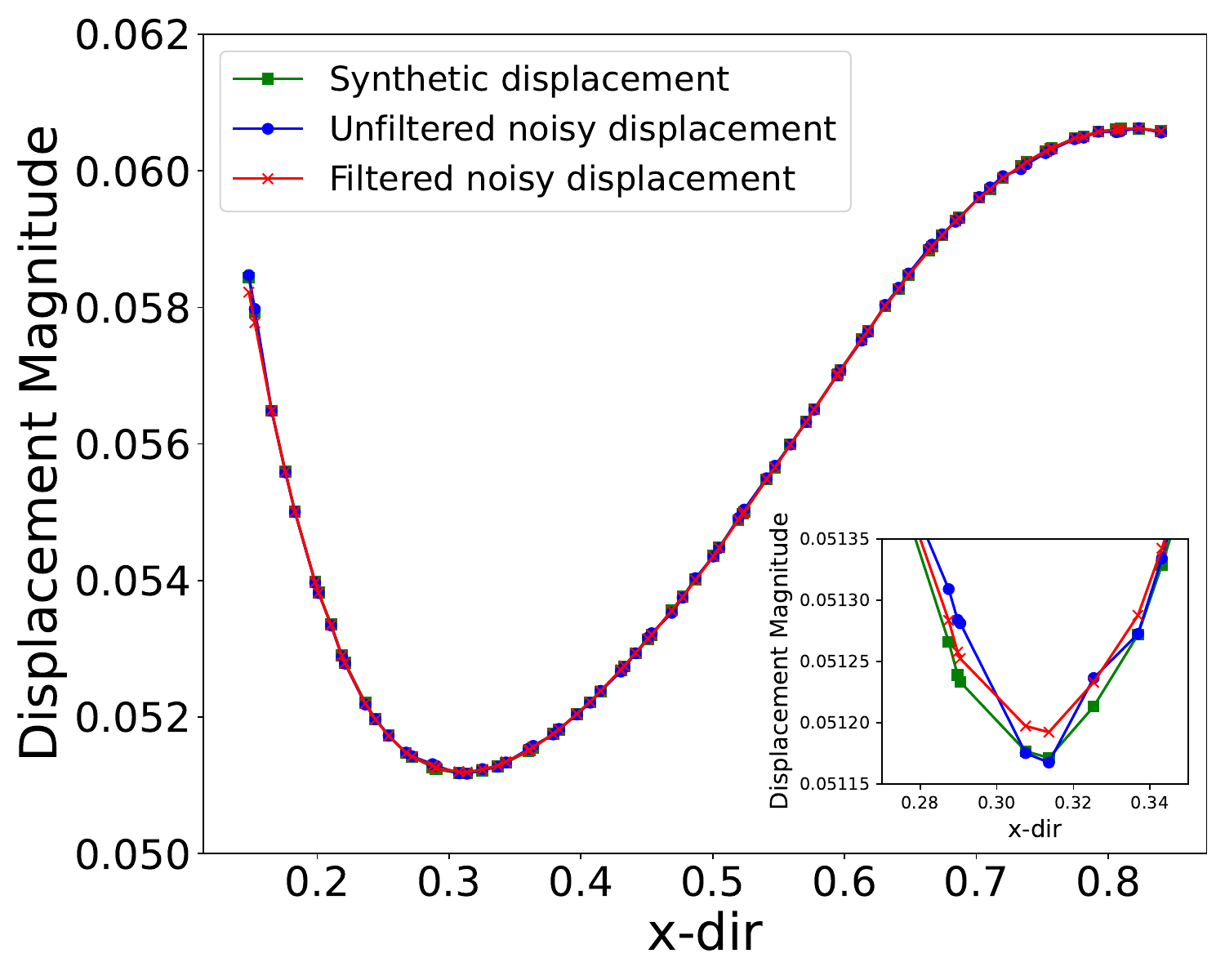}}
\caption{y = 0.7}
\label{fig:smoothing_load_disp_LNSF_0}
\end{subfigure}
\begin{subfigure}{0.49\textwidth}
\centering
\includegraphics[width=0.99\linewidth]{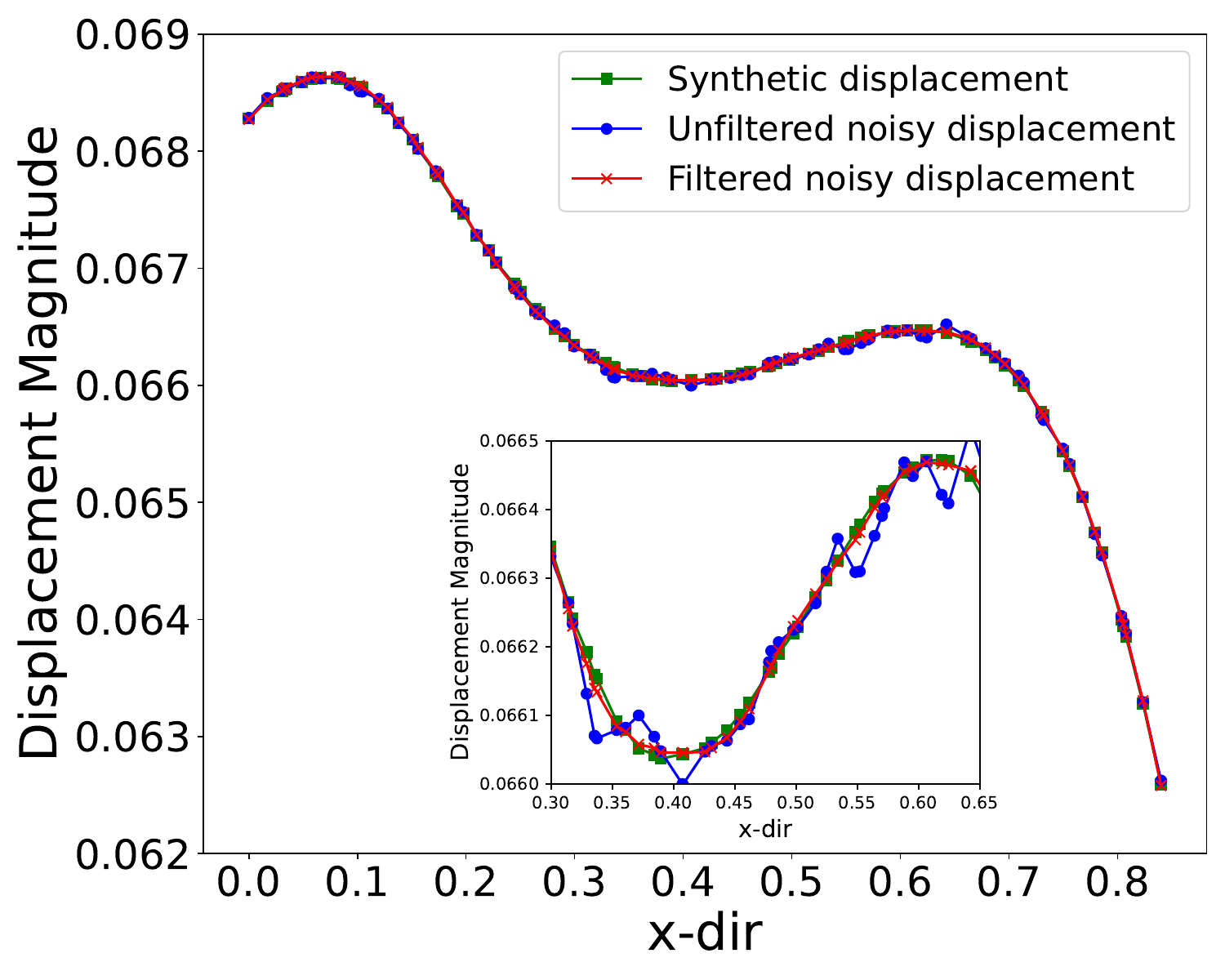}
\caption{y = 0.9}
\label{fig:smoothing_load_disp_LNSF_1}
\end{subfigure}

\caption{Plot of displacement data along the width (x-dir) at different heights (y), illustrating a comparison between filtered and unfiltered noisy data, alongside synthetic displacement. Displacement noise scale factor = 5x. All units are in mm.} 
\label{fig:smoothing}
\end{figure}

\end{appendices}

\clearpage
\bibliography{femuVfm}  
\end{document}